\documentclass[aps, prb, unsortedaddress,floatfix, nofootinbib, twocolumn, superscriptaddress,floatfix, longbibliography, showkeys]{revtex4-2}
\usepackage{rotating}
\usepackage[dvipsnames, pdftex]{xcolor}
\usepackage{graphicx, color, soul}
\graphicspath{{./}{../}}
\usepackage{tikz}
\usepackage{quantikz}
\providecommand{\wireoverride}[1]{}
\usepackage{amssymb}
\usepackage{ulem}
\usepackage{multirow}
\usepackage{bm}
\usepackage[mathscr]{eucal}
\usepackage{physics}
\usepackage{multirow}
\usepackage{subfiles}
\definecolor{NavyBlue}{HTML}{000080}
\usepackage[colorlinks, breaklinks,
linkcolor=NavyBlue,
citecolor=NavyBlue,
urlcolor=NavyBlue]{hyperref}
\usepackage{tabularx}
\usepackage[all]{hypcap}
\usepackage{tikz}
\usepackage[titletoc]{appendix}
\usepackage{titletoc}
\usepackage{float}
\usepackage[linesnumbered,ruled,vlined]{algorithm2e}
\usepackage{subcaption}
\usepackage{derivative}
\usepackage{amsmath,amssymb}
\usepackage{braket}
\usepackage[T1]{fontenc}
\usepackage{algorithm2e}
\usepackage{url}
\usepackage{hyperref}
\usepackage{orcidlink}
\usepackage{booktabs}
\usepackage{balance}
\usepackage[most]{tcolorbox}

\newcolumntype{M}[1]{>{\centering\arraybackslash}m{#1}}

\definecolor{C1}{HTML}{C49062}
\definecolor{C2}{HTML}{49798F}
\newtcolorbox{resultbox}[1]{
    breakable,
    colback=black!2,
    colframe=C2,
    colbacktitle=C2!12,
    coltitle=black,
    fonttitle=\bfseries,
    boxrule=0.5pt,
    arc=1mm,
    left=5pt,
    right=5pt,
    top=4pt,
    bottom=4pt,
    before skip=6pt,
    after skip=6pt,
    title={#1}
}
\newtcolorbox{conceptbox}{
    breakable,
    colback=black!2,
    frame hidden,
    borderline west={1.5pt}{0pt}{C2},
    colbacktitle=C2!12,
    coltitle=black,
    fonttitle=\bfseries,
    title={Key distinction},
    sharp corners,
    left=7pt,
    right=5pt,
    top=5pt,
    bottom=5pt,
    before skip=7pt,
    after skip=7pt
}

\let\oldnl\nl
\newcommand{\nonl}{\renewcommand{\nl}{\let\nl\oldnl}}

\newcommand{\red}[1]{#1}
\newcommand{\bs}[1]{\boldsymbol{#1}}
\newcommand{\g}[0]{{$\mathfrak{g}$-sim} }
\newcommand{\Ug}[0]{$U_{\mathfrak{g}}$}

\makeatletter
\def\@keys@name{\textbf{Keywords:} }
\makeatother

\begin{document}
\title{Reducing QPU Calls in Variational Quantum Algorithms via Classical Offloading of Gradient Computation}

\author{Rahul Bhowmick \orcidlink{0000-0002-5893-0847}}
\email{rahul.bhowmick@fujitsu.com}
\affiliation{Quantum Lab, Fujitsu Research of India, Bengaluru, India}

\author{Harsh Wadhwa \orcidlink{0009-0008-9537-692X}}
\affiliation{Quantum Lab, Fujitsu Research of India, Bengaluru, India}

\author{Avinash Singh \orcidlink{0009-0009-9116-2884}}
\affiliation{Quantum Lab, Fujitsu Research of India, Bengaluru, India}

\author{Tania Sidana \orcidlink{0000-0002-5336-0371}}
\affiliation{Quantum Lab, Fujitsu Research of India, Bengaluru, India}

\author{Quoc Hoan Tran \orcidlink{0000-0003-1652-2332}}
\affiliation{Quantum Laboratory, Fujitsu Research, Fujitsu Limited Japan}

\author{Krishna Kumar Sabapathy \orcidlink{0000-0003-3107-6844}}
\affiliation{Quantum Lab, Fujitsu Research of India, Bengaluru, India}
\begin{abstract}
Variational quantum algorithms (VQAs) are a promising approach for leveraging near-term quantum computers, but trainability and quantum-resource costs remain major challenges. Here we address these challenges with the \textbf{H}ardware \textbf{E}fficient and dynamical \textbf{LI}e algebra supported \textbf{A}nsatz (HELIA) and two training methods combining the classical-enhanced \g method with the quantum-based Parameter-Shift Rule (PSR). This distributes gradient estimation and training across classical and quantum hardware. We evaluate ground-state estimation of 6- to 18-qubit Hamiltonians using the Variational Quantum Eigensolver (VQE) and quantum phase classification for Hamiltonians up to 12 qubits. For VQE, our methods improve accuracy and success rates, with a mean reduction in estimated logical quantum evaluations of up to $60\%$ relative to Full-PSR training of the same ansatz. For classification, the largest improvement in mean peak test accuracy relative to Full-PSR is $2.8$ percentage points.  We also use these methods as a warm-start pretraining strategy for Hamiltonians with exponential-dimensional dynamical Lie algebras. Gradient diagnostics for the studied Hamiltonian configurations show slower gradient-variance decay than the selected deep hardware-efficient reference over the simulated system sizes, potentially paving the way for training large-qubit quantum models.
\end{abstract}
{\color{red}
}

\maketitle

\binoppenalty=10000
\relpenalty=10000

\section{Introduction}
Quantum computing holds great promise for tackling problems that are intractable for classical computers or would take them years to solve, such as simulating natural systems~\cite{daley2022practical}, prime factorization~\cite{shor1999polynomial}, solving linear equations~\cite{lloyd2010quantum}, machine learning tasks~\cite{riste2017demonstration, havlivcek2019supervised,bravyi2020quantum}, optimization~\cite{abbas2024challenges} and quantum chemistry~\cite{bauer2020quantum, motta2022emerging}.
Despite recent breakthroughs in implementing quantum error correction~\cite{acharya2024quantum}, it may still take many years~\cite{fujitsuroadmap, googleroadmap, ibmroadmap} to develop fault-tolerant quantum hardware. Current quantum devices face significant challenges, including low number of qubits, qubit coherence time and gates with limited fidelity~\cite{acharya2024quantum,ibm_qpu, loschnauer2407scalable, chen2024benchmarking,evered2023high,zimboras2025myths, AcceleratingQuantumComputing}.
Limitations of current quantum computing hardware have spurred a focus on variational quantum algorithms (VQA). These are hybrid quantum-classical approaches that utilize parameterized quantum circuits (PQC) to address challenges across diverse fields, including machine learning, optimization, and ground-state energy calculations. VQAs are especially suited for noisy intermediate-scale quantum (NISQ) devices, which have a limited qubit count and noisy (unprotected) operations. The core idea is to train quantum circuits with tunable parameters and optimize these parameters using a classical optimizer for an objective function related to the problem being solved.

In VQAs, the parameters of PQC are updated via either gradient-free or gradient-based optimization techniques. The gradient-free methods~\cite{spall1998overview, powell1994direct, conf,Ostaszewski2021structure, gasnikov2023randomized,lin2022gradient} are preferred when calculating gradients is challenging or the loss function is not differentiable, but their performance might not scale well with problem size and system noise~\cite{pellow2021comparison, hottois2023comparing}.
In contrast, the literature on gradient-based training of quantum machine learning (QML) primarily emphasizes the efficient computation of gradients of parameters of trainable quantum circuits on quantum hardware. The commonly employed techniques include the parameter-shift rule (PSR)~\cite{schuld2019evaluating} and its various generalizations or extensions~\cite{izmaylov2021analytic, wierichs2022general}. PSR requires running two or more quantum circuit evaluations with shifted parameter values for each trainable parameter in the circuit, and it runs exclusively on quantum hardware.

VQAs are often challenging due to several factors. A major hurdle being the barren plateau (BP) phenomenon~\cite{larocca2024review}, where gradients vanish exponentially as qubit count increases, making it difficult to find numerically meaningful updates for large circuits. Several BP-free models have been proposed in literature like quantum convolutional neural networks~\cite{cong2019quantum,chinzei2024splitting} and permutation-equivariant quantum neural networks~\cite{schatzki2024theoretical}. Recent research also points toward the possibility that BP-free quantum models might be efficiently classically simulable~\cite{cerezo2023does}. These issues need to be taken into consideration when designing new VQA, in order to avoid such pitfalls.

The optimization landscape in quantum circuits is also highly non-convex, with numerous local minima and saddle points that can trap algorithms and complicate the search for global minima~\cite{Cerezo2020CostFunctionDependentBP, anschuetz2024unified, anschuetz2022quantum, bittel2021training}.  Additionally, quantum measurements are inherently stochastic, requiring many runs to obtain accurate estimates and thus potentially slowing down optimization. Even for models that are trainable, computing gradients for quantum circuits is often quantum resource-intensive as methods such as the PSR or finite differences necessitate multiple circuit executions, which scale linearly~\cite{abbas2024quantum} with the number of parameters and the number of shots.
These factors, coupled with hardware limitations such as qubit connectivity and decoherence~\cite{acharya2024quantum,ibm_qpu, loschnauer2407scalable, chen2024benchmarking,evered2023high,zimboras2025myths, AcceleratingQuantumComputing}, make effective gradient-based optimization in quantum computing a complex task, frequently driving the need for alternative or hybrid methods.

Classical algorithms for simulating general quantum circuits do exist, but they suffer from exponential computational and memory overhead~\cite{QuantumHW}. Several methods have been developed for efficient classical simulation of certain classes of quantum circuits, leveraging special structures in the problems. These include matrix-product-state-based tensor-network methods that can simulate shallow noisy circuits for hundreds of qubits with limited gate fidelity~\cite{ayral2023density}. Clifford perturbation theory has been recently put forward to approximate operator expectation values in near-Clifford circuits~\cite{beguvsic2023simulating}. \g is another method that offers an efficient simulation algorithm, based on the study of the Lie group and the associated Lie algebra $\mathfrak{g}$,
which is generated by the PQC. This method is effective when both the generators of the ansatz and the measurement operator are within a dynamical Lie algebra (DLA) \cite{goh2023lie}, whose dimension scales polynomially with the number of qubits. Initially proposed by Somma et al.\,\cite{somma2005quantum}, the techniques were reframed in a modern context tailored to the quantum computing community by Goh et al.\,\cite{goh2023lie}.
However, recent work has clarified that, in the Pauli setting, polynomial-dimensional dynamical Lie algebras (DLAs) are highly constrained, with the free-fermionic case being essentially the only example \cite{aguilar2024full}. Consequently, polynomial-DLA Pauli benchmarks such as the XY and TFIM models should not be interpreted as demonstrating computational advantage over the best known classical methods. In this work, we use such models as controlled proof-of-principle settings for studying hybrid gradient evaluation and resource allocation. At the same time, the practical scope of Lie-algebraic simulation algorithms like \g is not exhausted by free-fermionic Pauli examples: recent works have identified additional nontrivial polynomial-dimensional DLA families through symmetry-adapted and non-Pauli constructions \cite{barligea2026enabling,kordonowy2026lie}. We will explain more about the \g method in a later section as it is central to our ansatz construction.

\begin{conceptbox}
Importantly, efficient Lie-algebraic simulation should not be conflated with efficient global optimization or ground-state estimation. A polynomial-dimensional DLA, together with an efficiently constructible representation and accessible initial expectation values, enables \g to evaluate selected observables and parameter derivatives using polynomial classical resources~\cite{goh2023lie,barligea2026enabling}. It does not guarantee that the resulting non-convex optimization problem can be solved efficiently or that the Hamiltonian ground state can be found efficiently. Indeed, variational training can remain NP-hard even for classically tractable free-fermionic systems~\cite{bittel2021training}. Thus, \g provides an efficient evaluation primitive rather than a general efficient algorithm for variational optimization or ground-state estimation.
\end{conceptbox}

Although simulating arbitrary quantum circuits is hard, recent advancements in classical simulation methods have significantly enhanced our ability to estimate quantum circuits in specific parameter regions and in the presence of noise.
Fontana et al. proposed low weight efficient simulation algorithm (LOWESA) to classically estimate expectation values of PQCs in noisy hardware~\cite{fontana2023classical}. The authors recreate a classical surrogate of the cost landscape in the presence of device noise by ignoring Pauli terms beyond a certain frequency threshold. Rudolph et al. extended this to noiseless quantum simulation and classically simulated the 127-qubit quantum utility experiment~\cite{rudolph2023classical, kim2023evidence} while  Angrisani et. al further demonstrated provable guarantees for most noiseless quantum circuits~\cite{angrisani2024classically}.
Recently, Lerch et al.  showed it is always possible to identify patches for which similar classical surrogate can be generated~\cite{lerch2024efficient}. These techniques should be accounted for when constructing variational models with potential for quantum utility. For heuristic algorithms such as VQAs, identifying quantum advantage therefore requires substantial end-to-end resource considerations. In particular, polynomial-time classical simulability alone does not determine the preferable implementation, since preprocessing, quantum-data acquisition, sampling, and classical and quantum runtimes can substantially shift the practical crossover between classical and quantum execution~\cite{chang2026cost}.

\begin{figure*}
          \includegraphics[width=0.9\linewidth]{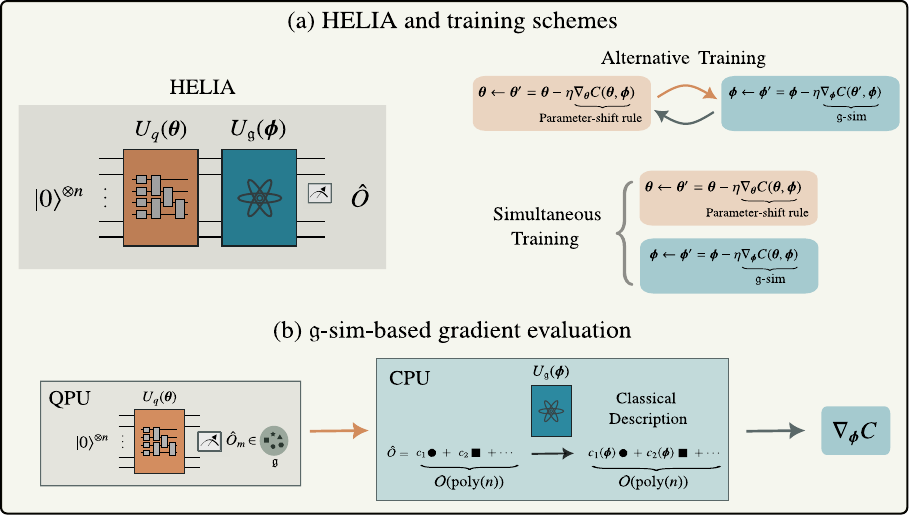}
    \caption{\raggedright \textbf{Overview of our main contributions} (a) We propose \textbf{H}ardware \textbf{E}fficient and dynamical \textbf{LI}e algebra supported \textbf{A}nsatz (HELIA) composed of two blocks of gates $U_q$ and $U_{\mathfrak{g}}$  whose gradients can be obtained using parameter-shift rule and \g respectively. The resources required for gradient evaluation of the full ansatz can hence be delegated to both quantum and classical hardware. We further propose two training methods: Alternate and Simultaneous which benefit from the hybrid gradient of HELIA. (b) To evaluate the gradients of $U_{\mathfrak{g}}$, we use a \g based method. The operators of a dynamical Lie algebra are measured after applying $U_q$ (using QPU in the leftmost block), which is then passed on to classical hardware that evaluates the cost function and gradients using \g (using CPU in the middle block). Further details of our contribution are elaborated in Sec.~\ref{our_contri_sec}. The numerical experiments in Sec.\,\ref{result} evaluate logical gradient-evaluation costs, optimization outcomes, and finite-size gradient behavior for the proposed ansatz and training procedures. The proposed procedure can also be used as a pretraining strategy for Hamiltonians with exponential-dimensional DLAs: training is first performed using a polynomial-DLA restriction and then continued using the complete Hamiltonian, as schematically illustrated in Fig.~\ref{fig:ltfim-workflow}.}
    \label{fig:overview}
\end{figure*}

We focus on enhancing quantum model training through tailored ansätze and customized training methods, improving accuracy in tasks such as ground-state estimation with the VQE and quantum phase classification using quantum neural networks (QNNs).
To mitigate the high quantum resource costs associated with training VQAs, we propose \textbf{H}ardware \textbf{E}fficient and dynamical \textbf{LI}e algebra supported \textbf{A}nsatz (HELIA) and two hybrid methods\text{---}Alternate and Simultaneous\text{---}that integrate \g and PSR  (Fig.~\ref{fig:overview}). Our gradient-estimation task is delegated to both classical and quantum hardware to reduce the number of Quantum Processing Unit (QPU) calls, thereby saving quantum resources compared to PSR. The purpose of this study is to evaluate hybrid gradient computation and training-resource allocation in controlled numerical benchmarks. We do not establish computational advantage over classical methods, hardware-level measurement savings, or asymptotic scalability. The results identify tested settings in which classical offloading reduces the estimated quantum-side gradient-evaluation workload.\\

\noindent {\bf Remark}\,: While this study does not evaluate the practical utility of these models or their performance relative to classical counterparts, it aims to explore \textit{the potential of scalable and accurate trainable quantum models}. Our findings suggest that increasing the scale and accuracy of trainable models could bridge the gap toward practical quantum utility.\, $\square$\\

The remaining sections are divided as follows. We introduce the relevant background literature in Sec.~\ref{overview_sec}. We then discuss our contributions in terms of the choice of quantum circuits and two training methods, namely Alternate and Simultaneous, in Sec.~\ref{our_contri_sec}. In Sec.~\ref{result} we demonstrate the improvement in various metrics through extensive numerical simulations. Finally, we end with the Conclusion and Outlook in Sec.~\ref{conclusion_sec}.

\section{Overview of Literature}\label{overview_sec}

\subsection{Variational Quantum Algorithms}
VQAs constitute a fundamental framework within hybrid quantum-classical computation, providing a viable approach to exploit the capabilities of quantum systems in the NISQ era. By integrating PQCs with classical optimization techniques, VQAs are specifically tailored to tackle computationally demanding problems across domains such as quantum chemistry, combinatorial optimization, and machine learning~\cite{cerezo2021variational,biamonte2017quantum,wittek2014quantum,schuld2018supervised, lanyon2010towards,farhi2014quantum}.

VQAs involve an $n$-qubit state $\rho$ (usually encoding problem data), acting on a Hilbert space
 $\mathcal{H}={(\mathbb{C}^2)}^{\bigotimes n},$ which undergoes evolution through a parameterized quantum unitary. The unitary transformation is expressed as a sequence of $L$ parameterised unitary gates
\begin{equation}\label{unitary}
U(\boldsymbol{\theta}) = \prod_{i=1}^{L} U_i(\bs\theta_i)W_i
\end{equation}
where $\boldsymbol{\theta}=[\bs\theta_{1},\bs\theta_{2},\ldots,\bs\theta_{L}]$ denotes a set of trainable real-valued parameters and $W_i$ are non-trainable gates. An example of such case is shown in Fig.\,\ref{yz_linear}, where $U_i(\bs\theta_i)$ is be composed of multiple trainable single-qubit gates, while the non-trainable block contains entangling CNOT gates.

The expectation value of a Hermitian observable
$\hat{O}$ is subsequently measured on the evolved quantum state, using a quantum device, given as
\begin{equation}\label{loss_func}
\ell_{\theta}(\rho,\hat{O}) = \operatorname{Tr}\!\left[ U(\theta)\rho U^\dagger(\theta)\hat{O} \right] = \operatorname{Tr}\!\left[ \rho U^\dagger(\theta)\hat{O}U(\theta) \right].
\end{equation}

Classical optimizers are utilized to perform the optimization task
\begin{equation}\label{loss_func_min}
    \underset{\bs\theta}{argmin}\>\>{ F(l_{\bs\theta}(\rho, \hat{O}))},
\end{equation}
where we minimize some loss function $F$ of the expectation value depending on the task.

The form of the loss function in Eq.~\eqref{loss_func_min} can vary depending on the task. For example, when estimating ground states the loss function can be the expectation value of a Hamiltonian, and finding the ground state is equivalent to minimizing this expectation value. For classification task, the loss function can be mean squared error between predicted labels from a quantum model and the true labels obtained from a dataset. In both cases, the algorithm iteratively optimizes a cost function using the output of a quantum circuit, enabling the exploration of complex solution spaces that are potentially
intractable for classical approaches.

Despite their promise, VQA face several critical challenges, including barren plateaus in the optimization landscape, limited scalability with increasing problem size, and the detrimental effects of noise and decoherence on circuit fidelity~\cite{larocca2024review,cerezo2021variational, wang2021noise}. Overcoming these obstacles necessitates advancements in algorithmic design, optimization techniques, and the development of robust quantum hardware.

\begin{figure}
\scalebox{0.8}{
    \centering
    \begin{quantikz}
    \lstick{} & \gate{R_y(\theta_0)}\gategroup[5,steps=2,style={dashed,rounded
corners,fill=C1!40, inner
xsep=2pt},background,label style={label
position=below,anchor=north,yshift=-0.2cm}]{{\sc
Trainable Unitaries}} & \gate{R_z(\theta_1)} & \ctrl{1}\gategroup[5,steps=8,style={dashed,rounded
corners,fill=C2!40, inner
xsep=2pt},background,label style={label
position=below,anchor=north,yshift=-0.2cm}]{{\sc
Non-Trainable Unitaries}} &  &  &  &  &  & & &  \\
    \lstick{} & \gate{R_y(\theta_2)} & \gate{R_z(\theta_3)} & \targ{} & \ctrl{1} &  &  &  &  & & &  \\
    \lstick{} & \gate{R_y(\theta_4)} & \gate{R_z(\theta_5)} &  & \targ{} & \ctrl{1} &  &  &  & &  &  \\
    \lstick{} & \vdots \wireoverride{n} & \wireoverride{n} \vdots \wireoverride{n} & \ldots \wireoverride{n} & \wireoverride{n} & \targ{} & &  & \wireoverride{n} \ldots \wireoverride{n} & \wireoverride{n} & \ctrl{1} &  \\
    \lstick{} & \gate{R_y(\theta_{n-1})} & \gate{R_z(\theta_n)} &  &  &  &  &  & & & \targ{} &
    \end{quantikz}
}

    \caption{\raggedright Quantum circuit block for YZ linear (HEA) ansatz. The parametrized $R_y$ and $R_z$ rotations are applied on each qubit followed by $CNOT$ between neighboring qubits in a linear fashion. The blocks are often repeated to increase entanglement and expressivity.}\
    \label{yz_linear}
\end{figure}

The available options for PQC in VQAs is quite vast and it is not a priori clear which circuit structure is ideal for the task at hand.
One common choice is Hardware Efficient Ansatz (HEA), which consists of blocks of single-qubit rotations followed by entangling two-qubit gates repeated for a chosen number of layers~\cite{kandala2017hardware}.
An example within this class is the $YZ$ linear ansatz shown in Fig.~\ref{yz_linear}, where each block consists of a $Y$-rotation and a $Z$-rotation applied on each qubit, followed by CNOT gates between neighbouring qubits in a linear fashion.
Commonly selected problem-motivated ansatz include the Unitary Coupled Cluster (UCC)~\cite{Anand_2022}, which is widely utilized for investigating ground states of fermionic molecular Hamiltonians, and the Quantum Alternating Operator Ansatz (QAOA)~\cite{Hadfield_2019,kremenetski2021quantum, golden2023quantum}, often applied to combinatorial optimization problems.

Another interesting problem-motivated example is the Hamiltonian Variational Ansatz (HVA)~\cite{wecker2015progress}. Given the task of finding ground state of a Hamiltonian $H = \sum_i H_i$, the ansatz is composed of unitary blocks of the form
\begin{equation}
    U(\bs{\theta}) = \prod_i e^{-i\theta_i H_i},
\end{equation}
and can repeated for a chosen number of layers. Building on the insights from Ref.~\cite{goh2023lie}, this approach can be extended to include terms from the dynamical Lie Algebra (DLA) of $\{iH_i\}_{H_i \in H}$, provided the DLA has a dimension that scales polynomially with the number of qubits. We elaborate on the details of DLA in Sec.\,\ref{sec_gsim}. In this study, we will combine the HVA ansatz with the HEA ansatz. For a detailed overview of the ansatz options explored in the literature, we refer to Ref.~\cite{cerezo2021variational}.

\subsection{Training Methods}
After selecting an appropriate ansatz, the subsequent step involves determining the optimal parameters of the ansatz to address the problem being solved.
This is an iterative procedure, with each iteration comprising a loss function evaluation followed by a parameter update step. Although several gradient-free methods exist for parameter update~\cite{spall1998overview, powell1994direct, conf,Ostaszewski2021structure}, here we focus on gradient-based techniques, which require efficient and accurate algorithms to compute the partial derivatives of the cost function.

Contemporary classical machine learning models, which often contain billions of parameters, efficiently compute gradients using the backpropagation algorithm~\cite{rojas1996backpropagation}. This method requires only a single forward and backward pass to compute the gradients for all parameters simultaneously.
Since classical backpropagation cannot be directly applied to quantum circuits~\cite{abbas2024quantum}, the PSR~\cite{mitarai2018quantum,schuld2019evaluating} is typically employed to compute circuit gradients. Alternatively, when the quantum circuit can be efficiently simulated classically, automatic differentiation frameworks, such as PyTorch~\cite{ansel2024pytorch} and Tensorflow~\cite{Abadi_TensorFlow_Large-scale_machine_2015}, can be utilized for gradient evaluation. A notable case within this second scenario is \g, where placing certain restriction on quantum gates and measurement operators enables efficient classical simulation of the circuit.

\subsubsection{Parameter-shift Rule (PSR)}
PSR is commonly used for accurately estimating gradients in a PQC~\cite{schuld2019evaluating} using quantum hardware. Consider the loss function $\ell$ in Eq.~\eqref{loss_func}, where the unitary $U_i(\theta_i)$ is of the form $U_i(\theta_i) = \exp{-i\theta_i\hat{G}_i}$ with $G_i$ having eigenvalues $\pm r$. The derivative of the loss function with respect to $\theta_i$ can be expressed as (Appendix \ref{psr_append} for a detailed derivation):
\begin{equation}\label{eqn:PSR}
     \frac{\partial{\ell}}{\partial{\theta_i}} = r \Big(\ell(\bs\theta + \frac{\pi}{4r}\hat{e}_i) - \ell(\bs\theta - \frac{\pi}{4r} \hat{e}_i)\Big).
\end{equation}

By measuring the loss function at two shifted values of the parameter $\theta_i$, the partial derivative can be estimated accurately up to shot noise~\cite{schuld2019evaluating}. For more general eigenspectrum with equispaced eigenvalues, the parameter shift-rule can be generalized as shown in~\cite{crooks2019gradients,izmaylov2021analytic,wierichs2022general}. Markovich et.al. have further extended this procedure to irregular eigenspectrum as well~\cite{markovich2024parameter}.

The computation of a gradient for a single parameter in a quantum circuit requires executing the circuit at least twice, apart from the measurement overhead to obtain expectation value accurately.
This limitation is illustrated through a specific example. Assuming each circuit iteration takes 1.48 milliseconds\,\cite{somoroff2023millisecond}, equivalent to the latest coherence time of superconducting qubits, one iteration for a circuit with a billion parameters would take approximately \( 2 \times 10^9 \times 1.48 \) ms, or 2.96 million seconds ( $\sim 34.26$ days). This demonstrates the poor scalability of PSR for large-scale systems.

For certain examples of PQC construction, the training may scale more favorably. One such example is proposed by Bowles et al.  to address the scalability issue by introducing structured circuits consisting of blocks of commuting parametrized quantum gates\,\cite{bowles2023backpropagation,chinzei2025-pa}.
Brnović et al. further extended this approach using layerwise-commuting PQC\,\cite{brnovic2024efficient}.
However, the utility of such structured PQCs still need to be fully understood.

\subsubsection{\texorpdfstring{\g}{$\mathfrak{g}$-sim} Method}\label{sec_gsim}
In this section, we briefly explain the details of $\mathfrak{g}$-sim method proposed by  Goh et al.\,\cite{goh2023lie}, that is fully classical in nature. Consider a problem in which an $n$-qubit state $\rho$, acting on a Hilbert space $\mathcal{H}={(\mathbb{C}^2)}^{\bigotimes n},$  is sent through a PQC $U(\boldsymbol{\theta})$ of the form
\begin{equation}\label{eq_anz}
U(\boldsymbol{\theta}) = \prod_{\ell=1}^{L} \prod_{k=1}^{K} e^{i \theta_{\ell k} H_k},
\end{equation}
where $\boldsymbol{\theta}=[\theta_{11},\theta_{12},\ldots,\theta_{LK}]$ is a set of trainable real-valued parameters, and $\{H_1,H_2,\ldots,H_K\}$ are  $K$ Hermitian operators that are the gate generators of the circuit.

The vector space spanned by all possible nested commutators of $\{iH_1, \dots, iH_K\}$, obtained by repeatedly taking the commutator between all elements until no new linearly independent element emerge is known as the DLA. For the ansatz of the form in Eq. \eqref{eq_anz}, this DLA, denoted as $\mathfrak{g}$, is used to describe the system's structure.

The dynamical Lie group $\mathcal{G}$ of a circuit of the form Eq. \eqref{eq_anz} is defined as
$$\mathcal{G}=\{e^{iA}: iA \in \mathfrak{g}\},$$ that is, the dynamical Lie group is obtained via the exponentiation of DLA.

The significance of the DLA lies in the fact that all unitaries of the form in Eq. \eqref{eq_anz} belong to a Lie group $\mathcal{G}$. Specifically, for any  $V \in \mathcal{G}$, there exists (at least) one choice of parameter values $\boldsymbol{\theta}$ such that for a sufficiently large, but finite, number of layers $L$, we have $U(\boldsymbol{\theta}) = V$~\cite{d2021introductiontoquantumcontrolanddynamics}.    That is, the dynamical Lie group $\mathcal{G}$  determines  all possible unitaries that can be implemented by circuits of the form in Eq. \eqref{eq_anz}.

The $\mathfrak{g}$-sim method is applicable only under two assumptions~\cite{goh2023lie}:
\begin{itemize}
    \item either the measurement operator or the initial state of a variational quantum circuit belong to the  DLA of the ansatz,
    \item the dimension of DLA scales polynomially with the number of qubits.
\end{itemize}

For more details on implementation and theory behind \g we refer the reader to Appendix \ref{gsim_append}.

The limitations of the PSR and the efficiency of \g provide motivation for combining them into a hybrid approach to achieve efficient gradient evaluation.
Apart from scalability and resource challenges in gradient evaluation, the training of quantum models remains challenging, primarily due to the vanishing gradient phenomenon, also known as the BP problem, where gradients decay exponentially as the number of qubits increases, making larger models effectively untrainable. For a survey of BP phenomenon and proposed mitigation techniques, refer to Appendix \ref{BP_app}.

Finally, using classical processing to improve the results from quantum device is not unique to our work, and has been used in Refs.~\cite{takeshita2020increasing, jozsa2008matchgates, parrish2019quantum, stair2020multireference, huggins2020non, bharti2021iterative, bharti2021quantum, yuan2021quantum, mazzola2019nonunitary,fujii2022deep, kim2017robust, heidari2023quantum}. In fact, a recent paper also deals with the idea of processing different parts of the quantum circuit using classical and quantum hardware with Clifford Perturbation Theory~\cite{fuller2025improved}.
However, our approach combines \g and PSR in a hybrid iterative procedure that transfers part of the gradient-evaluation workload to classical hardware. This makes it a promising area for further research and development in quantum computing. The quantum-side cost is compared with Full-PSR training of the same ansatz. The presence of a hardware-efficient block ensures that it is not simulable by \g but does not, by itself, establish classical intractability of the complete circuit or computational advantage over classical alternatives.

\section{Our Contributions}\label{our_contri_sec}

We investigate four aspects of hybrid variational training:
\begin{itemize}
\item HELIA ansatz construction that permits classical evaluation of gradients for a structured circuit block;
\item Alternate and Simultaneous update procedures that reduce logical quantum-evaluation costs relative to Full-PSR;
\item warm-start pretraining for exponential-DLA Hamiltonians using a polynomial-DLA restriction; and
\item finite-size optimization, resource, and gradient benchmarks across the studied configurations.
\end{itemize}

Our overall results are summarized in Table~\ref{tab:numerical_results_summary}. The HELIA architecture and update procedures are developed in this section, while warm-start pretraining is detailed in Sec.~\ref{gen_hamil}. The finite-size benchmarks are presented in Sec.~\ref{result} and the Appendices. The gradient analysis is empirical and regime-specific; it does not provide a general guarantee of barren-plateau avoidance.

\subsection*{Algorithms considered in this work}
In this study, we focus on two applications of VQAs, illustrating their potential to address complex quantum problems. The first application investigates the Variational Quantum Eigensolver (VQE), a hybrid quantum-classical algorithm widely used for determining the ground state of quantum systems and calculating their corresponding energies~\cite{peruzzo2014variational}. VQE has become a cornerstone in quantum chemistry and materials science, enabling the efficient simulation of molecular systems and facilitating the discovery of new materials. Recent research has also incorporated VQE into Quantum Error Detection pipeline~\cite{urbanek2020error, gowrishankar2024logical}. In this work, we employ VQE to find ground states of XY, Transverse Field Ising Model (TFIM), Longitudinal-Transverse Field Ising Model (LTFIM) and LiH Hamiltonian.

The second application delves into quantum phase classification, where VQAs are utilized to identify and distinguish different quantum phases of matter~\cite{cong2019quantum,monaco2023quantum,herrmann2022realizing} of the bond-alternating spin-$1/2$ Heisenberg chain~\cite{PhysRevB.87.054402}. This application leverages the unique capabilities of quantum circuits to encode and process quantum states, enabling the precise detection of phase transitions and the systematic classification of quantum phases.

\subsection{PQC Selection}\label{pqcs}

Our ansatz construction is motivated primarily by reducing the quantum-side workload of gradient evaluation as well as avoiding BPs. In general, simulating the full density matrix requires keeping track of all the $n$-qubit Pauli operators which scales as $4^n -1$. However, to effectively utilize \g without incurring exponential computational and memory overhead, it is essential to restrict the training ansatz and measurement operator for a QML task to a \textit{poly}-DLA. A polynomial-dimensional DLA permits a reduced operator-space description, but does not imply that the associated variational family can approximate every target state. For VQE, the ground state may be poorly approximated by the states reachable from the chosen initialization using $U_\mathfrak{g}$ alone (as shown in Table \ref{vqe_xy_error} in Appendix \ref{xy_numerical}). Similarly, the restricted variational family may limit classification accuracy (as shown in Table~\ref{class_all}).

\noindent \textbf{Choice of unitary blocks\,:} To elevate \g to a higher expressibility, we propose a hybrid ansatz consisting of a parameterized non-DLA block of gates $U_q$,
which cannot be classically simulated within the \g framework, followed by a parameterized DLA block of gates $U_\mathfrak{g}$ that can be classically simulated. A specific ansatz falling within this proposal is the \textbf{H}ardware \textbf{E}fficient and dynamical \textbf{LI}e algebra Supported \textbf{A}nsatz (HELIA), shown in Fig.~\ref{fig:overview}, where $U_q$ represents the hardware efficient ansatz (HEA).
 $U_\mathfrak{g}$ is composed of multi-qubit Pauli rotations $\exp{-i\theta P}$, where the Pauli operators $P$ form an orthonormal basis of a chosen \textit{poly}-DLA.
In our numerical experiments, we use the YZ linear ansatz, shown in Fig.~\ref{yz_linear} (which belongs to the HEA class), for the non-DLA circuits, while the Lie-algebra-supported ansatz is motivated by the problem being solved.\\

\noindent \textbf{Ordering of unitary blocks\,:} The specific ordering of gates is important for \g to be used efficiently in our proposal. This is because we require the gates and measurement operator chosen from the \textit{poly}-DLA to be adjacent in the quantum circuit.
 To elaborate, let us consider the operator $\hat{H}$, chosen from a \textit{poly}-DLA. If we look at the Heisenberg-evolved operator created by applying $U$, we obtain the following $H' = U^\dagger H U$. When the generators of $U$ are chosen from the \textit{poly}-DLA the operator $H'$ is efficiently computable using \g as we only need to track polynomially many operators. This is the case when we consider a unitary of type $U_{\mathfrak{g}}(\phi)$. On the other hand, if $U$ is composed of a general set of gates similar to $U_q(\theta)$, the expression for $H'$ may not be classically tractable as it can contain exponentially many operators. \\

Both HEA~\cite{kandala2017hardware} and the Lie-algebra-supported ansatz~\cite{fontana2023adjoint} have been studied in the literature. Our proposal, however, aims at combining them into a single ansatz, which we will show has several important advantages. The primary advantage of HELIA lies in its ability to use PSR for computing gradients in $U_q$, while utilizing \g for the $U_\mathfrak{g}$ from measured intermediate-state expectation values. Additionally, we introduce two training methods and demonstrate their superior performance compared with standard PSR or \g independently. Finally, we investigate the BP phenomenon and demonstrate that our ansatz exhibits a slower decay of gradients with increasing qubit count than a deep hardware-efficient circuit, thereby enabling the effective training of larger qubit models.

\subsection{Training Method}\label{training_scheme}
For the selected PQC, two distinct training methods are employed: \textbf{Alternate} and \textbf{Simultaneous}. We evaluate the proposed methods on two different VQA applications: ground state estimation using VQE and Quantum Phase Classification. The core idea in both the methods is to exploit the ansatz structure and distribute the resources required for gradient evaluation to both classical and quantum hardware.
This delegation significantly reduces the number of QPU calls relative to PSR, while shifting part of the computational burden to classical \g processing, and also increases the accuracy as shown in the results of VQE in Table \ref{vqe_xy_successqpu} and \ref{vqe_xy_error} of Appendix \ref{xy_numerical}. Throughout the paper, our resource claims are therefore restricted to quantum-side resource usage, not total end-to-end training cost.

\subsubsection{Alternate Training}\label{alternate_sec}
\begin{figure}[!tbh]
    \centering
    \includegraphics[width=1\linewidth]{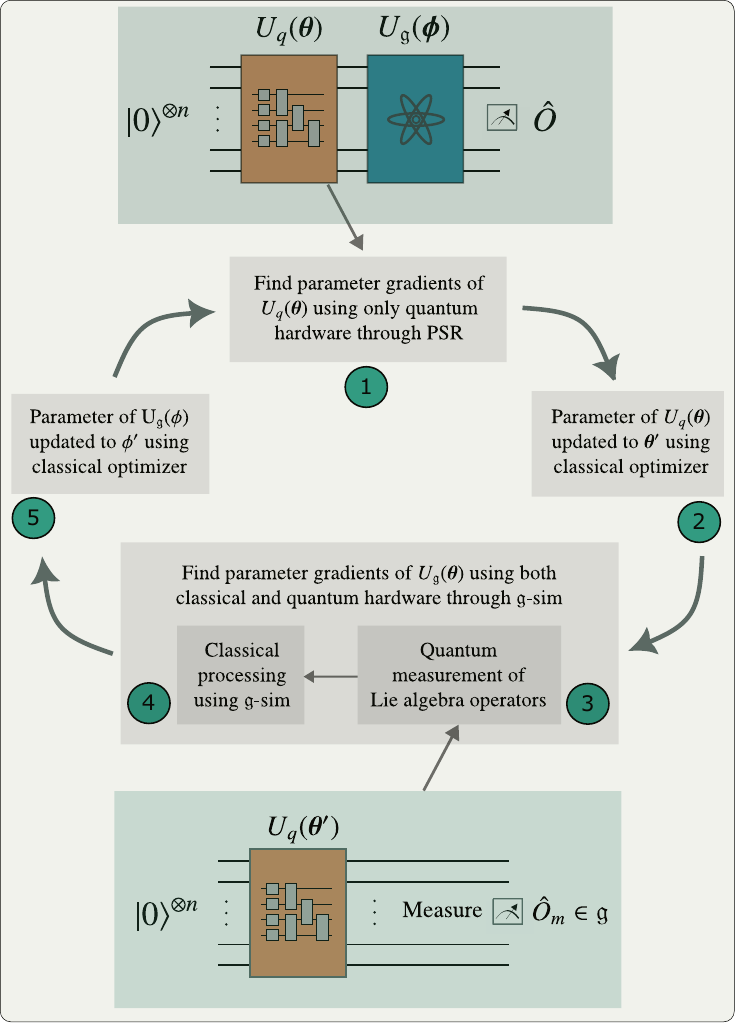}
    \caption{Workflow of Alternate training method based on Alg.\ref{alternate}}
    \label{alternate_fig}
\end{figure}

This section outlines a training method that integrates both the PSR and \g to develop a hybrid algorithm for gradient evaluation and quantum model training. As described in Algorithm~\ref{alternate} and depicted in Fig.~\ref{alternate_fig}, this method alternates the gradient evaluation process between classical and quantum hardware. Although we use VQE to explain our training methods, it can be easily extended to the classification task without changing the core idea.

To make the workflow explicit, we describe here one full optimization iteration and clarify the role of the two circuit blocks. The HELIA ansatz is written as
\begin{equation}
U(\theta,\phi) = U_{\mathfrak{g}}(\phi)\,U_q(\theta),
\end{equation}
where $U_q(\theta)$ denotes the hardware-efficient block and $U_{\mathfrak{g}}(\phi)$ denotes the dynamical-Lie-algebra-supported block. The parameters $\theta$ are updated using the parameter-shift rule (PSR), whereas the parameters $\phi$ are updated using the $\mathfrak{g}$-sim procedure.

Considering an initial state of $\ket{\psi}$, in the $i$-th iteration of the Alternate method, the update proceeds in two stages:
\begin{enumerate}
    \item \textit{Update of the $U_q$ block:}
    Keeping $\phi$ fixed, we evaluate the gradient with respect to $\theta$ using PSR on the full circuit
    \begin{equation}
    U(\theta,\phi)=U_{\mathfrak{g}}(\phi)\,U_q(\theta),
    \end{equation}
    and then apply the optimizer step
    \begin{equation}
    \theta_i \longrightarrow \theta_{i+1}.
    \end{equation}

    \item \textit{Update of the $U_{\mathfrak{g}}$ block:}
    Using the updated parameters $\theta_{i+1}$, we prepare the intermediate state
    \begin{equation}
    \ket{\psi(\theta_{i+1})} = U_q(\theta_{i+1})\ket{\psi}
    \end{equation}
    on the quantum device and measure the expectation values of a basis of DLA operators
    \begin{equation}
    e_m(\theta_{i+1}) = \bra{\psi(\theta_{i+1})} O_m \ket{\psi(\theta_{i+1})},
    \quad iO_m \in \mathfrak{g},
    \end{equation}
    where $\{O_m\}_{m=1}^{\dim(\mathfrak{g})}$ are the elements of a Hermitian basis of the operator space associated with the skew-Hermitian Lie algebra $\mathfrak{g}$. These expectation values are then passed to the classical $\mathfrak{g}$-sim routine, which evaluates the cost function and its gradient with respect to $\phi$, followed by the optimizer step
    \begin{equation}
    \phi_i \longrightarrow \phi_{i+1}.
    \end{equation}
\end{enumerate}

Thus, one Alternate iteration first performs a PSR-based update of the hardware-efficient block and then performs a $\mathfrak{g}$-sim-based update of the Lie-algebra block using measurements on the intermediate state after $U_q$. Fig.\ref{alternate_fig} represents this workflow of the Alternate training method.

The key property enabling the classical update of the $U_{\mathfrak{g}}$ block is that $U_{\mathfrak{g}}(\phi)$ is generated by operators belonging to the dynamical Lie algebra $\mathfrak{g}$.
Hence, under the adjoint action of $U_{\mathfrak{g}}(\phi)$, any operator in $\mathfrak{g}$ is mapped to a linear combination of operators in the same algebra. Therefore, once the expectation values of a basis $\{O_m\}\subset\mathfrak{g}$ are known on the state entering $U_{\mathfrak{g}}$, namely $U_q(\theta)\ket{\psi}$,
the effect of $U_{\mathfrak{g}}(\phi)$ on the relevant observables can be computed entirely within the $\mathfrak{g}$-sim framework. In this way, the cost function and the derivatives with respect to $\phi$ can be evaluated
classically without performing PSR over the $U_\mathfrak{g}$ parameters on the quantum device. We emphasize that the measurements required for the $\mathfrak{g}$-sim update are performed on the intermediate state after the application of $U_q(\theta)$ and before the action of $U_{\mathfrak{g}}(\phi)$. By contrast, the PSR gradients for $\theta$ are evaluated using shifted executions of the full circuit $U_{\mathfrak{g}}(\phi)U_q(\theta)$.

Let $p$ and $m$ denote the numbers of trainable parameters in $U_q$ and $U_\mathfrak{g}$, respectively, and let $d$ denote the number of operator expectation values used by the classical \g routine. For the Pauli-basis constructions used in the XY experiments, the chosen design has $m=d=\dim(\mathfrak{g})$. Considering the case where $U_q$ is composed of YZ linear ansatz (forming a unitary $U_q(\theta_{i})$), and $U_{\mathfrak{g}}$ is composed of DLA gates (forming a unitary $U_{\mathfrak{g}}(\phi_i)$), and the initial state is $|\psi\rangle $. Obtaining the gradients of $U_q$ will require at least $2p$ different runs of the quantum circuit for PSR. Using the gradients, we update $U_q$ to $U_q(\theta_{i+1})$.

\SetKwComment{Comment}{/* }{ */}
\begin{algorithm}[!tbh]
\DontPrintSemicolon
\SetKwFunction{PSR}{\textbf{PSR Step}}
\SetKwFunction{gsim}{\textbf{\g Step}}

\caption{Alternate Training for VQE}\label{alternate}
\nonl\textbf{Input}: $T_{max}, \eta, U_q(\mathbf{\theta_0}), U_\mathfrak{g}(\mathbf{\phi_0}), \mathfrak{g}_{DLA}, H $\;
\nonl\textbf{Initialize: } $\mathbf{\phi}_0,\mathbf{\theta}_0 \sim Norm(0,1); i=0$\;
\nonl\textbf{Define: } $C(\theta,\phi) = \langle 0^{\otimes n}|U_q^\dagger(\theta)U_{\mathfrak{g}}^\dagger(\phi) H U_\mathfrak{g}(\phi) U_q(\theta)|0^{\otimes n}\rangle$\;
\nonl\Begin{
\nonl\While{$i < T_{max}$}{\nonl
    \;
    \nonl\PSR\;
    Obtain gradients $\grad_{\theta_i} C(\mathbf{\theta_i},\mathbf{\phi_i})$, w.r.t $\theta_i$ via PSR \label{Alt_PSR}\;
    $\mathbf{\theta}_{i+1} \gets \theta_i - \eta \grad_{\theta_i} C(\theta_i,\phi_i) $\;\nonl
    \;
    \nonl\gsim\;
    Prepare $|\psi(\theta_{i+1})\rangle = U_q(\theta_{i+1})|0^{\otimes n}\rangle$\;
    \For{$m=1,\ldots,\dim(\mathfrak{g}_{DLA})$\label{Alt_DLA_measure}}{
        Measure and store $o_m = \langle \psi(\theta_{i+1}) | \hat{O}_m|\psi(\theta_{i+1})\rangle $
        }
    Obtain gradients $\grad_{\phi_i} C(\theta_{i+1},\phi_i)$, w.r.t $\phi_i$ via \g using $\{o_m\}_{\hat{O}_m \in \mathfrak{g}_{DLA}}$\label{Alt_gsim}\;
    $\mathbf{\phi}_{i+1} \gets \phi_i - \eta \grad_{\phi_i} C(\theta_{i+1},\phi_i) $\;\nonl
    \;
    $i\gets i+1$\;
    }
}
\end{algorithm}

To evaluate gradients of $U_{\mathfrak{g}}$, we measure the expectation value of all $g$ Pauli operators on the state $|\psi(\theta_{i+1})\rangle = U_q(\theta_{i+1})|\psi\rangle$ on quantum hardware.
The measured values are fed into the \g algorithm to evaluate the gradients entirely on classical hardware using an automatic differentiation framework~\cite{ansel2024pytorch, Abadi_TensorFlow_Large-scale_machine_2015}. Finally, the parameters of $U_{\mathfrak{g}}$ are updated from $\phi_i$ to $\phi_{i+1}$, finishing one iteration. Hence, each of these iterations costs us $2p + g$ circuit evaluations (ignoring the factor of number of shots required for estimating the expectation values. Since the measurement outcome is processed differently for gradient evaluation of $U_q$ and $U_\mathfrak{g}$, the number of shots required in each case might differ). This is in contrast with the $2p + 2g$ circuit evaluations required per iteration to train the entire circuit via PSR alone.

Importantly these are logical evaluation counts, not counts of physical shots or hardware jobs. Each shifted cost evaluation and each required intermediate-state operator expectation is counted as one logical evaluation. The accounting excludes the shots needed for a target precision, observable grouping, basis changes, circuit compilation, and classical processing time. In particular, evaluating a Hamiltonian expectation may itself require multiple measurement settings.

\subsubsection{Simultaneous Training}
In this section, we propose a modification to the Alternate training method that enables the simultaneous updating of parameters in each iteration.

The pseudo-algorithm is shown in Algorithm \ref{simultaneous}. The main difference from the Alternate training method is that here we update $U_{\mathfrak{g}}$ based on $\theta_{i}$, rather than the updated $\theta_{i+1}$. Both parameter blocks are therefore updated from gradients evaluated at the same parameter pair $(\theta_i,\phi_i)$, as in a simultaneous full-ansatz update. The gradient-estimation procedures remain different: PSR is used for $\theta$, whereas $\mathfrak{g}$-sim is used for $\phi$. This reduces the QPU-call requirements relative to Full-PSR by delegating part of the gradient estimation to classical hardware. The required circuit evaluations per iteration are the same as for the Alternate method; the only difference is the use of the non-updated unitary $U_q(\theta_i)$ in the measurement step before \g.
In practice, we find that using the Alternate method for a fixed number of iterations followed by Simultaneous training till convergence provides the best results.

\SetKwComment{Comment}{/* }{ */}
\begin{algorithm}
\DontPrintSemicolon
\SetKwFunction{PSR}{\textbf{PSR Step}}
\SetKwFunction{gsim}{\textbf{\g Step}}

\caption{Simultaneous Training for VQE}\label{simultaneous}
\nonl\textbf{Input}: $T_{max}, \eta, U_q(\mathbf{\theta_0}), U_\mathfrak{g}(\mathbf{\phi_0}), \mathfrak{g}_{DLA}, H $\;
\nonl\textbf{Initialize: } $\mathbf{\phi_0},\mathbf{\theta_0} \sim Norm(0,1); i=0$\;
\nonl\textbf{Define: } $C(\theta,\phi) = \langle 0^{\otimes n}|U_q^\dagger(\theta)U_{\mathfrak{g}}^\dagger(\phi) H U_\mathfrak{g}(\phi) U_q(\theta)|0^{\otimes n}\rangle$\;
\nonl\Begin{
\nonl\While{$i<T_{max}$}{\nonl
\;
    \nonl\PSR\;
    Obtain gradients $\grad_{\theta_i} C(\theta_i,\phi_i)$, w.r.t $\theta_i$ via PSR\;
    $\mathbf{\theta}_{i+1} \gets \theta_i - \eta \grad_{\theta_i} C(\theta_i,\phi_i) $\;
    \nonl\;
    \nonl\gsim\;
    Prepare $|\psi(\theta_{i})\rangle = U_q(\theta_{i})|0^{\otimes n}\rangle$\;
    \For{$m=1,\ldots,\dim(\mathfrak{g}_{DLA})$}{
        Measure and store $o_m = \langle \psi(\theta_{i}) | \hat{O}_m|\psi(\theta_{i})\rangle $
        }
    Obtain gradients $\grad_{\phi_i} C(\theta_{i},\phi_i)$, w.r.t $\phi_i$ via \g using $\{o_m\}_{\hat{O}_m \in \mathfrak{g}_{DLA}}$\;
    $\mathbf{\phi}_{i+1} \gets \phi_i - \eta \grad_{\phi_i} C(\theta_{i},\phi_i) $\;
    \nonl\;
    $i\gets i+1$\;
}
}
\end{algorithm}

To evaluate the gradient for $U_q$, our approach is to implement the full circuit on quantum hardware for the shifted parameter values, as done in standard PSR. This requires $2p$ unique circuit evaluations. In another approach for gradient evaluation, one can prepare the states $|\psi(\theta_{i})\rangle = U_q(\theta_{i})|\psi\rangle$, as described in Sec.\ref{alternate_sec}, for each of the shifted values of the parameters $\theta_i$ followed by measuring the operators of the DLA. This will incur a total of $2p\times dim(\mathfrak{g})$ unique circuit evaluations for $U_q$ with $p$ parameters and $U_{\mathfrak{g}}$ with $dim(\mathfrak{g})$ operators. Because it requires fewer unique circuit evaluations for the full circuit implementation, we use it for our numerical simulations.
However in practice, one might have to make a choice between these two methodologies based on circuit depth and number of unique circuit evaluations.

\section{Results}\label{result}

We demonstrate the effectiveness of our proposals on two main tasks without having any particular restriction on the DLA sizes of the operators involved: VQE and Classification using QNNs. In the VQE task where the Hamiltonian has \textit{poly}-DLA, we use HELIA and our proposed training methods. If one wishes to apply our methods, for the case of \textit{exponential}-DLA Hamiltonian, it cannot be applied directly owing to limitations of $\mathfrak{g}$-sim. Instead, we propose a pre-training procedure before Full-PSR optimization of the complete Hamiltonian. The corresponding LTFIM results are presented in Sec.~\ref{gen_hamil} and Appendix~\ref{ltfim_finite_size}.

For Classification of quantum phases using QNNs, we consider the bond-alternating spin-$1/2$ Heisenberg chain which spans an \textit{exponential}-DLA. To construct HELIA out of the given Hamiltonian, we choose multiple \textit{poly}-DLA options for $U_{\mathfrak{g}}$ of the quantum circuit (as described in Sec.~\ref{pqcs}). In these numerical experiments, our proposed methods show improvement in accuracy.

Table~\ref{tab:numerical_results_summary} presents the finite-size benchmark suite for HELIA and the proposed training protocols.

\begin{table*}[!t]
\caption{\textbf{Finite-size benchmark suite for HELIA and hybrid training protocols.} Overview of the numerical experiments used to assess HELIA and the proposed training protocols. \textbf{All results are finite-size numerical studies, and no claim of quantum advantage is made.} For each benchmark, the table reports the physical model, simulated system sizes, principal observations relative to Full-PSR or the relevant reference, and the qualifications governing their interpretation. QPU-call reductions count logical shifted-cost evaluations and required intermediate-state observable expectations up to the best iterate; they exclude shot budgets, observable grouping, basis changes, circuit compilation, and classical processing. The gradient observations compare the displayed variances with the HEA50 reference and do not imply asymptotic barren-plateau avoidance.}
\label{tab:numerical_results_summary}
\scriptsize
\setlength{\tabcolsep}{3pt}
\renewcommand{\arraystretch}{1.18}
\begin{tabularx}{\textwidth}{>{\raggedright\arraybackslash}p{0.13\textwidth} >{\raggedright\arraybackslash}p{0.13\textwidth} >{\centering\arraybackslash}p{0.06\textwidth} X X}
\toprule
\textbf{Experiment type} & \textbf{\shortstack{Hamiltonian/\\model}} & \textbf{Qubits} & \textbf{Main improvements or observations} & \textbf{Comments} \\
\midrule
\hyperref[numerical]{XY VQE benchmark} & XY spin Hamiltonian; polynomial DLA & 6--18 & \textbf{Optimization:} Up to approximately $60\%$ fewer QPU calls, a 26.57-percentage-point increase in success, and an order-of-magnitude reduction in relative error. \newline\textbf{Gradients:} HELIA gradient variance decays more slowly than the HEA50 reference over the tested sizes and depths. & Improvements depend on the configuration. The XY model is classically tractable, and the gradient comparison is finite-size and not complexity-matched. \\
\hyperref[tfim_vqe]{TFIM VQE benchmark} & Transverse-field Ising model; polynomial DLA & 8--14 & \textbf{Optimization:} Alt+Sim generally provides comparable or lower error and higher success, with up to $52.11\%$ fewer QPU calls. \newline\textbf{Gradients:} HELIA maintains larger gradient variance than HEA50 over the displayed system sizes. & Improvements depend on the configuration; although generally more efficient, one setting increases QPU calls. No asymptotic gradient-scaling claim is made. \\
\hyperref[gen_hamil]{LTFIM trajectory pretraining} & Longitudinal-transverse-field Ising model; exponential DLA & 10, 12 & QPU-call reductions of $10.71\%$--$16.18\%$, together with lower average relative error. & Uses fixed pretraining schedules with comparable total iteration counts. \\
\hyperref[ltfim_finite_size]{LTFIM finite-size warm start} & LTFIM with $\lambda\in\{0.5,1,2\}$ & 6--12 & Mean QPU-call reductions of $26.29\%$--$29.94\%$ across couplings, with a maximum of $46.2\%$; slightly improved success and median-error ratios. \newline\textbf{Gradients:} The response is block-specific and heterogeneous. & Includes pretraining calls and conditions resource comparisons on paired successful trials. No barren-plateau mitigation claim is made. \\
\hyperref[lih_vqe]{LiH molecular VQE} & LiH in the STO-3G basis; exponential DLA & 12 & QPU calls are reduced by $9.09\%$; median relative error decreases from $0.02586$ to $0.02179$, approximately $16\%$. & Interpreted as comparable accuracy with reduced QPU usage because the trial-to-trial spread is substantial. \\
\hyperref[classification_results]{Quantum-phase classification} & Bond-alternating Heisenberg chain & 12 & The best Alt+Sim accuracy is $0.911\pm0.008$ for the ZX DLA, 2.8 percentage points above Full-PSR and 3.8 points above $\mathfrak{g}$-sim. & XY and YZ results are comparable to Full-PSR. \\
\hyperref[collective_spin]{Collective-spin VQE} & Permutation-invariant collective-spin orbit Hamiltonian & 6, 8 & Alt+Sim matches or exceeds Full-PSR success in all eight qubit-depth settings, improves it in six, and reduces QPU calls by $17.5\%$-$38.0\%$. & Logical-observable accounting only; physical measurement-grouping costs and hardware validation are excluded. \\
\bottomrule
\end{tabularx}
\end{table*}

\subsection{Ground State Estimation }\label{ground_state_VQE}

\begin{figure}[!thp]
    \centering
          \includegraphics[width=0.9\linewidth]{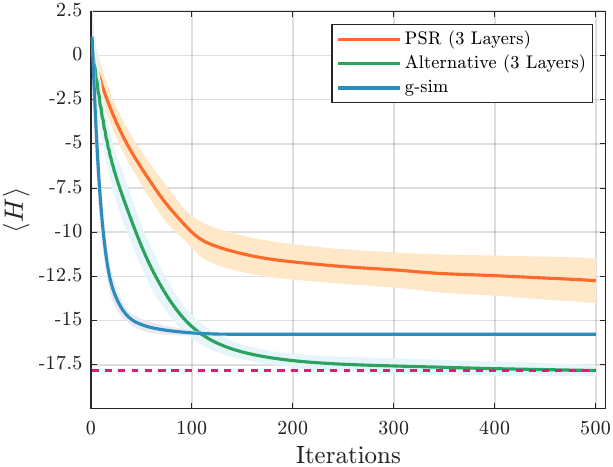}
    \caption{ \raggedright VQE performed for a 6-qubit TFIM Hamiltonian. All coefficients of the Hamiltonian terms are either $1$ or $-1$ chosen randomly. Only using HEA with PSR (shown in orange) or multi-qubit Pauli rotation gates with \g (shown in blue) is not enough to reach exact ground state (shown in red). But combining both methods into an Alternate training procedure (in green) allows us to reach the correct solution.}
    \label{simple_VQE}
\end{figure}

We demonstrate the benefit of using HELIA as compared to $U_q$ and $U_{\mathfrak{g}}$ separately through a simple example, illustrated in Fig.~\ref{simple_VQE}. To find ground state of a 6-qubit Transverse Field Ising Model (TFIM) Hamiltonian (with coefficients either $1$ or $-1$ chosen randomly) using VQE, a quantum circuit with only $U_q$ (HEA via PSR) or $U_{\mathfrak{g}}$ (multi-qubit Pauli gates via \g) fails to reach the target energy in 500 iterations. However, combining the two blocks using an Alternate training method (discussed in Sec.~\ref{training_scheme}) successfuly converges to the exact ground state energy. Although the combined ansatz has more trainable parameters than either of $U_q$ and $U_{\mathfrak{g}}$ individually, the quantum resources required to train the full ansatz are reduced significantly, compared to standard PSR, by using our training methods. We use this as a motivating example that the combined search space can solve problems that the individual circuits cannot. However, in the subsequent numerics, we keep HELIA as the trainable circuit and only vary the training methods to demonstrate the advantage of our methods.

To rigorously test the performance of our proposal, we employ several metrics.\\

\noindent\textbf{Relative Error}:
\begin{equation}
    \text{ Relative Error} = \Big|\frac{E_g^t - E_g^*}{E_g^*}\Big|\label{error}.
\end{equation}
Here, we define $E_g^t$ as the lowest possible energy achieved in the current trial and $E_g^*$ is an estimate of the true ground state energy of the Hamiltonian. For large Hamiltonians, it is computationally impractical to numerically diagonalize it and obtain the true ground state energy. Instead, taking inspiration from Ref.~\cite{goh2023lie}, we use the lowest energy reached across all trials for the corresponding Hamiltonian as an estimate of the true ground state energy $E_g^*$, whenever exact diagonalization is prohibitively expensive. \\

\begin{figure*}[tbh!]
    \includegraphics[width = 1\linewidth]{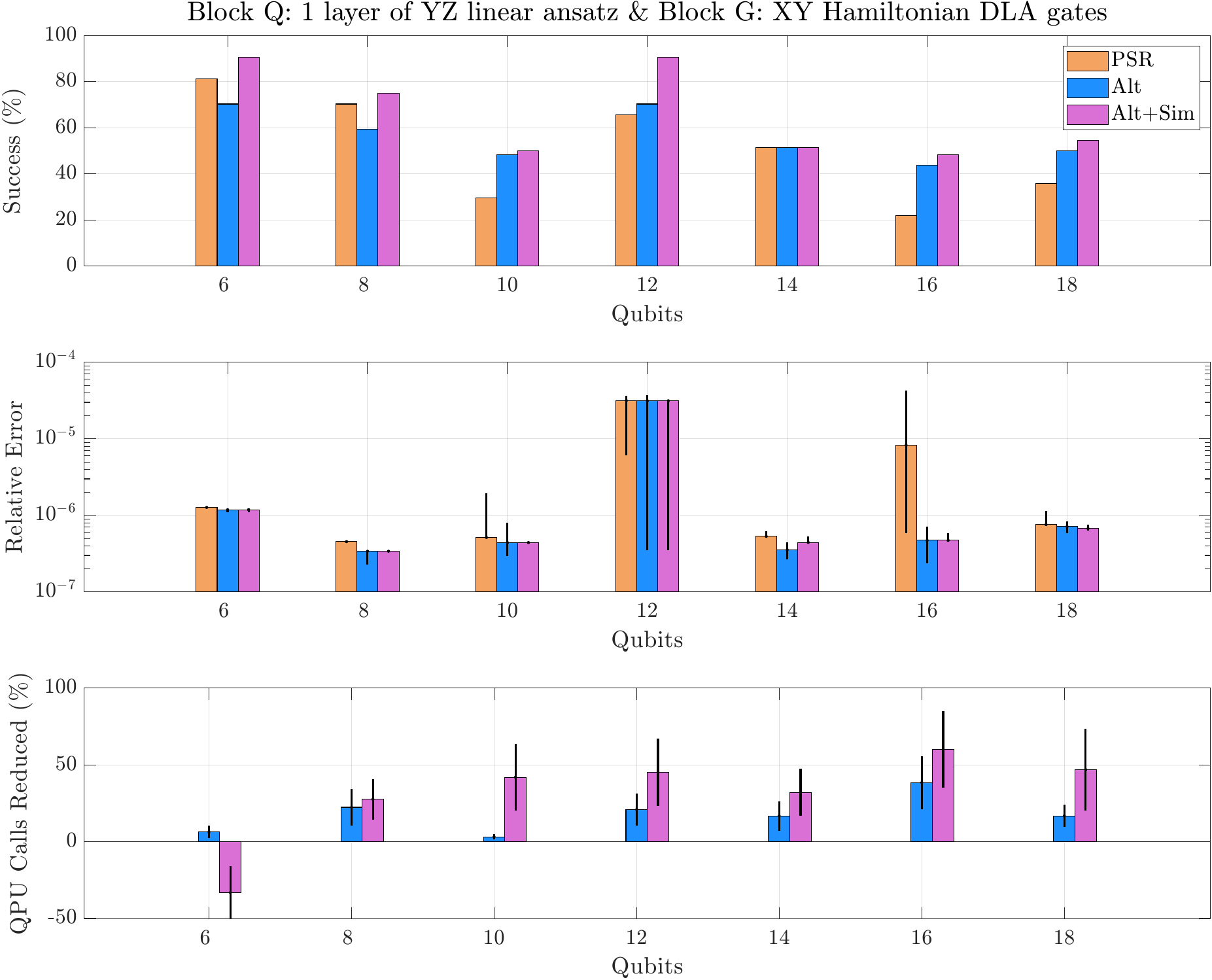}

    \caption{\raggedright \textbf{XY Hamiltonian VQE (1 YZ linear layer)}: The success (Row 1), relative error (Row 2) and QPU calls reduction  (Row 3) are plotted for configurations with 1 layer of YZ linear ansatz in $U_q$ (referred to as Block Q) and Hamiltonian DLA gates in $U_{\mathfrak{g}}$ (referred to as Block G)  for 6 to 18 qubits. For each qubit count and metric, PSR (orange), Alternate (Alt, blue) and Alternate+Simultaneous (Alt+Sim, purple) training methods are compared. \textbf{On the first row}, we show that Alt+Sim attains equal or higher success rate as Full-PSR for every tested qubit count in this configuration, whereas Alternate improves over Full-PSR for several, but not all, qubit counts. \textbf{For relative error (Row 2)}, we plot the median and the $25\%$ and $75\%$ quantiles (in black). The Alt method performs well for Layer 1 configuration, which is improved by Alt+Sim, giving lower relative error than PSR for all qubits. Most importantly, \textbf{on the lowermost row} we plot the percentage of reduction in QPU calls for Alt and Alt+Sim compared to PSR.
    The spread in standard deviation is shown in black. Except for 6 qubit case, we consistently see reduction in QPU calls using our methods. Alt+Sim generally lowers successful-trial relative error, while its success rate generally improves relative to Full-PSR.}
    \label{xy_layer1}
\end{figure*}
\begin{figure}[!tbh]

    \includegraphics[width = 0.9\linewidth]{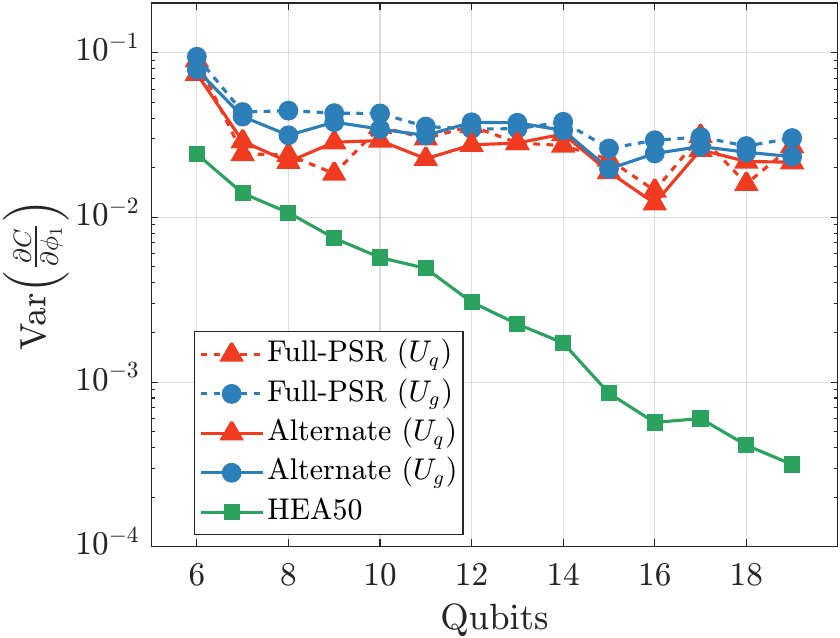}
        \caption{\raggedright \textbf{Gradients for XY Hamiltonian VQE:} Variance of the first parameter from $U_q$ (red) and $U_{\mathfrak{g}}$ (blue) are plotted with increasing qubits for both Full-PSR and our Alternate method for finding the ground state of the XY Hamiltonian using VQE. The diagrams show the plots for 1 YZ linear layer in $U_q$. The same task is performed using a deep HEA (referred to as HEA50) with standard PSR and the variance of its first partial derivative is plotted in green, to serve as a reference for how BP shows up in practice. In both $U_q$ and \Ug, we notice a slower decay of gradients as compared to HEA50 over the displayed system sizes.}
    \label{vanishing_fig1}
\end{figure}

\noindent \textbf{Success}:
Fraction of the trials that are able to reach below a relative error threshold of $10^{-3}$, which has been chosen ad-hoc. Success rates are computed over all trials. Relative-error statistics and QPU-call reductions are computed only over trials satisfying the relative-error success threshold of $10^{-3}$. This gives an estimate of how often the method can get close to the true solution. The \g summaries are reported separately for comparison.\\

\noindent \textbf{QPU Calls Reduction}:
Fraction of quantum circuit evaluations, referred to as QPU calls, reduced in current method as compared to standard training with PSR. We use all the parameter-shifted circuit evaluations for Full-PSR in each iteration to measure quantum hardware usage resulting in $2(p+g)$ QPU calls per iteration. For Alternate and Simultaneous protocols, we use all shifted circuit evaluations when obtaining gradients of $U_q$ (i.e, $2p$) and the number of DLA generators (i.e, $g$) that need to be measured independently for obtaining gradient of \Ug using $\mathfrak{g}$-sim resulting in $(2p +g)$ QPU calls per iteration to obtain the gradients of the full circuit.

Importantly these are logical evaluation counts where each shifted cost evaluation and each required intermediate-state operator expectation is counted as one logical evaluation which excludes the shots needed for a target precision, observable grouping, basis changes, circuit compilation, and classical processing time, as they can vary vastly from problem to problem. We use the number of QPU calls required to reach the point of lowest energy in the corresponding trial as our estimate.

\subsubsection{Improved VQE for XY Spin Hamiltonian (\textit{poly}-DLA)}\label{numerical}

We apply our chosen ansatz and proposed training methods on VQE, and demonstrate its effectiveness for this task in comparison to PSR and $\mathfrak{g}$-sim. We consider a XY Spin Hamiltonian given by:
\begin{equation}\label{xy_hamil}
    H_{XY} = \sum_{i=0}^{N-2} \left(\alpha_i\hat{X}_i\hat{X}_{i+1} + \beta_i\hat{Y}_i\hat{Y}_{i+1}\right),
\end{equation}
which has a \textit{poly}-DLA with the dimension given by $dim(\mathfrak{g}) = 2n^2 - n$.

Since all of DLA elements of the XY spin Hamiltonian produce zero expectation value with the zero initial state $|0^{\otimes n}\rangle$, we added a layer of Hadamard gates $H^{\otimes n}$ to facilitate training . In order to maintain consistency we have added the $H^{\otimes n}$ layer before $U_{\mathfrak{g}}$ in all of the examples for the Hamiltonian. Hence, via \g we optimize the following cost function,
\begin{equation}
    C(\phi) = \langle 0| H^{\otimes n} U_\mathfrak{g}^\dagger(\phi) H_{XY} U_\mathfrak{g}(\phi) H^{\otimes n}  |0\rangle.
\end{equation}

For the Full-PSR (all parameters of $U_q$ and $U_{\mathfrak{g}}$ are optimized using PSR), Alternate (Alt) and Alternate + Simultaneous (Alt+Sim, using Alternate training followed by Simultaneous training) methods, we consider a larger ansatz by adding $U_q$, and minimize the following cost function:

\begin{equation}\label{cost_func}
    C(\theta,\phi) = \langle 0| U_q^\dagger(\theta) H^{\otimes n} U_\mathfrak{g}^\dagger(\phi) H_{XY} U_\mathfrak{g}(\phi) H^{\otimes n} U_q(\theta)|0\rangle.
\end{equation}

Now we are not restricted to only searching in the same polynomial space but can search a much larger state space, by virtue of the gates in $U_q$. For visualization, we plot the success, relative error and QPU reduction metrics comparing Full-PSR, Alt and Alt+Sim using 1 YZ linear layer in $U_q$ in Fig. \ref{xy_layer1}. In the plots we interchangeably refer to $U_{\mathfrak{g}}$ and $U_q$ as Block G and Q respectively. The plots for other configurations are provided in Fig. \ref{xy_vqe_all} of Appendix \ref{xy_numerical},  while the actual numbers along with comparison with \g is provided in Table \ref{vqe_xy_error} and \ref{vqe_xy_successqpu}.  The relative error for \g is significantly higher in the chosen examples, as compared to other methods. Hence, we ignore it in the plots, and only mention the values in Table \ref{vqe_xy_error}. In Sec. \ref{gen_hamil}, we also discuss how this method can be used as a starting point for finding ground states in more general Hamiltonians without the restriction of having a \textit{poly}-DLA.

Although we are exploring a larger state space, the ground state might lie close to the space searched by $\mathfrak{g}$-sim, rendering a quantum method unnecessary for the task. This happens in the odd qubit examples for XY Hamiltonian, where \g is able to get very low relative errors without additional gates.
A possible explanation for the observed odd/even asymmetry is that, for finite XY chains, the Jordan-Wigner mapping leads to a free-fermionic description in which the spectrum is naturally decomposed into distinct parity sectors, and the physical ground state is obtained by comparing the lowest energies across these sectors \cite{de2009xy}. At the same time, the $\mathfrak{g}$-sim update is driven by the expectation-value vector of DLA observables for the chosen input state, so its effectiveness can depend on how strongly the initialization overlaps with the symmetry sector most relevant to the target state. We therefore interpret the present odd/even behavior as an empirical feature of the initialization adopted in our simulations, rather than as a universal property of the method.
We mention this here, but in the numerics we only focus on even qubit cases to demonstrate where our method can be effective and standard \g fails. Some of the odd qubit examples are demonstrated in Appendix \ref{odd_gsim}.

The XY Hamiltonian provides a convenient controlled benchmark for our purposes because its associated DLA is polynomially sized and the corresponding $\mathfrak{g}$-sim workflow can be implemented explicitly. However, we stress that the XY Hamiltonian is classically tractable, since it can be mapped via the Jordan-Wigner transformation to a quadratic free-fermionic Hamiltonian. Accordingly, the role of this benchmark in the present work is not to demonstrate quantum advantage, but to isolate and evaluate the behavior of the proposed HELIA training strategy in particular.

The XY and TFIM benchmarks studied here are free-fermionic and therefore admit independent classical ground-state methods. Their role is to provide controlled settings for evaluating hybrid gradient computation and resource allocation. To test portability beyond this class, Appendix~\ref{collective_spin} extends the HELIA updates to a non-free-fermionic, permutation-invariant collective-spin orbit algebra based on Ref.~\cite{barligea2026enabling}. This establishes applicability of the hybrid update procedure beyond individual-Pauli free-fermionic constructions. Because the particular collective-spin Hamiltonian is fully permutation invariant, however, we do not claim that its ground-state problem is classically hard. Its role is to test whether symmetry-adapted observable propagation can support HELIA training and reduce logical QPU evaluations in a non-free-fermionic setting.

\textbf{Experiment Details}: All parameters associated with the Hamiltonian terms are initialized independently according to a uniform distribution over the interval $[-1,1]$. We present the results for 4 different configurations for each qubit, with the number of YZ linear layers in $U_q$: 1, 3, 6 and 9 changing between the configurations. The results are given in Table \ref{vqe_xy_successqpu} and \ref{vqe_xy_error} of Appendix \ref{xy_numerical}, and visualized in Fig.~\ref{xy_layer1} for layer 1 and Fig. \ref{xy_vqe_all} of Appendix \ref{xy_numerical} for layer 3, 6 and 9.
Here, in the Alt+Sim method, we train the ansatz using Alternate for 500 iterations followed by Simultaneous till convergence.
In order to keep the comparison fair, we use an Adam optimizer with initial learning rate $0.01$ and initial state $\ket{0}^{\otimes n}$, and keep it consistent across all experiments.

In Fig.~\ref{xy_layer1}, we plot the mean and standard deviation (in black) over the successful trials when measuring the QPU calls reduction for the configuration with 1 YZ linear layer in $U_q$. We find the relative error is often asymmetrically spread around the mean, owing to being lower-bounded by $0$ (by definition Eq.~\eqref{error}) and upper bounded by $10^{-3}$ (success threshold). Hence, we resort to the median and show the spread from $25\%$ to $75\%$ quantiles in black instead of mean and standard deviation. This is due to the fact that median is a more robust metric when the data is distributed asymmetrically. This unequal distribution is also explicit from the $25\%$ to $75\%$ quantiles which extend unequally from the median. The spread from $25\%$ to $75\%$ quantiles in relative error often varies a lot across qubits due to initialization sensitivity with broad spreads occuring where some trials converge to good solutions and others stall, and narrow spreads occur where successful trials cluster tightly. The corresponding plots for 3,6 and 9 YZ linear layers in $U_q$ are provided in Appendix \ref{xy_numerical} (Fig. \ref{xy_vqe_all}).

Across the tested configurations, our method often reduces the relative error of ground-state estimation (by up to an order of magnitude), improves success rates by up to 26.57 percentage points, and reduces QPU calls by up to approximately $60\%$ relative to standard PSR training of the same ansatz. The improvements are configuration-dependent rather than uniform across all settings.

\begin{resultbox}{Key result}
Across 6--18 qubits and the tested circuit depths, the proposed training methods improve success and successful-trial relative error in many configurations while reducing logical QPU calls by up to approximately $60\%$ relative to Full-PSR. The positive trends observed across much of this finite-size range motivate evaluation at larger qubit counts, where similar resource reductions may persist. The XY model is classically tractable, so these results do not demonstrate quantum advantage.
\end{resultbox}

\subsubsection*{Gradient behavior in the studied XY configurations}

We analyze the parameter gradients for the given VQE task with varying number of qubits. We track the gradient of the first parameter from each of $U_q$ and $U_\mathfrak{g}$, and plots its variance during the training (Fig.\ref{vanishing_fig1}). Each datapoint is obtained after evaluating the variance from the full training of 64 independent trials.

For comparison, we also perform the same VQE task using a 50 layer HEA (referred to as HEA50), which is known to exhibit BP~\cite{mcclean2018barren}, and plot the variance of the first parameter of this circuit. This circuit serves as a selected deep-circuit reference rather than a baseline matched in depth, parameter count, or gate count. HELIA shows a clear improvement in terms of slow decay in gradients in Fig.~\ref{vanishing_fig1} for the configuration with 1 YZ linear layer in $U_q$. The gradients for $U_q$ and $U_{\mathfrak{g}}$ demonstrate a much slower decay compared to the HEA50 circuit. However, we highlight that HEA50 is only supposed to serve as a reference for BP in literature in order to demonstrate that HELIA is indeed able to preserve non-exponentially decaying gradients while training. Additionally, the improved gradient behavior is a result of HELIA and not the training methods as they will obtain the same gradients as PSR as can be seen from Fig.~\ref{vanishing_fig1}. Similar trend was observed for increasing YZ linear layers up to 9, as shown in Fig.~\ref{vanishing_fig2}. However, these observations do not determine an asymptotic scaling law or establish that HELIA is generally barren-plateau-free.

The BP condition in Refs.~\cite{ragone2023unified,fontana2023adjoint} is not applicable here because we choose gates from a \textit{poly}-DLA and a shallow HEA circuit, preventing us from forming a 2-design on $SU(2^n)$. To understand this further, we consider the variance of parameters of $U_{\mathfrak{g}}$. Since we do not require shallow circuits for this block, we can focus on the case where the gates form a 2-design on $e^{i\mathfrak{g}}$ (not on full $SU(2^n)$). Using the result from Ref.~\cite{ragone2023unified} and considering $\mathfrak{g}$ to be simple, we can write the following,
\begin{equation}\label{BP_eq}
    \text{Var}_{\phi} [C(\theta,\phi)] = \frac{\mathcal{P}_\mathfrak{g}(\rho_q(\theta)) \mathcal{P}_\mathfrak{g}(\hat{O})}{dim(\mathfrak{g})},
\end{equation}
where $\rho_q(\theta) = \Tilde{U}_q(\theta)\rho_0 \Tilde{U}_q^\dagger(\theta)$. $\mathcal{P_\mathfrak{g}(\cdot)}$ denotes the $\mathfrak{g}$-Purity of an operator and $dim(\cdot)$ denotes the dimension of an algebra, as defined in \cite{ragone2023unified}. We elaborate on these functions in Appendix \ref{purity_sec}.

By design of HELIA and choice of operators $\hat{O}$, $\mathcal{P}_\mathfrak{g}(\hat{O})= \Tr (\hat{O}^\dagger\hat{O}) = poly(n)$ and $dim(\mathfrak{g}) = poly(n)$. Hence we only need to ensure that $\mathcal{P}_\mathfrak{g}(\rho_q(\theta))$ is not decaying exponentially. By numerically testing at varying qubit counts using $\mathfrak{g}_{XY}$, we find that $\mathcal{P}_\mathfrak{g}(\rho_q(\theta))$ decays sub-exponentially for constant and logarithmic depth circuits, and exponentially for linear depth circuits. The details of the experiment are shown in Appendix \ref{purity_sec}.

Using Eq.\eqref{BP_eq}, we can see that subexponential decay of average purity implies that the variance of loss function with respect to $U_{\mathfrak{g}}$ parameters also decays sub-exponentially. Importantly, this does not guarantee substantial gradients for every $\theta$ value in $U_q$, but only on average.

\begin{figure*}[!htp]
    \centering
    \scalebox{0.85}{
    \includegraphics[width=1\linewidth]{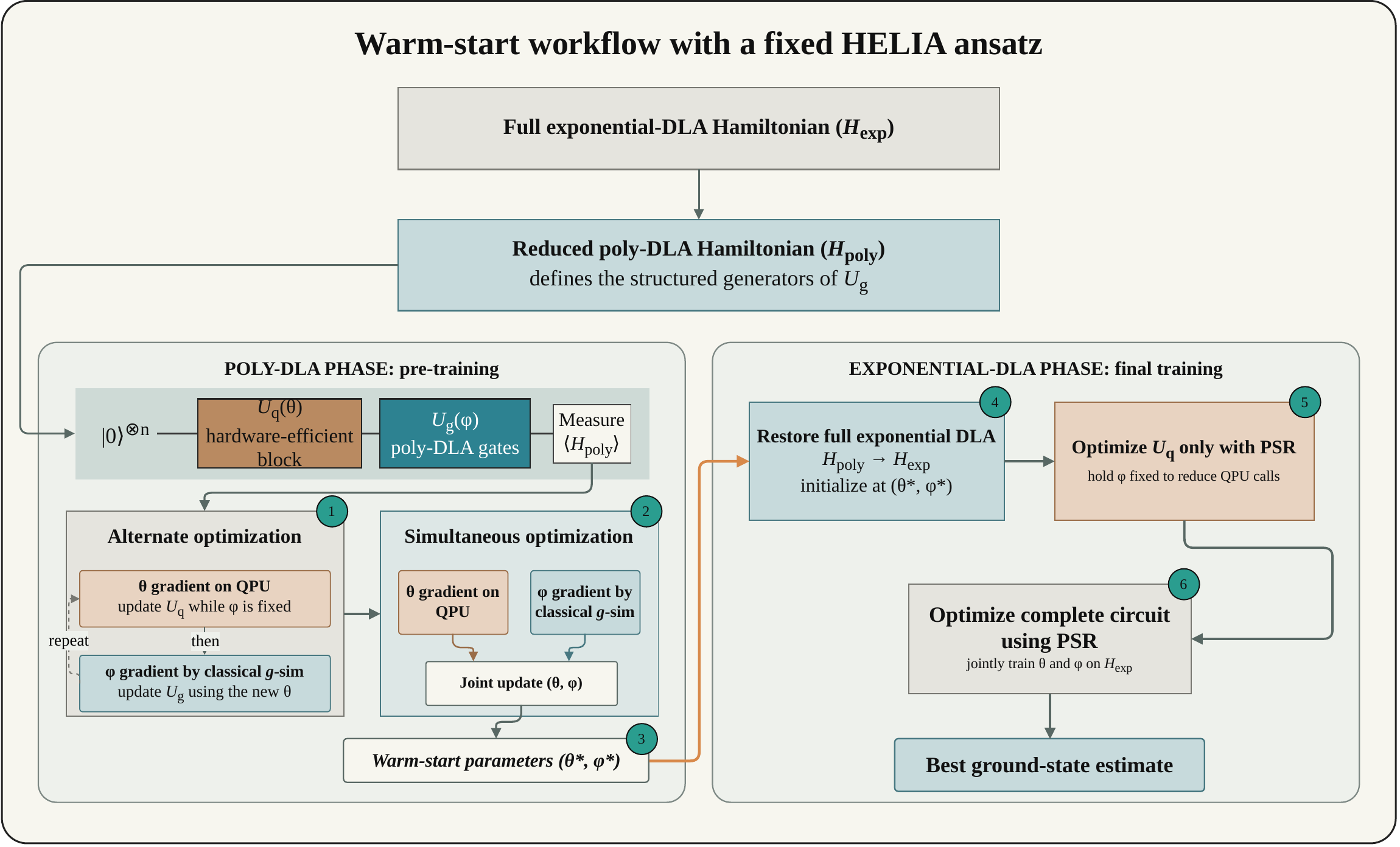}}
\caption{\raggedright
\textbf{Warm-start pretraining workflow for exponential-DLA Hamiltonians.}
The HELIA ansatz is first optimized on a polynomial-DLA restriction of the target Hamiltonian using Alternate and Simultaneous training. The resulting parameters initialize optimization of the complete exponential-DLA Hamiltonian, first by updating $U_q$ and subsequently by training the full circuit with PSR. The circuit architecture remains unchanged throughout the procedure; only the Hamiltonian and gradient-evaluation protocol change. For the tested LTFIM configurations, this warm start achieves lower average relative error and reduces logical QPU calls by $11.9\%$--$46.2\%$ relative to Full-PSR, including the pretraining cost. Detailed finite-size results across system sizes, circuit depths, and longitudinal coupling strengths are provided in Appendix~\ref{ltfim_finite_size}.}
    \label{fig:ltfim-workflow}
\end{figure*}

As shown in Fig.\,\ref{vanishing_fig1}, numerical tests on the full circuit created by composing $U_q$ and $U_{\mathfrak{g}}$, shows polynomial decay of gradients in both. However, here we only provide an outline of proof for $U_{\mathfrak{g}}$, and show that polynomial gradients can be seen for sub-linear circuit depth of block $U_q$. We believe that the polynomial decay in $U_q$ also stems from the shallowness of the HEA circuit, and might be preserved for sub-linear circuit depths. We leave the mathematical proof for this intuition as future work.

The architecture introduces a trainability-expressivity trade-off. Restricting the Lie-algebra block permits efficient classical gradient evaluation, while limiting the hardware-efficient block can preserve favorable finite-size gradient behavior. These restrictions also constrain the accessible variational family. Increasing the hardware-efficient depth can enlarge that family, but may reduce the favorable gradient behavior observed at shallower depths. The present results identify useful tested configurations rather than a depth-independent balance between trainability and expressivity.

\begin{resultbox}{Trainability diagnostic}
HELIA gradients decay more slowly than the HEA50 reference over the tested qubit counts and circuit depths. This positive finite-size trend suggests that favorable gradient behavior may persist at larger qubit counts. However, HEA50 is not complexity-matched, and larger-scale studies are required before drawing conclusions about asymptotic scaling or general barren-plateau avoidance.
\end{resultbox}

\subsubsection{Pre-training in LTFIM Hamiltonian (\textit{exponential}-DLA)}\label{gen_hamil}

The ability to combine \g and PSR efficiently in our proposal depends on the \textit{poly}-DLA of $U_{\mathfrak{g}}$ as well as the associated measurement operator. This limits the form of Hamiltonians on which we can perform the VQE task. However, our method can serve as a pretraining strategy for \textit{exponential}-DLA Hamiltonians within a fixed HELIA framework, as demonstrated in Appendix\,\ref{ltfim}, Algorithm \ref{gen_hamil_alg} and schematically in Fig.~\ref{fig:ltfim-workflow}. We first illustrate this warm-start strategy through representative 10- and 12-qubit optimization trajectories and then evaluate it over system sizes $n=6$-$12$, three circuit-depth families, and longitudinal couplings $\lambda\in\{0.5,1,2\}$. The finite-size results are summarized in Table~\ref{tab:ltfim_coupling_summary} and detailed in Appendix~\ref{ltfim_finite_size}.

\begin{table*}[]
\centering
\caption{Coupling-wise summary over all qubit-depth configurations. Values are mean
$\pm$ sample standard deviation across configurations. The success and relative-error ratios are
$R_{\mathrm{succ}}=p_{\text{pre-trained}}/p_{\mathrm{Full\text{-}PSR}}$ and
$R_{\mathrm{err}}=\tilde{\epsilon}_{\text{pre-trained}}/
\tilde{\epsilon}_{\mathrm{Full\text{-}PSR}}$, respectively. QPU reduction is with respect to full PSR on paired successful trials, and includes pre-training QPU calls as well.}
\label{tab:ltfim_coupling_summary}
\small
\begin{tabular}{c| c |c |c}
\hline
$\lambda$ &
$R_{\mathrm{succ}}$ &
$R_{\mathrm{err}}$ &
QPU Calls reduction \\
\hline
0.5 & $1.0093 \pm 0.0272$ & $0.9085 \pm 0.1921$ & $29.9430\% \pm 10.2848\%$ \\
1.0 & $1.0110 \pm 0.0405$ & $0.8883 \pm 0.2122$ & $28.2384\% \pm 8.9714\%$ \\
2.0 & $1.0175 \pm 0.0418$ & $0.8852 \pm 0.2249$ & $26.2868\% \pm 6.7164\%$ \\
\hline
\end{tabular}
\end{table*}

Although similar to the pre-training approach used in Ref.~\cite{goh2023lie}, \emph{our method is unique in its use of the full ansatz over the complete training procedure, with only the Hamiltonian changing between the two phases of training}. Changing the ansatz structure between pre-training and full-training phases as discussed in~\cite{goh2023lie} can potentially alter the optimization landscape significantly. On the other hand, only updating the Hamiltonian with \textit{exponential}-DLA terms will introduce changes in the landscape of the order of magnitude of the new terms, which is generally much smaller than the initial Hamiltonian magnitude.  The first phase can be thought of as a warm-start method, while the second phase trains the circuit to find ground-state of the full Hamiltonian starting from the parameters of the first phase. We expand this point further in the following paragraphs and highlight this important difference through a numerical example.

To start the pre-training method, we need to choose a subset of operators that form a \textit{poly}-DLA. In the \textit{poly}-DLA phase, we perform Alternate, followed by Simultaneous training on the reduced Hamiltonian. During the \textit{exponential}-DLA phase of training in Alg.\,\ref{gen_hamil_alg} (Appendix \,\ref{ltfim}), we train only $U_q$ for some iterations before training the full block. This is done with the aim of keeping the overall QPU calls low, as $U_q$ has fewer trainable parameters and hence needs less circuit evaluations to obtain gradients during training. Finally, all of the parameters are trained together using standard PSR. The experimental details are provided in Appendix \ref{ltfim}.

We demonstrate the results of this proposal for the longitudinal-transverse-field Ising model (LTFIM),
\begin{equation}
    H_{LTFIM} = \sum_{j=0}^{N-2} \alpha_j X_j X_{j+1} + \sum_{j=0}^{N-1} (\beta_j Z_j + \gamma_j X_j)
\end{equation}
which has an \textit{exponential}-DLA. One choice of \textit{poly}-DLA can be constructed by dropping the longitudinal terms
\begin{equation}
    \mathfrak{g}_{TFIM} = \langle \{iX_j X_{j+1}, iZ_j\} \rangle_{Lie}.
\end{equation}

\begin{figure*}
    \begin{subfigure}{0.45\linewidth}
        \includegraphics[width=1\linewidth]{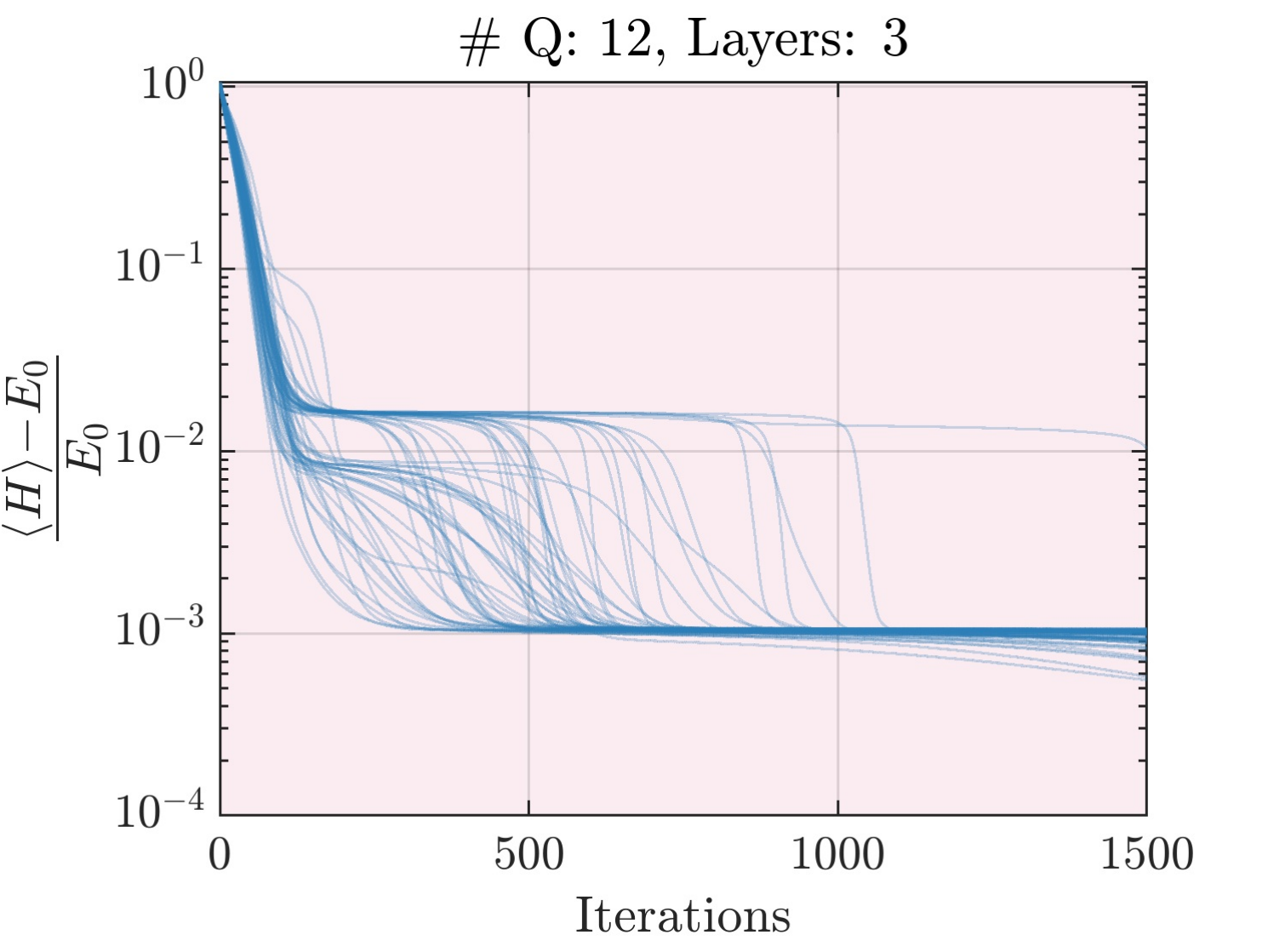}
        \caption{Trained using standard PSR}
    \end{subfigure}
    \begin{subfigure}{0.45\linewidth}
        \includegraphics[width=1\linewidth]{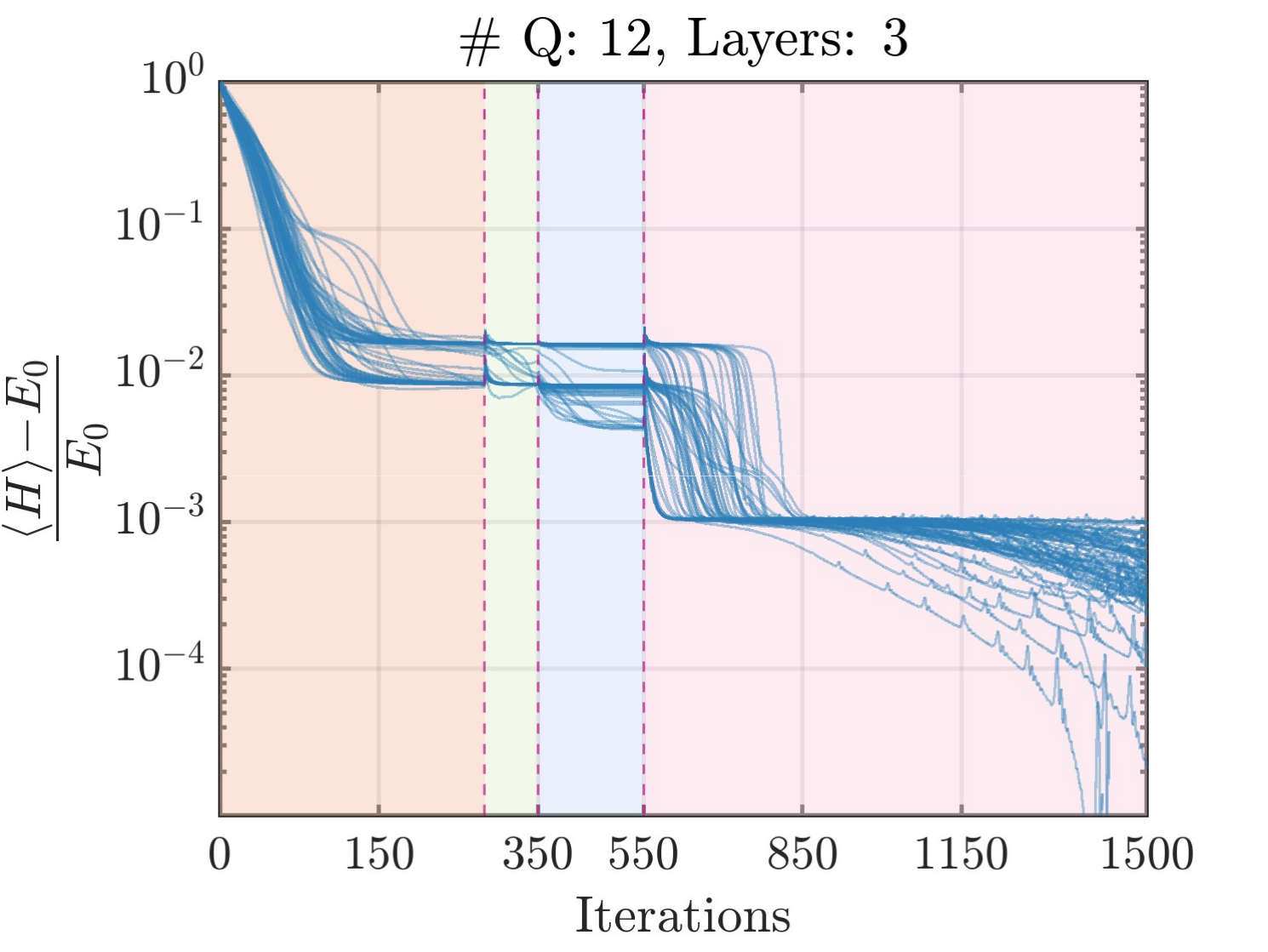}
        \caption{Pre-training using our method followed by PSR}
    \end{subfigure}
    \caption{ \raggedright Demonstration of PSR (a) and our method (b) to run VQE for a 12-qubit LTFIM Hamiltonian (with an \textit{exponential}-DLA) using 3 YZ-linear layers in $U_q$. The orange and green regions show Alternate and Simultaneous training, respectively, for the restricted Hamiltonian, followed by training $U_q$ and the complete circuit ($U_q$ and \Ug) in the blue and pink regions, respectively, for the full LTFIM Hamiltonian. Our training methods in the initial phase allow VQE to reach much closer to the target ground state during the final training than using PSR for the full training, for similar total iteration counts. For the overall training, the QPU calls are reduced by approximately $16.18\%$.}
    \label{general_hamil_img1}
\end{figure*}

We plot the relative error metric in Eq.~\eqref{error} for the 12 qubit LTFIM Hamiltonian with 3 YZ linear layers in $U_q$ in Fig.~\ref{general_hamil_img1}. Importantly, although the initial training phase is based on the reduced Hamiltonian, the plotted energy values correspond to expectation values of the full Hamiltonian. As shown in Fig.~\ref{general_hamil_img1}, the algorithm is indeed able to converge to good solutions by using  our methods in the first phase of the training.
Compared with PSR, our proposal shows lower relative error value on average. The QPU calls required for PSR is $6.12e+5$ while our method requires $5.13e+5$ ($\sim 16.18\%$ reduction) respectively. Overall, our method shows lower relative error while reducing the QPU calls relative to Full-PSR for a variational task involving operators with exp-DLA.
The corresponding plots for other configuration and qubit counts are shown in Appendix \ref{ltfim} (Fig.~\ref{gen_hamil_img2}).

\red{We extend the trajectory-level analysis in Appendix~\ref{ltfim_finite_size} to system sizes $n=6$--$12$, considering constant-, logarithmic-, and linear-depth circuit families across different longitudinal couplings. Table~\ref{tab:ltfim_coupling_summary} summarizes the success-rate ratio, median relative error ratio and QPU Calls reduction achieved by the proposed pre-training scheme and Full-PSR for $\lambda\in\{0.5,1,2\}$. Across all three coupling strengths, pre-training achieves a slightly higher average success rate than Full-PSR, and lower median relative error. Furthermore, among successful trials, the proposed approach reduces QPU calls by $11.9\%$--$46.2\%$ across the tested settings, including QPU calls incurred during the initial HELIA-based training phase where applicable. These finite-size comparisons however do not establish asymptotic savings or barren-plateau mitigation for the full LTFIM Hamiltonian.}

\begin{resultbox}{Warm-start result}
For the exponential-DLA LTFIM, reduced-Hamiltonian pretraining lowers average relative error and reduces logical QPU calls by $11.9\%$--$46.2\%$ across the tested settings. The persistence of positive reductions across multiple sizes, depths, and coupling strengths suggests that pretraining may remain beneficial at larger system sizes. These finite-size results do not establish asymptotic savings or barren-plateau mitigation.
\end{resultbox}

\subsubsection{Pre-training in LiH Hamiltonian (\textit{exponential}-DLA)}\label{lih_vqe}

The pre-training method is useful for ground-state estimation of molecules as well, where the Hamiltonian can span an exponential-DLA. To demonstrate that, we consider the example of LiH molecule.

The electronic structure of the LiH molecule was modeled using the Slater-type orbital (STO-3G) minimal basis set, resulting in a second-quantized fermionic Hamiltonian~\cite{sibaev2020molecular, Whitfield10032011}. This Hamiltonian was then mapped to a qubit representation using the Jordan–Wigner transformation implemented using OpenFermion~\cite{somma2002simulating, openfermion}. Without applying any orbital freezing or symmetry-based reductions, the full system comprises 12 spin orbitals, corresponding to a 12-qubit Hamiltonian. The resulting qubit Hamiltonian consists of a linear combination of 631 Pauli strings, each expressed as a tensor product over the Pauli operators ${I,X,Y,Z}$. An interatomic distance of $1.5~\text{\AA}$ is considered for the application of the VQE algorithm to this Hamiltonian. This uncompressed configuration serves as a benchmark for evaluating both the standard VQE approach and the proposed method.

Since we are considering a chosen fixed Hamiltonian, we first need to decide what portion of the computation can be delegated to classical hardware. For this, we choose the threshold of $n^2$ for an $n$-qubit Hamiltonian, to be the maximum allowed DLA size. This ensures we always use a reasonable amount of classical resource, and it does not become exponentially large in the number of qubits. Next, we arrange the Hamiltonian terms in descending order of magnitude and choose the first $k$ operators which have a DLA size below the stated threshold. In our case, this creates a DLA of size 78. Ignoring the identity, the operators considered within the DLA comprise $99.47\%$ of the magnitude of the full Hamiltonian. These operators form the generators for $U_{\mathfrak{g}}$ in the pre-training method shown in the previous example. In practice, it might be possible to increase the DLA size further based on available classical hardware.

\begin{table}[!tbh]
    \centering
    \begin{tabular}{c|c|c|c}
        \toprule
         \multicolumn{2}{c|}{} & \textbf{Full PSR} & \textbf{Our Proposal} \\
         \midrule
        \multirow{3}{*}{\shortstack{Relative\\Error}}
        & $q_{25}$ & 0.01009 & 0.00503 \\
        & Median & 0.02586 & \textbf{0.02179} \\
        & $q_{75}$ & 0.03770 & 0.02573 \\
        \midrule
        \multicolumn{2}{c|}{QPU Reduction} & - & \textbf{9.09}\% \\
        \bottomrule
    \end{tabular}
    \caption{Relative error quartiles and QPU reduction for Full-PSR vs our proposal.}
    \label{vqe_lih}
\end{table}

We compare Full PSR and our proposal in Table \ref{vqe_lih} in terms of median relative error and total QPU calls required in the corresponding methods. In the above experiment, we are able to reduce median relative error by $\sim 16\%$ and QPU calls by 9.09\% . However, as the spread across trials is non-negligible, we interpret this result as showing comparable accuracy with reduced QPU usage, rather than a statistically significant improvement in error. The details of the experiment are provided in Appendix \ref{liH_class_app}.

\begin{resultbox}{Molecular benchmark}
For 12-qubit LiH, pretraining reduces logical QPU calls by $9.09\%$, while the median relative error decreases from $0.02586$ to $0.02179$. This provides encouraging evidence that the pretraining strategy may extend beyond spin models to larger molecular systems. Because the trial-to-trial spread is substantial, the present result is interpreted as comparable accuracy with reduced QPU usage rather than a statistically significant error improvement.
\end{resultbox}

\subsection{Classification of Quantum Phases}\label{classification_results}

Our training methods are also useful for QML tasks such as classification. For numerical examples, we consider the bond-alternating spin-$1/2$ Heisenberg chain~\cite{PhysRevB.87.054402}:
\begin{equation}\label{bond-alt}
    H = J\sum_{i=0}^{N/2-1} \vec{S}_{2i} \cdot\vec{S}_{2i+1} + J'\sum_{i=0}^{N/2-1} \vec{S}_{2i+1} \cdot\vec{S}_{(2i+2)\bmod N},
\end{equation}
where $\vec{S} = (X,Y,Z)$ are the Pauli matrices and site indices are taken modulo $N$, corresponding to periodic boundary conditions. The model undergoes a second-order phase transition at $J/J' = 1$~\cite{PhysRevB.87.054402}.

We consider the task of learning the quantum phase of the ground state as a classification problem for a 12-qubit case. Since quantum phase transition only happens in the limit $N\rightarrow\infty$, it is not accurate to refer to these states as different phases. Rather, we just assign two separate labels $\pm1$ to the ground states for $J<J'$ and $J>J'$.
As the qubit counts are small enough, the training and test data (100 train + 100 test dataset) are generated by sampling $J,J'$ uniformly from $[-1,1]$,    numerically diagonalizing the Hamiltonian in Eq.~\eqref{bond-alt} and obtaining the ground state. This ground state is used as an initial state in the PQC and the ansatz is trained to generate the correct labels.

For designing the PQC, we encounter the choice of gates for $U_q$ and $U_\mathfrak{g}$. For $U_q$, we stick to the YZ linear ansatz choice with 9 layers. For $U_{\mathfrak{g}}$ however, we do not have a unique choice like we had for previous examples. Hence, we explore three different DLA choices to construct the gates:
\begin{align}
    \mathfrak{g}_{XY} &= \langle\{iX_jX_{j+1}, iY_j Y_{j+1} \}\rangle_{Lie} \\
    \mathfrak{g}_{YZ} &= \langle\{iZ_jZ_{j+1}, iY_j Y_{j+1} \}\rangle_{Lie} \\
    \mathfrak{g}_{ZX} &= \langle\{iX_jX_{j+1}, iZ_j Z_{j+1} \}\rangle_{Lie}.
\end{align}
Interestingly, the Hamiltonian is symmetric under the group action that maps $X_i,Y_i,Z_i$ Paulis cyclically into one another, which also maps the aforementioned DLA choices into one another.  However, in practice we notice that choice of DLA still impacts the peak test accuracies. The details of the experiment are provided in Appendix \ref{liH_class_app}. The results are tabulated in Table \ref{class_all}.

We observe different peak accuracies for each of the DLA choices with $i\mathfrak{g}_{ZX}$ providing the highest accuracy in \g($0.873\pm0.020$) . In each case, the Alt+Sim and Full-PSR methods provide a higher peak accuracy on average than \g. Furthermore, Alt+Sim is able to reach higher average test accuracies than Full-PSR for $\mathfrak{g}_{ZX}$, although only by up to $2.8$ percentage points.

\begin{table}[h]

    \centering
    \renewcommand{\arraystretch}{1.2}
    \begin{tabular}{c|c|c|c}
        \toprule
        \textbf{DLA}  & \textbf{g-sim} & \textbf{Full PSR} & \textbf{Alt+Sim} \\
        \midrule
        XY & 0.863 $\pm$ 0.005 & 0.884 $\pm$ 0.016 & \textbf{0.888 $\pm$ 0.021} \\
        YZ & 0.858 $\pm$ 0.006 & 0.887 $\pm$ 0.025 & \textbf{0.889 $\pm$ 0.020} \\
        ZX & 0.873 $\pm$ 0.020 & 0.883 $\pm$ 0.015 & \textbf{0.911 $\pm$ 0.008} \\
        \bottomrule
    \end{tabular}

\caption{\raggedright Classification task using $\mathfrak{g}$-sim, Full-PSR and Alt+Sim training (with 9 layers in $U_q$) for 12 qubits. Three different choices of \textit{poly}-DLA (referred to as $\mathfrak{g}_{XY}, \mathfrak{g}_{YZ}$ and $\mathfrak{g}_{ZX}$) are explored for $\mathfrak{g}$-sim, and also to construct $U_{\mathfrak{g}}$ for Alternate and Full-PSR method. Average peak test accuracy is reported for each of the configuration. Alt+Sim attains performance comparable to Full-PSR for the $\mathfrak{g}_{XY}$ and $\mathfrak{g}_{YZ}$ choices, while the most distinct improvement is observed for $\mathfrak{g}_{ZX}$ of up to 3.8 and 2.8 percentage points over the accuracy of $\mathfrak{g}$-sim and Full-PSR respectively.}
\label{class_all}
\end{table}

\begin{resultbox}{Classification result}
For the ZX DLA, Alt+Sim achieves a mean peak test accuracy of $0.911\pm0.008$, improving over Full-PSR by 2.8 percentage points. This positive result motivates testing the hybrid training strategy on larger classification problems and additional datasets. Results for the XY and YZ choices remain comparable to Full-PSR.
\end{resultbox}

\section{Conclusion and Outlook}\label{conclusion_sec}

Achieving quantum advantage requires deploying larger, more complex algorithms on quantum computers while ensuring classical intractability. For VQAs, which are inherently heuristic, determining thresholds for quantum utility or advantage is challenging. Consequently, a practical approach of implementing and rigorously testing larger-scale algorithms may yield critical insights into their potential for quantum advantage.

In this work, we provide a step in that direction by an informed ansatz design and combining \g with PSR, thus delegating the gradient-estimation task to both classical and quantum hardware carefully. We rigorously evaluate our proposals using VQE benchmarks spanning 6--18 qubits and a 12-qubit quantum-phase-classification benchmark. Our results demonstrate substantial reductions in QPU calls (even up to 60\% for VQE on a 16-qubit XY Hamiltonian), particularly at higher qubit counts, alongside improved relative error and success metrics for VQE and enhanced accuracy for classification. Notably, across the finite system sizes studied, HELIA exhibits slower gradient-variance decay than the selected deep hardware-efficient reference. While this comparison does not establish asymptotic barren-plateau avoidance, it indicates that the structured ansatz can retain favorable gradient behavior in the investigated regimes. Together with the observed reductions in logical quantum evaluations, these results motivate further study of hybrid gradient-evaluation strategies at larger scales and on quantum hardware.

A limitation of the present polynomial-DLA benchmarks is that, in the Pauli setting, recent classification results show that small DLAs are strongly constrained and essentially reduce to the free-fermionic case \cite{aguilar2024full}. This limits any direct interpretation of our XY results as evidence of computational advantage over classical algorithms. We therefore view HELIA and the associated training methods here primarily as a methodological contribution on hybrid training and resource allocation, showing how part of the gradient-estimation burden can be transferred from QPU-side parameter-shift evaluations to classical $\mathfrak{g}$-sim updates. At the same time, recent works suggest that Lie-algebraic simulation is not restricted in practice to free-fermionic Pauli examples: symmetry-adapted and non-Pauli constructions have identified additional nontrivial polynomial-dimensional DLA families, while restricted polynomial-DLA ans\"atze have also been used as warm-starting stages for larger variational circuits \cite{barligea2026enabling,kordonowy2026lie}. These developments motivate extending HELIA-style hybrid strategies beyond the specific Pauli/free-fermionic settings considered here.

\red{The collective-spin results in Appendix~\ref{collective_spin} provide an initial symmetry-adapted example of this extension. They demonstrate finite-size portability of the hybrid update rule together with a reduction in estimated QPU calls. Scaling, end-to-end classical cost, and hardware validation remain open questions.}

We now discuss several open questions for further research that can improve our proposed methods:

\begin{itemize}
    \item The Pauli-based benchmarks exploit a particularly simple adjoint-action formula. Appendix~\ref{collective_spin} provides an initial extension using averaged collective-spin orbit generators. Further extensions should quantify the costs of constructing the operator representation, applying the adjoint evolution, evaluating gradients, and measuring the required observables.

\item  The PSR method, although invaluable for training quantum models, has a high quantum overhead. Our proposed method is aimed at reducing the necessity of using PSR to only a limited number of parameters, but leaves open the possibility of improving PSR itself.

\item As noted in App. \ref{odd_gsim}, for specific Hamiltonians \g by itself is quite effective in obtaining the ground state. In fact, it may be possible to get accurate solutions with \g by choosing an appropriate initial state which can be appropriately mapped to the required ground state within the polynomial search space of the DLA. However, finding such an initial state for general Hamiltonians appropriate for \g can itself be a challenge and to our knowledge has not been explored in the literature. We therefore leave extensive study of this direction for future work.

\item ADAPT-VQE is an important and interesting approach in the VQE literature because it adaptively constructs compact, problem-tailored ans\"atze by selecting operators from a pool using energy-gradient information \cite{grimsley2019adaptive}. This makes it a strong alternative to fixed-ansatz methods when ansatz compactness is a primary goal. By contrast, the pretraining role considered here is that of warm initialization within a fixed HELIA framework, where a polynomial-DLA stage is used to reduce the QPU-side gradient-estimation burden through a hybrid PSR + $\mathfrak{g}$-sim strategy. A direct comparison is therefore not possible unless one also matches the target accuracy, final ansatz expressivity, operator pool, and overall measurement budget \cite{grimsley2019adaptive,anastasiou2023really}. Hence, HELIA and ADAPT-VQE can act as complementary approaches addressing different aspects of variational design, and we leave a detailed quantitative comparison to future work.

\item  Practical quantum utility potentially lies at much higher qubit numbers than simulated in this article and would require a demonstration on actual quantum hardware. Whether the observed logical-evaluation savings persist at larger system sizes and on quantum hardware remains open. Fewer logical evaluations do not, by themselves, establish greater noise robustness or lower total measurement cost. Future studies should include shot allocation, observable grouping, hardware noise, compilation, and classical processing costs.

\item Another promising direction is to broaden the practical scope of the $\mathfrak{g}$-sim framework~\cite{goh2023lie} through symmetry-adapted representations of operator spaces. In particular, decomposing the space of linear operators acting on $n$ qubits into invariant subspaces under the adjoint action of the circuit's dynamical group can permit the propagation of selected observables within the relevant subspaces, rather than the full operator space. Such an approach may reduce computational overhead when these subspaces have manageable dimensions and their representations can be constructed and applied efficiently. Representation theory provides a natural framework for identifying this structure, but the existence of an invariant-subspace decomposition alone does not guarantee efficient simulation.

Constructing useful decompositions and explicit bases remains a system-dependent challenge. Relevant tasks include identifying the invariant components, determining their dimensions, expressing the observables in suitable bases, and computing the action of the circuit generators within those components. For example, Diaz et al.~\cite{diaz2023showcasing} characterize invariant operator modules for parameterized matchgate circuits using products of Majorana operators, extending the analysis beyond observables confined to the circuit's DLA. This provides a concrete illustration of how representation-theoretic structure can reveal operator spaces relevant to both trainability and classical simulability. Further background on group representations, invariant subspaces, and symmetry-based constructions in quantum machine learning is provided by Ragone et al.~\cite{ragone2022representation}. Developing efficient, explicit representations for additional circuit families is therefore a useful direction for extending HELIA-style hybrid gradient evaluation.

\item Another direction would be to move away from the standard PQC structure to include non-unitary or general Completely Positive Trace Preserving (CPTP) maps such as in Ref.~\cite{heredge2024non}.

\end{itemize}
\section{Acknowledgments}
The authors thank the quantum team at Fujitsu Research India, especially Naipunnya Raj, Ruchira Bhat, and Rajiv Sangle, for their discussions and input. The paper benefited greatly from discussions with Hannes Leipold and Bibhas Adhikari from Fujitsu Research America. The authors are immensely grateful to Yasuhiro Endo, Hirotaka Oshima, Shintaro Sato, and the entire Robust Quantum Computing Department at Fujitsu Limited for their strategic and technical support. The authors also thank Masayoshi Hashima for his valuable assistance with the state-of-the-art Fujitsu Quantum Simulator. A preliminary version of this paper was presented at the Fujitsu-IISc Quantum Workshop on January 23--24, 2025, in Bengaluru, India, and at the Fujitsu Quantum Day event on March 28, 2025, in Kawasaki, Japan. The results were also presented as a poster at TQC 2025. This research received no external funding and was conducted as part of the authors' employment with Fujitsu Limited.

\section{Conflict of Interest}
The authors declare no conflict of interest.

\section{Data Availability}
The data and code that support the findings of this study are available from the corresponding author upon reasonable request.

\newpage
\sloppy

\balance
\bibliographystyle{unsrt}
\bibliography{references}
\sloppy
\onecolumngrid
\newpage
\appendix
\appendixpage

\section{Parameter-shift Rule} \label{psr_append}
The parameter-shift rule (PSR) is commonly used for accurately estimating gradients of a PQC. For a loss function of the form given in Eq.\eqref{loss_func}, where each unitary is defined as $U_i(\theta_i) = e^{-i\theta_i G_i}$ for some Hermitian operator $G_i$, the partial derivatives are given by

\begin{align}\label{loss_deriv}
    \frac{\partial{\ell}}{\partial{\theta_k}} &= i \bra{0}{U}^{\dagger}_{\geq k}  G_k ({U}^{\dagger}_{<k}{\hat{O}}U_{<k})U_{\geq k} \ket{0}\nonumber\\
    &-i \bra{0}U^\dagger_{\geq k}(U^\dagger_{<k}\hat{O}{U}_{<k})  G_k{U}_{\geq k}\ket{0}
\end{align}

where we defined $ U_{<k} = U_1 W_1 U_2W_2\hdots U_{k-1}W_{k-1} $ and $U_{\geq k} = U_{k}W_{k}U_{k+1}W_{k+1}\hdots U_nW_n$.
PSR can be employed to obtain a closed-form expression of Eq.\eqref{loss_deriv} when the eigenspectrum of the generator $G_k$ either has only two unique eigenvalues or is symmetric and evenly spaced\,\cite{mitarai2018quantum,schuld2019evaluating,vidal2018calculus,crooks2019gradients,wierichs2022general}.

Considering the case where $G_k$ with only two unique eigenvalues $\pm r$ i.e., \( G_k^2 = r^2 I \) the corresponding unitary can be rewritten as
\begin{equation}\label{eigen_gen}
U_k(\theta) = \cos{r\theta} I - i r^{-1}G_k\sin{r\theta}
\end{equation}

Following\,\cite{schuld2019evaluating}, through some simple algebra the above equation can be turned into
\begin{align}
    \frac{\partial{\ell}}{\partial{\theta_k}} &= \frac{r}{2}\Big(\langle \psi | (I - ir^{-1}G)^{\dagger} Q  (I - ir^{-1}G) | \psi \rangle\nonumber \\
    & -\langle \psi | (I + ir^{-1}G)^{\dagger} Q  (I + ir^{-1}G) | \psi \rangle\Big)
\end{align}
where we have defined $|\psi\rangle = U_{\geq k}|0\rangle$ and $Q = {U}^{\dagger}_{<k}{\hat{O}}U_{<k}$. Finally using Eq.\eqref{eigen_gen} and substituting $\theta = \pi/4r$, the partial derivative simplifies to a difference between the loss functions evaluated on either side of the parameter with a specific eigenvalue-related shift
\begin{equation}
     \frac{\partial{\ell}}{\partial{\theta_k}} = r \Big(\ell(\bs\theta + \frac{\pi}{4r}\hat{e}_k) - \ell(\bs\theta - \frac{\pi}{4r} \hat{e}_k)\Big).
\end{equation}
It is important to note that, despite its similarity to the finite-difference approximation, the above formula is in fact an exact estimation of the partial derivative up to shot noise.

\section{\texorpdfstring{\g}{$\mathfrak{g}$-sim} Method}\label{gsim_append}
In order to understand, how $\mathfrak{g}$-sim  method works, we need to know how the elements of the DLA  $\mathfrak{g}$ transform under commutation with elements of the DLA $\mathfrak{g}$, and under conjugation with elements of the dynamical Lie group $\mathcal{G}$. As $\mathfrak{g}$ is closed under commutation, the commutator between any two elements in  $\mathfrak{g}$ is an element of $\mathfrak{g}$ and can be decomposed as a linear combination of a Hilbert--Schmidt orthonormal basis $\{ iG_\alpha: \alpha \in \{1,2,\ldots,\text{dim}(g)\} \}$ of $\mathfrak{g}$.
 In particular,
\begin{equation}\label{structure}
    \comm{iG_\alpha}{iG_\beta} = \sum_{\gamma=1}^{\text{dim}(\mathfrak{g})} f_{\alpha\beta}^{\gamma}iG_\gamma
\end{equation}

 where $ f_{\alpha\beta}^{\gamma} = \text{Tr}[(iG_\gamma)^\dagger [iG_\alpha,iG_\beta]] \in \mathbb{R}$  is computed via the standard projection in vector space $\mathfrak{g}$. The coefficients  $f_{\alpha\beta}^{\gamma}$ are called the \textit{structure constants} of the DLA $\mathfrak{g}.$

\begin{algorithm}[t] \caption{g-sim Implementation for VQE} \label{alg:gsim_vqe} \KwIn{ $T_{\max}$, $\eta$, Hermitian basis $\{G_l\}_{l=1}^{g_{\max}}$, $H=\sum_{l=1}^{g_{\max}}a_lG_l$, $U(\vec{\theta}) =e^{-iG_1\theta_1}\cdots e^{-iG_n\theta_n}$ } \KwOut{Optimized parameters $\vec{\theta}^{\,*}$} Initialize $\vec{\theta}\sim\mathcal{N}(0,1)$\; Define \[C(\vec{\theta}) = \langle 0^{\otimes q}| U^\dagger(\vec{\theta})HU(\vec{\theta}) |0^{\otimes q}\rangle \] \For{$l=1,\ldots,g_{\max}$}{ $g_l\leftarrow \langle 0^{\otimes q}|G_l|0^{\otimes q}\rangle$\; } $i\leftarrow 0$\; \While{$i<T_{\max}$}{ \tcp{Reset the Hamiltonian coefficients for this iteration} \For{$k=1,\ldots,g_{\max}$}{ $a_k^{n}\leftarrow a_k$\; } $H^{n}\leftarrow H$\; $m\leftarrow n$\; \While{$m\geq 1$}{ \tcp{Initialize the output coefficient vector} \For{$j=1,\ldots,g_{\max}$}{ $a_j^{m-1}\leftarrow 0$\; } \For{$k=1,\ldots,g_{\max}$}{ $H_k^{m}\leftarrow a_k^{m}G_k$\; Expand in the chosen basis: \[ e^{iG_m\theta_m}H_k^{m}e^{-iG_m\theta_m} = \sum_{j=1}^{g_{\max}} b_{jk}^{m-1}G_j \] \For{$j=1,\ldots,g_{\max}$}{ $a_j^{m-1}\leftarrow a_j^{m-1}+b_{jk}^{m-1}$\; } } $H^{m-1}\leftarrow \sum_{j=1}^{g_{\max}}a_j^{m-1}G_j$\; $m\leftarrow m-1$\; } $C\leftarrow \sum_{j=1}^{g_{\max}}a_j^{0}g_j$\; Obtain $\nabla_{\vec{\theta}}C$ classically using automatic differentiation through the coefficient-propagation steps\; $\vec{\theta}\leftarrow \vec{\theta}-\eta\nabla_{\vec{\theta}}C$\; $i\leftarrow i+1$\; } $\vec{\theta}^{\,*}\leftarrow\vec{\theta}$\; \Return{$\vec{\theta}^{\,*}$}\; \end{algorithm}

Using the above DLA structure we demonstrate an implementation of the full \g sub-routine in action for VQE in Alg.~\ref{alg:gsim_vqe}. We choose a Hamiltonian with a DLA of polynomial dimension. The ansatz is chosen as $U(\bs\theta) = e^{-i\theta_1G_1}\hdots e^{-i\theta_nG_n}$, where the generators are from the same DLA, allowing us to efficiently use the DLA structure in our computation.

The crucial factor facilitating the use of DLA structure is the Baker-Campbell-Hausdorff (BCH) formula\,\cite{bch} in the coefficient-propagation step of Alg.~\ref{alg:gsim_vqe}, which is given by
\begin{equation}\label{bch_form}
    e^{X} Y e^{-X} = Y + \comm{X}{Y} + \frac{1}{2!}\comm{X}{\comm{X}{Y}} + \hdots
\end{equation}

For operators $G_i,G_j$ belonging to a chosen DLA, the expression for $e^{i\theta G_i} G_j e^{-i\theta G_i}$  consists of only nested commutators. Since the DLA is closed under commutation operation, all of the terms in the infinite series in Eq.\eqref{bch_form} belong to the DLA itself. By choosing a DLA scaling polynomially in the number of qubits, we ensure the expansion only has polynomially many unique operators, whose co-efficients are functions of the \textit{structure constants} $f_{\alpha\beta}^\gamma$ as shown in Eq.\eqref{structure}.

However, it might still be difficult to obtain a closed form expression for Eq.\eqref{bch_form}. Hence, we further simplify it to the scenario when all the operators are n-qubit Pauli strings. This gives a very simple expression
\begin{equation}
e^{i\theta P_i} P_j e^{-i\theta P_i} = \begin{cases} P_j, & [P_i,P_j]=0, \\[6pt] \cos(2\theta)\,P_j +\dfrac{i}{2}\sin(2\theta)\,[P_i,P_j], & \{P_i,P_j\}=0. \end{cases}
\end{equation}
where $P_i,P_j$ are n-qubit Pauli strings. We further make our numerics computationally efficient by leveraging the $2n \times  2n$ binary symplectic matrix representation of Pauli, rather than their $2^n\times2^n$ unitary matrix representation (Section 2.1 of \cite{hsieh2008entanglement}). As a result, we can replace matrix multiplication by bit-wise addition along with phase-tracking, which is much faster.

Fast application of the BCH formula due to choosing Pauli strings as well a DLA of polynomial dimension allows to calculate the cost function efficiently with polynomial compute resources. The gradients can then be efficiently calculated using an automatic-differentiation framework such as Pytorch\,\cite{ansel2024pytorch}, Tensorflow\,\cite{Abadi_TensorFlow_Large-scale_machine_2015}.

\section{Overview of Barren Plateaus (BPs) and proposed mitigation techniques}\label{BP_app}

BPs are characterized by regions in the parameter space where the gradient of the loss function becomes exponentially small, severely impeding the efficient optimization of quantum circuits~\cite{mcclean2018barren,ragone2023unified, fontana2023adjoint}. The presence of a BP landscape implies that, for most parameter configurations, an exponentially large number of measurement samples is required to reliably estimate the loss gradient, which is essential for effective optimization.

VQA is believed to exhibit a BP when variations in the model parameters $\theta$ lead to only exponentially small changes in the loss function $l_{\theta}(\rho, \hat{O})$, or magnitude of gradients $\partial l_{\theta}(\rho, \hat{O})/ \partial \theta_\mu$
(where $\theta_\mu \in \theta$)~\cite{larocca2024review,ragone2023unified, fontana2023adjoint}.
Cerezo et al., in their investigation of cost function locality, established that the inclusion of global observables \((\hat{O})\) in the cost function results in the emergence of exponentially vanishing gradients, irrespective of the circuit's shallow depth~\cite{Cerezo_2021}. Additional factors contributing to this phenomenon include the degree of entanglement~\cite{Patti_2021, ortiz2021entanglement, CerveroMartin2023barrenplateausin} and the presence of noise~\cite{Wang_2021}. A unified theoretical framework, employing group theoretical principles, covering above factors were presented in Refs.~\cite{ragone2023unified, fontana2023adjoint}.

Beyond theoretical insights, substantial efforts have been devoted to developing practical strategies for mitigating BPs~\cite{cunningham2024investigating}. These include advanced initialization techniques, such as layer-wise training~\cite{ Skolik_2021}, Bayesian-inspired parameter settings~\cite{rad2022surviving}, meta-learning for initializing parameters~\cite{verdon2019learning}, initializing circuits with sequences of identity blocks~\cite{Grant_2019}, and transfer-learning-based initialization approaches~\cite{Liu_2023}. Alternative optimization frameworks have also been introduced, including adaptive learning rate strategies~\cite{Haug2021OptimalTO} and novel line search methods designed to facilitate navigation through flat loss landscapes~\cite{nadori2024promising}.  Mhiri et al. provided a unified analysis on warm-starting approaches in BP-affected loss landscape~\cite{mhiri2025unifying}. Classical neural network-based methodologies have also been explored, such as generating parameters for PQCs using neural networks~\cite{PhysRevA.106.042433, yi2024enhancing}.

Empirical studies, such as those conducted in Ref.~\cite{kulshrestha2022beinit}, have validated the effectiveness of these approaches in reducing the prevalence of BPs in practical applications, including QNNs. Enrique et al. have shown that the optimization landscapes of certain quantum circuit architectures inspired by classical tensor network structures such as, tree tensor networks (qTTN) and the multiscale entanglement renormalization ansatz (qMERA) are free of BPs~\cite{CerveroMartin2023barrenplateausin}. Furthermore, the HVA and parameterized matchgate circuits have also been shown to effectively mitigate or entirely circumvent BPs~\cite{diaz2023showcasing, Larocca2022diagnosingbarren}.

Recent research has shed light on the classical simulability of BP-free models, offering valuable insights into their underlying structure. Cerezo et al. suggested that BP-free models might be classically simulable when classical data is gathered during an initial data acquisition phase from the quantum device~\cite{cerezo2023does}. This hypothesis stems from the observation that BPs arise from the curse of dimensionality, with mitigation strategies effectively reducing the problem to subspaces that are classically tractable.

\section{Numerical Experiments on XY Hamiltonian}\label{xy_numerical}

The success, relative error and QPU calls reduction have been plotted for the HELIA configuration with 1 layer of YZ linear ansatz in $U_q$ in Fig.\,\ref{xy_layer1}. For the configurations with 3, 6, and 9 YZ linear layers we plot the corresponding metrics in Fig.\ref{xy_vqe_all}.

\begin{figure*}[tbh!]
    \begin{subfigure}{1\linewidth}
        \includegraphics[width=1\linewidth]{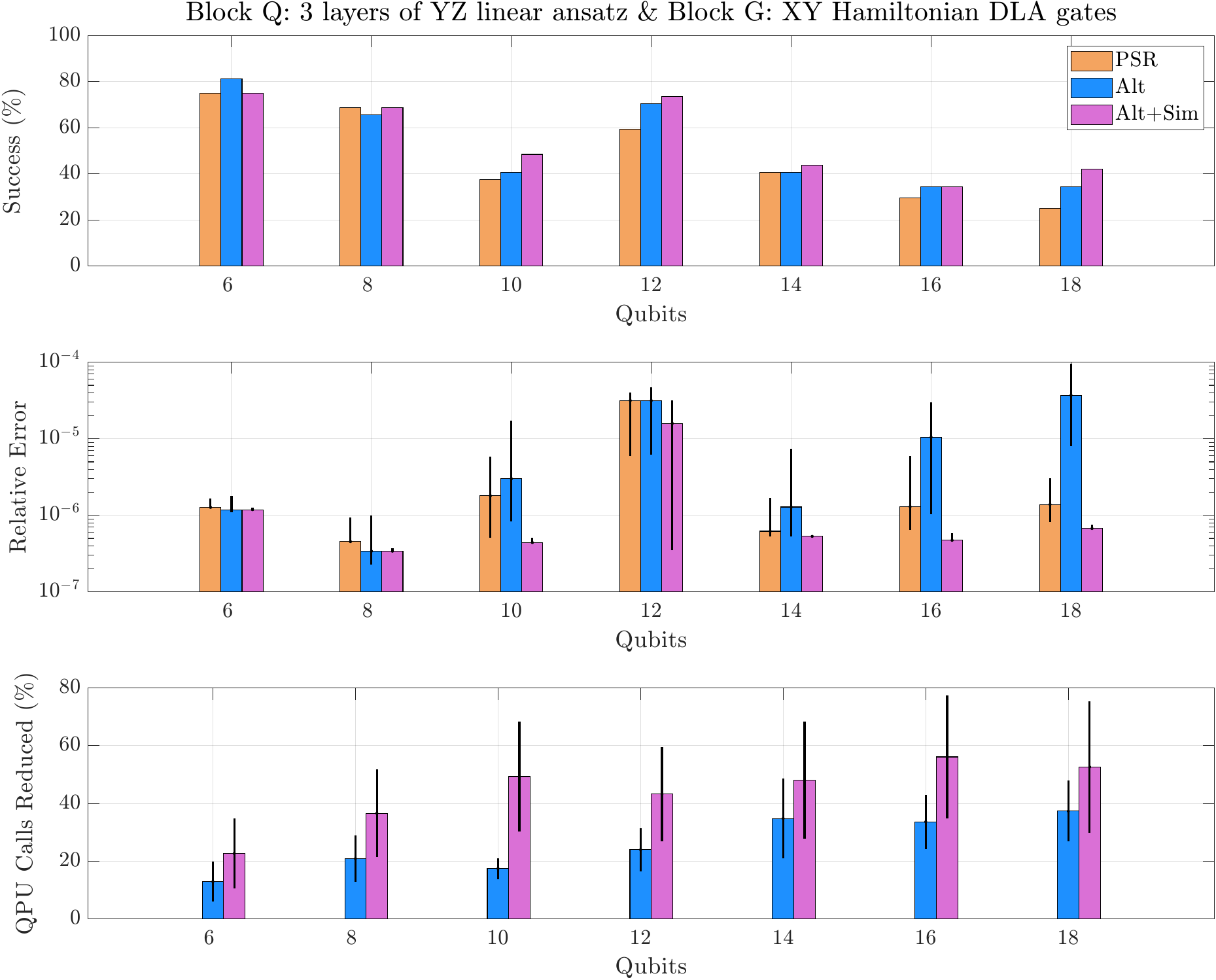}
        \caption{Success, Relative Error and QPU Call Reduction for 3 YZ linear layers in $U_q$}
        \label{xy_layer3}
    \end{subfigure}
\end{figure*}

\begin{figure*}\ContinuedFloat
    \begin{subfigure}{1\linewidth}
        \includegraphics[width=1\linewidth]{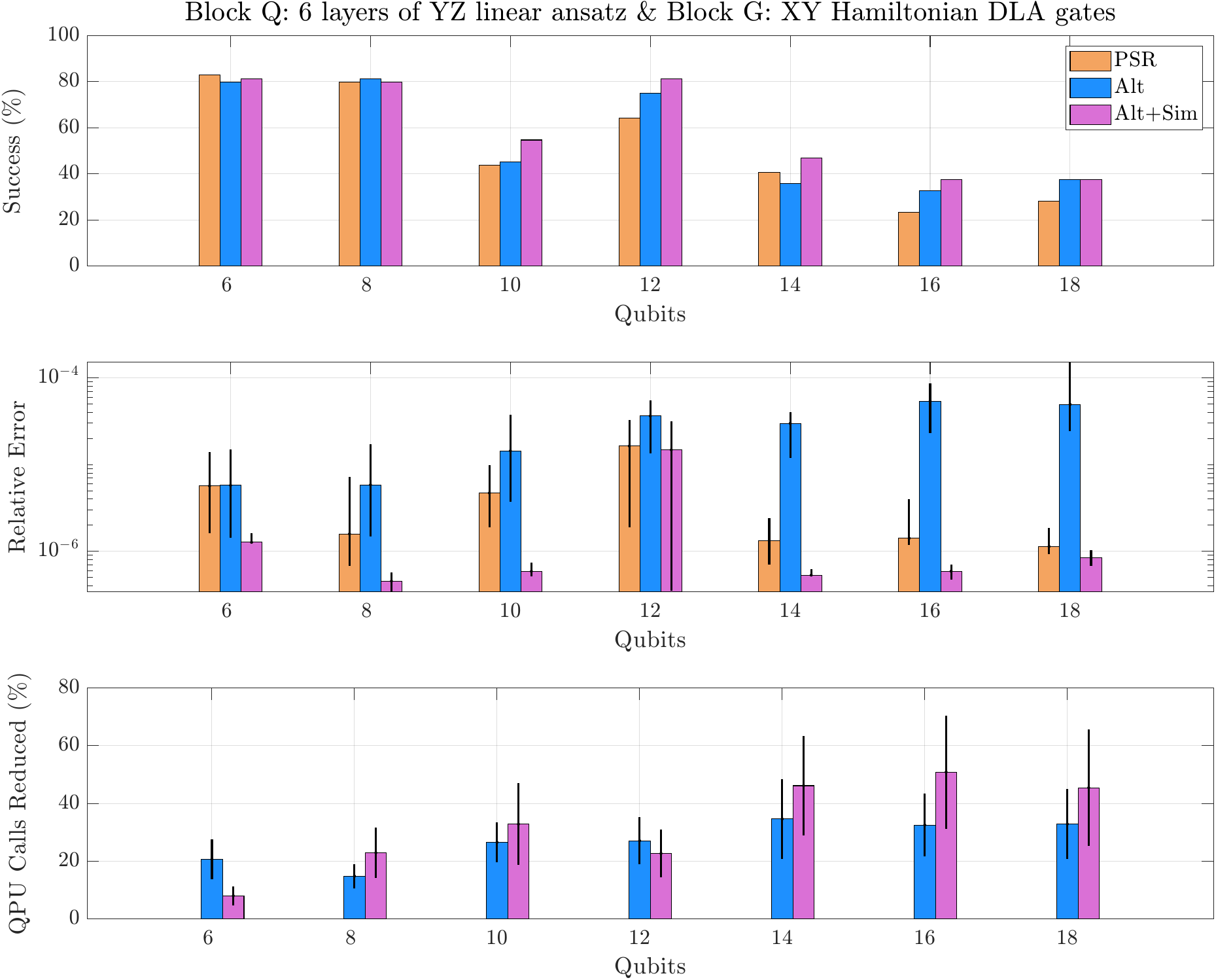}
        \caption{Success, Relative Error and QPU Call Reduction for 6 YZ linear layers in $U_q$}
        \label{xy_layer6}
    \end{subfigure}
\end{figure*}

\begin{figure*}\ContinuedFloat
    \begin{subfigure}{1\linewidth}
        \includegraphics[width=1\linewidth]{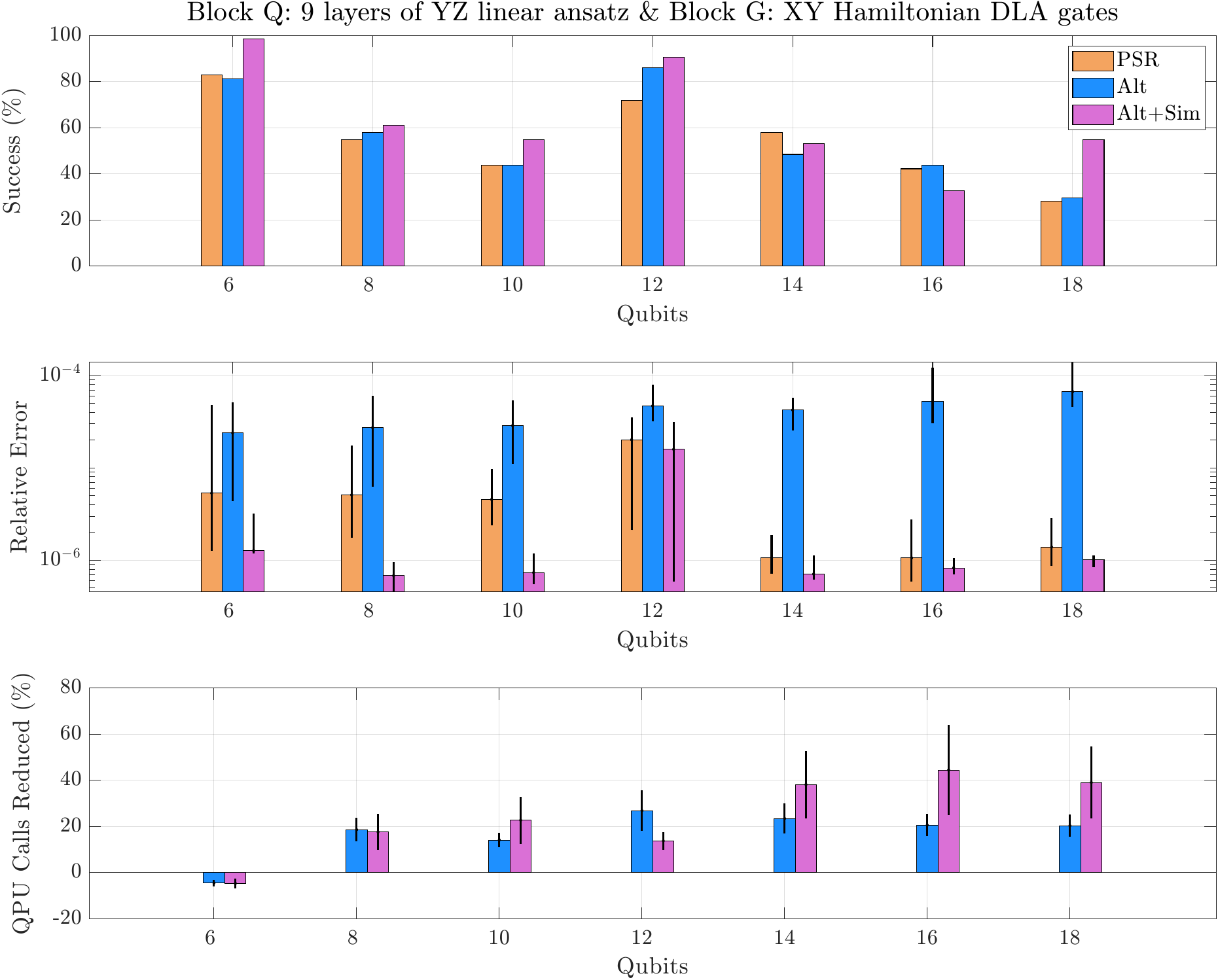}
        \caption{Success, Relative Error and QPU Call Reduction for 9 YZ linear layers in $U_q$}
        \label{xy_layer9}
    \end{subfigure}
    \caption{ \raggedright \textbf{XY Hamiltonian VQE (3, 6, and 9 YZ linear layers)}: Success (Row 1), relative error (Row 2), and QPU-call reduction (Row 3) are plotted for configurations with 3 (a), 6 (b), and 9 (c) layers of the YZ linear ansatz in $U_q$ (referred to as Block Q) and Hamiltonian DLA gates in \Ug (referred to as Block G) for 6 to 18 qubits. For each qubit count and metric, PSR (orange), Alternate (blue), and Alternate+Simultaneous (purple) training methods are compared. \textbf{On the first row}, we show that the Alternate and Alternate+Simultaneous protocols generally have better success rates than PSR, except in some depth--qubit configurations. \textbf{For relative error (Row 2)}, we plot the median and the $25\%$ and $75\%$ quantiles (in black). The Alternate method struggles in these examples, but Alternate+Simultaneous gives lower relative error than PSR for all qubits. Most importantly, \textbf{in the lowermost row}, QPU-call reduction is plotted only for Alternate and Alternate+Simultaneous because it is measured relative to PSR (which coincides with 0\% QPU-call reduction). The standard deviation is shown in black. Logical-evaluation reductions are positive in most configurations, but success-rate improvements are configuration-dependent. }
    \label{xy_vqe_all}
\end{figure*}

We also tabulate the results of the numerical experiments comparing \g, standard PSR, Alternate and Alternate+Simultaneous method of training on the XY Hamiltonian for various qubit counts in Table \ref{vqe_xy_successqpu} and \ref{vqe_xy_error}, as described in Sec. \ref{numerical} and visualized in Fig. \ref{xy_layer1} and \ref{xy_vqe_all}. We run VQE on XY Hamiltonian for qubits ranging from 6 to 18, for the configurations with 1, 3, 6 and 9 layers of YZ linear ansatz in $U_q$ and XY Hamiltonian DLA gates in \Ug.

Across the tested configurations, our method often reduces the relative error of ground-state estimation (by up to an order of magnitude), improves success rates by up to 26.57 percentage points, and reduces QPU calls by up to approximately $60\%$ relative to standard PSR training. The improvements are configuration-dependent rather than uniform across all settings.

\subsection{Resource accounting for the XY experiments}
To make the training-resource tradeoff more transparent, we explicitly account for both the quantum-side workload and the classical-side $\mathfrak{g}$-sim workload for the XY VQE experiments. For the HELIA ansatz, the \Ug block contains $g = dim(\mathfrak{g})$ parameters and DLA generators while the hardware-efficient block $U_q$ contributes $p$ parameters. The XY DLA dimension scales as $g = \dim(\mathfrak g) = n(2n-1)$.

\paragraph{Quantum-side accounting.}
We use all the parameter-shifted circuit evaluations for Full-PSR in each iteration to measure quantum hardware usage resulting in $2(p+g)$ calls. For Alternate and Simultaneous protocols, we use all shifted circuit evaluations when obtaining gradients of $U_q$ (i.e, $2p$) and the number of DLA generators that need to be measured independently for obtaining gradient of \Ug using $\mathfrak{g}$-sim resulting in $(2p +g)$ QPU calls to obtain the gradients of the full circuit in each iteration. This count treats each required DLA-observable expectation estimate as one logical evaluation. It excludes the shots required to achieve a target precision and does not model observable grouping, basis changes, or hardware-specific measurement constraints.

\paragraph{Classical-side accounting.}
The $\mathfrak g$-sim branch evolves a $g$-dimensional expectation vector rather than a $2^n$-dimensional state vector. The minimal working memory is therefore $\mathcal O(g)$, while storing dense adjoint-representation matrices for a generators requires $\mathcal O(g^2)$ memory.

If each $\mathfrak g$-sim layer is implemented using a dense matrix exponential in adjoint space, then the worst-case forward cost is
\begin{equation}
\mathcal O(g^3),
\end{equation}
The adjoint matrices corresponding to Pauli generators are generally sparse reducing the overhead significantly.

\begin{table}
\begin{tabular}{cc}
    \begin{minipage}{.45\linewidth}

\begin{subtable}{\linewidth}

\begin{tabular}{|c|ccc|}
\hline
\multirow{2}{*}{\textbf{Qubits}} & \multicolumn{3}{c|}{\textbf{Success(\%)}}                                                              \\ \cline{2-4}
                                 & \multicolumn{1}{c|}{\textbf{Full-PSR}} & \multicolumn{1}{c|}{\textbf{Alternate}} & \textbf{Alt+Sim} \\ \hline
6                                & \multicolumn{1}{c|}{81.25}             & \multicolumn{1}{c|}{70.31}              & \textbf{90.62}  \\
8                                & \multicolumn{1}{c|}{70.31}             & \multicolumn{1}{c|}{59.38}              & \textbf{75.00}  \\
10                               & \multicolumn{1}{c|}{29.69}             & \multicolumn{1}{c|}{48.44}              & \textbf{50.00}  \\
12                               & \multicolumn{1}{c|}{65.62}             & \multicolumn{1}{c|}{70.31}              & \textbf{90.62}  \\
14                               & \multicolumn{1}{c|}{\textbf{51.56}}    & \multicolumn{1}{c|}{\textbf{51.56}}     & \textbf{51.56}  \\
16                               & \multicolumn{1}{c|}{21.88}             & \multicolumn{1}{c|}{43.75}              & \textbf{48.44}  \\
18                               & \multicolumn{1}{c|}{35.94}             & \multicolumn{1}{c|}{50.00}              & \textbf{54.69}  \\ \hline
\end{tabular}
\caption{Success rate for 1 layer YZ linear ansatz in $U_q$ and XY Hamiltonian DLA gates in \Ug}
\end{subtable}

\begin{subtable}{\linewidth}

\begin{tabular}{|c|ccc|}
\hline
\multirow{2}{*}{\textbf{Qubits}} & \multicolumn{3}{c|}{\textbf{Success(\%)}}                                                                                   \\ \cline{2-4}
                                 & \multicolumn{1}{c|}{\textbf{Full-PSR}} & \multicolumn{1}{c|}{\textbf{Alternate}} & \multicolumn{1}{c|}{\textbf{Alt+Sim}} \\ \hline
6                                & \multicolumn{1}{c|}{75.00}             & \multicolumn{1}{c|}{\textbf{81.25}}     & 75.00                                \\
8                                & \multicolumn{1}{c|}{\textbf{68.75}}    & \multicolumn{1}{c|}{65.62}              & \textbf{68.75}                       \\
10                               & \multicolumn{1}{c|}{37.50}             & \multicolumn{1}{c|}{40.62}              & \textbf{48.44}                       \\
12                               & \multicolumn{1}{c|}{59.38}             & \multicolumn{1}{c|}{70.31}              & \textbf{73.44}                       \\
14                               & \multicolumn{1}{c|}{40.62}             & \multicolumn{1}{c|}{40.62}              & \textbf{43.75}                       \\
16                               & \multicolumn{1}{c|}{29.69}             & \multicolumn{1}{c|}{\textbf{34.38}}     & \textbf{34.38}                       \\
18                               & \multicolumn{1}{c|}{25.00}             & \multicolumn{1}{c|}{34.38}              & \textbf{42.19}                       \\ \hline
\end{tabular}

\caption{Success rate for 3 layer YZ linear ansatz in $U_q$ and XY Hamiltonian DLA gates in \Ug}
\end{subtable}

\begin{subtable}{\linewidth}

\begin{tabular}{|c|ccc|}
\hline
\multirow{2}{*}{\textbf{Qubits}} & \multicolumn{3}{c|}{\textbf{Success(\%)}}                                                                                   \\ \cline{2-4}
                                 & \multicolumn{1}{c|}{\textbf{Full-PSR}} & \multicolumn{1}{c|}{\textbf{Alternate}} & \multicolumn{1}{c|}{\textbf{Alt+Sim}} \\ \hline
6                                & \multicolumn{1}{c|}{\textbf{82.81}}    & \multicolumn{1}{c|}{79.69}              & 81.25                                \\
8                                & \multicolumn{1}{c|}{79.69}             & \multicolumn{1}{c|}{\textbf{81.25}}     & 79.69                                \\
10                               & \multicolumn{1}{c|}{43.75}             & \multicolumn{1}{c|}{45.31}              & \textbf{54.69}                       \\
12                               & \multicolumn{1}{c|}{64.06}             & \multicolumn{1}{c|}{75.00}              & \textbf{81.25}                       \\
14                               & \multicolumn{1}{c|}{40.62}             & \multicolumn{1}{c|}{35.94}              & \textbf{46.88}                       \\
16                               & \multicolumn{1}{c|}{23.44}             & \multicolumn{1}{c|}{32.81}              & \textbf{37.50}                       \\
18                               & \multicolumn{1}{c|}{28.12}             & \multicolumn{1}{c|}{\textbf{37.50}}     & \textbf{37.50}                       \\ \hline
\end{tabular}

\caption{Success rate for 6 layer YZ linear ansatz in $U_q$ and XY Hamiltonian DLA gates in \Ug}
\end{subtable}

\begin{subtable}{\linewidth}

\begin{tabular}{|c|ccc|}
\hline
\multirow{2}{*}{\textbf{Qubits}} & \multicolumn{3}{c|}{\textbf{Success(\%)}}                                                                                   \\ \cline{2-4}
                                 & \multicolumn{1}{c|}{\textbf{Full-PSR}} & \multicolumn{1}{c|}{\textbf{Alternate}} & \multicolumn{1}{c|}{\textbf{Alt+Sim}} \\ \hline
6                                & \multicolumn{1}{c|}{82.81}             & \multicolumn{1}{c|}{81.25}              & \textbf{98.44}                                \\
8                                & \multicolumn{1}{c|}{54.69}             & \multicolumn{1}{c|}{57.81}              & \textbf{60.94}                       \\
10                               & \multicolumn{1}{c|}{43.75}             & \multicolumn{1}{c|}{43.75}              & \textbf{54.69}                       \\
12                               & \multicolumn{1}{c|}{71.88}             & \multicolumn{1}{c|}{85.94}              & \textbf{90.62}                       \\
14                               & \multicolumn{1}{c|}{\textbf{57.81}}    & \multicolumn{1}{c|}{48.44}              & 53.12                                \\
16                               & \multicolumn{1}{c|}{42.19}             & \multicolumn{1}{c|}{\textbf{43.75}}     & 32.81                                \\
18                               & \multicolumn{1}{c|}{28.12}             & \multicolumn{1}{c|}{29.69}              & \textbf{54.69}                       \\ \hline
\end{tabular}

\caption{Success rate for 9 layer YZ linear ansatz in $U_q$ and XY Hamiltonian DLA gates in \Ug}

\end{subtable}
\end{minipage}

&
\begin{minipage}{0.45\linewidth}

\begin{subtable}{\linewidth}
\begin{tabular}{|c|cc|}
\hline
\multirow{2}{*}{\textbf{Qubits}} & \multicolumn{2}{c|}{\textbf{QPU Reduction (\%)}}           \\ \cline{2-3}
                                 & \multicolumn{1}{c|}{\textbf{Alternate}} & \textbf{Alt+Sim} \\ \hline
6                                & \multicolumn{1}{c|}{6.24 $\pm$ 4.12}      & $-33.27$ $\pm$ 17.07 \\
8                                & \multicolumn{1}{c|}{22.30 $\pm$ 11.77}    & 27.53 $\pm$ 13.20  \\
10                               & \multicolumn{1}{c|}{3.07 $\pm$ 1.66}      & 41.88 $\pm$ 21.81  \\
12                               & \multicolumn{1}{c|}{20.81 $\pm$ 10.40}    & 45.25 $\pm$ 21.95  \\
14                               & \multicolumn{1}{c|}{16.70 $\pm$ 9.60}     & 32.07 $\pm$ 15.24  \\
16                               & \multicolumn{1}{c|}{38.39 $\pm$ 17.18}    & 60.00 $\pm$ 25.11  \\
18                               & \multicolumn{1}{c|}{16.71 $\pm$ 7.34}     & 46.81 $\pm$ 26.52  \\ \hline
\end{tabular}

\caption{QPU Reduction for 1 layer YZ linear ansatz in $U_q$ and XY Hamiltonian DLA gates in \Ug}
\end{subtable}

\begin{subtable}{\linewidth}

\begin{tabular}{|c|cc|}
\hline
\multirow{2}{*}{\textbf{Qubits}} & \multicolumn{2}{c|}{\textbf{QPU Reduction (\%)}}           \\ \cline{2-3}
                                 & \multicolumn{1}{c|}{\textbf{Alternate}} & \textbf{Alt+Sim} \\ \hline
6                                & \multicolumn{1}{c|}{12.87 $\pm$ 6.92}     & 22.62 $\pm$ 12.21  \\
8                                & \multicolumn{1}{c|}{20.84 $\pm$ 8.16}     & 36.58 $\pm$ 15.15  \\
10                               & \multicolumn{1}{c|}{17.42 $\pm$ 3.64}     & 49.32 $\pm$ 18.98  \\
12                               & \multicolumn{1}{c|}{23.94 $\pm$ 7.51}     & 43.24 $\pm$ 16.37  \\
14                               & \multicolumn{1}{c|}{34.77 $\pm$ 13.93}    & 47.99 $\pm$ 20.32  \\
16                               & \multicolumn{1}{c|}{33.62 $\pm$ 9.45}     & 56.12 $\pm$ 21.31  \\
18                               & \multicolumn{1}{c|}{37.31 $\pm$ 10.55}    & 52.63 $\pm$ 22.83  \\ \hline
\end{tabular}

\caption{QPU Reduction for 3 layer YZ linear ansatz in $U_q$ and XY Hamiltonian DLA gates in \Ug}
\end{subtable}

\begin{subtable}{\linewidth}

\begin{tabular}{|c|ll|}
\hline
\multirow{2}{*}{\textbf{Qubits}} & \multicolumn{2}{c|}{\textbf{QPU Reduction (\%)}}                                \\ \cline{2-3}
                                 & \multicolumn{1}{c|}{\textbf{Alternate}} & \multicolumn{1}{c|}{\textbf{Alt+Sim}} \\ \hline
6                                & \multicolumn{1}{l|}{20.65 $\pm$ 6.97}     & 7.93 $\pm$ 3.25                         \\
8                                & \multicolumn{1}{l|}{14.76 $\pm$ 4.21}     & 22.87 $\pm$ 8.77                        \\
10                               & \multicolumn{1}{l|}{26.49 $\pm$ 6.88}     & 32.82 $\pm$ 14.19                       \\
12                               & \multicolumn{1}{l|}{27.05 $\pm$ 8.22}     & 22.63 $\pm$ 8.37                        \\
14                               & \multicolumn{1}{l|}{34.59 $\pm$ 13.91}    & 46.14 $\pm$ 17.34                       \\
16                               & \multicolumn{1}{l|}{32.50 $\pm$ 10.90}    & 50.88 $\pm$ 19.61                       \\
18                               & \multicolumn{1}{l|}{32.84 $\pm$ 12.20}    & 45.45 $\pm$ 20.12                       \\ \hline
\end{tabular}

\caption{QPU Reduction for 6 layer YZ linear ansatz in $U_q$ and XY Hamiltonian DLA gates in \Ug}
\end{subtable}

\begin{subtable}{\linewidth}

\begin{tabular}{|c|cc|}
\hline
\multirow{2}{*}{\textbf{Qubits}} & \multicolumn{2}{c|}{\textbf{QPU Reduction (\%)}}           \\ \cline{2-3}
                                 & \multicolumn{1}{c|}{\textbf{Alternate}} & \textbf{Alt+Sim} \\ \hline
6                                & \multicolumn{1}{c|}{$-4.49$ $\pm$ 1.42}   & $-4.70$ $\pm$ 2.13 \\
8                                & \multicolumn{1}{c|}{18.62 $\pm$ 5.19}     & 17.64 $\pm$ 7.73   \\
10                               & \multicolumn{1}{c|}{14.12 $\pm$ 3.01}     & 22.70 $\pm$ 10.14  \\
12                               & \multicolumn{1}{c|}{26.85 $\pm$ 8.70}     & 13.74 $\pm$ 3.93   \\
14                               & \multicolumn{1}{c|}{23.44 $\pm$ 6.56}     & 38.17 $\pm$ 14.61  \\
16                               & \multicolumn{1}{c|}{20.59 $\pm$ 4.85}     & 44.47 $\pm$ 19.66  \\
18                               & \multicolumn{1}{c|}{20.32 $\pm$ 4.83}     & 39.07 $\pm$ 15.71  \\ \hline
\end{tabular}

\caption{QPU Reduction for 9 layer YZ linear ansatz in $U_q$ and XY Hamiltonian DLA gates in \Ug}
\end{subtable}

\end{minipage}
\end{tabular}

\caption{\raggedright \textbf{XY Hamiltonian (Success and QPU Reduction):} Success rates and QPU-call reductions are shown for configurations with 1, 3, 6, and 9 YZ-linear layers in $U_q$ and XY Hamiltonian DLA gates in \Ug using Full-PSR, Alternate, and Alternate followed by Simultaneous training (Alt+Sim). QPU-call reductions are reported as the mean $\pm$ sample standard deviation over successful trials only, where success is defined by a relative-error threshold of $10^{-3}$. Reductions are measured relative to Full-PSR. The corresponding successful-trial relative-error statistics are reported in Table~\ref{vqe_xy_error}.}
\label{vqe_xy_successqpu}
\end{table}

\begin{table}[]
\begin{subtable}{\linewidth}

\begin{tabular}{|c|l|ccccccccccccccc|}
\hline
\multirow{3}{*}{\textbf{Qubits}} &  & \multicolumn{15}{c|}{\textbf{Relative Error (units of $10^{-5}$)}}                                                                                                                                                                                                                                                                                                                                                                                                                                                                  \\ \cline{3-17}
                                 &  & \multicolumn{3}{c|}{\textbf{\g}}                                                                             & \multicolumn{1}{l|}{} & \multicolumn{3}{c|}{\textbf{Full-PSR}}                                                                       & \multicolumn{1}{l|}{} & \multicolumn{3}{c|}{\textbf{Alternate}}                                                                      & \multicolumn{1}{l|}{} & \multicolumn{3}{c|}{\textbf{Alt+Sim}}                                                   \\ \cline{3-5} \cline{7-9} \cline{11-13} \cline{15-17}
                                 &  & \multicolumn{1}{c|}{\textbf{Median}} & \multicolumn{1}{c|}{\textbf{Q25}} & \multicolumn{1}{c|}{\textbf{Q75}} & \multicolumn{1}{l|}{} & \multicolumn{1}{c|}{\textbf{Median}} & \multicolumn{1}{c|}{\textbf{Q25}} & \multicolumn{1}{c|}{\textbf{Q75}} & \multicolumn{1}{l|}{} & \multicolumn{1}{c|}{\textbf{Median}} & \multicolumn{1}{c|}{\textbf{Q25}} & \multicolumn{1}{c|}{\textbf{Q75}} & \multicolumn{1}{l|}{} & \multicolumn{1}{c|}{\textbf{Median}} & \multicolumn{1}{c|}{\textbf{Q25}} & \textbf{Q75} \\ \hline
6                                &  & \multicolumn{1}{c|}{12862.78}        & \multicolumn{1}{c|}{12856.2}      & \multicolumn{1}{c|}{12889.49}     & \multicolumn{1}{c|}{} & \multicolumn{1}{c|}{0.1272}          & \multicolumn{1}{c|}{0.1272}       & \multicolumn{1}{c|}{0.1272}       & \multicolumn{1}{c|}{} & \multicolumn{1}{c|}{\textbf{0.1187}} & \multicolumn{1}{c|}{0.1102}       & \multicolumn{1}{c|}{0.1187}       & \multicolumn{1}{c|}{} & \multicolumn{1}{c|}{\textbf{0.1187}} & \multicolumn{1}{c|}{0.1102}       & 0.1187       \\
8                                &  & \multicolumn{1}{c|}{7111.32}         & \multicolumn{1}{c|}{7110.25}      & \multicolumn{1}{c|}{7114.96}      & \multicolumn{1}{c|}{} & \multicolumn{1}{c|}{0.0454}          & \multicolumn{1}{c|}{0.0454}       & \multicolumn{1}{c|}{0.0454}       & \multicolumn{1}{c|}{} & \multicolumn{1}{c|}{\textbf{0.0340}} & \multicolumn{1}{c|}{0.0227}       & \multicolumn{1}{c|}{0.0340}       & \multicolumn{1}{c|}{} & \multicolumn{1}{c|}{\textbf{0.0340}} & \multicolumn{1}{c|}{0.0340}       & 0.0340       \\
10                               &  & \multicolumn{1}{c|}{359.87}          & \multicolumn{1}{c|}{358.91}       & \multicolumn{1}{c|}{362.73}       & \multicolumn{1}{c|}{} & \multicolumn{1}{c|}{0.0515}          & \multicolumn{1}{c|}{0.0515}       & \multicolumn{1}{c|}{0.1949}       & \multicolumn{1}{c|}{} & \multicolumn{1}{c|}{\textbf{0.0441}} & \multicolumn{1}{c|}{0.0294}       & \multicolumn{1}{c|}{0.0809}       & \multicolumn{1}{c|}{} & \multicolumn{1}{c|}{\textbf{0.0441}} & \multicolumn{1}{c|}{0.0441}       & 0.0441       \\
12                               &  & \multicolumn{1}{c|}{315.23}          & \multicolumn{1}{c|}{314.98}       & \multicolumn{1}{c|}{319.3}        & \multicolumn{1}{c|}{} & \multicolumn{1}{c|}{3.1575}          & \multicolumn{1}{c|}{0.6097}       & \multicolumn{1}{c|}{3.6670}       & \multicolumn{1}{c|}{} & \multicolumn{1}{c|}{\textbf{3.1457}} & \multicolumn{1}{c|}{0.0353}       & \multicolumn{1}{c|}{3.7230}       & \multicolumn{1}{c|}{} & \multicolumn{1}{c|}{\textbf{3.1457}} & \multicolumn{1}{c|}{0.0353}       & 3.1457       \\
14                               &  & \multicolumn{1}{c|}{940.51}          & \multicolumn{1}{c|}{933.67}       & \multicolumn{1}{c|}{956.65}       & \multicolumn{1}{c|}{} & \multicolumn{1}{c|}{0.0532}          & \multicolumn{1}{c|}{0.0532}       & \multicolumn{1}{c|}{0.0621}       & \multicolumn{1}{c|}{} & \multicolumn{1}{c|}{\textbf{0.0355}} & \multicolumn{1}{c|}{0.0266}       & \multicolumn{1}{c|}{0.0443}       & \multicolumn{1}{c|}{} & \multicolumn{1}{c|}{0.0443}          & \multicolumn{1}{c|}{0.0443}       & 0.0532       \\
16                               &  & \multicolumn{1}{c|}{426.07}          & \multicolumn{1}{c|}{423.47}       & \multicolumn{1}{c|}{433.41}       & \multicolumn{1}{c|}{} & \multicolumn{1}{c|}{0.8264}          & \multicolumn{1}{c|}{0.0590}       & \multicolumn{1}{c|}{4.3119}       & \multicolumn{1}{c|}{} & \multicolumn{1}{c|}{\textbf{0.0472}} & \multicolumn{1}{c|}{0.0236}       & \multicolumn{1}{c|}{0.0708}       & \multicolumn{1}{c|}{} & \multicolumn{1}{c|}{\textbf{0.0472}} & \multicolumn{1}{c|}{0.0472}       & 0.0590       \\
18                               &  & \multicolumn{1}{c|}{118.38}          & \multicolumn{1}{c|}{117.59}       & \multicolumn{1}{c|}{119.3}        & \multicolumn{1}{c|}{} & \multicolumn{1}{c|}{0.0758}          & \multicolumn{1}{c|}{0.0758}       & \multicolumn{1}{c|}{0.1137}       & \multicolumn{1}{c|}{} & \multicolumn{1}{c|}{0.0716}          & \multicolumn{1}{c|}{0.0590}       & \multicolumn{1}{c|}{0.0842}       & \multicolumn{1}{c|}{} & \multicolumn{1}{c|}{\textbf{0.0674}} & \multicolumn{1}{c|}{0.0632}       & 0.0758       \\ \hline
\end{tabular}

\caption{Relative Error for 1 layer YZ linear ansatz in $U_q$ and XY Hamiltonian DLA gates in \Ug}
\end{subtable}

\begin{subtable}{\linewidth}
\begin{tabular}{|c|l|ccccccccccccccc|}
\hline
\multirow{3}{*}{\textbf{Qubits}} &  & \multicolumn{15}{c|}{\textbf{Relative Error (units of $10^{-5}$)}}                                                                                                                                                                                                                                                                                                                                                                                                                                                                  \\ \cline{3-17}
                                 &  & \multicolumn{3}{c|}{\textbf{\g}}                                                                             & \multicolumn{1}{l|}{} & \multicolumn{3}{c|}{\textbf{Full-PSR}}                                                                       & \multicolumn{1}{l|}{} & \multicolumn{3}{c|}{\textbf{Alternate}}                                                                      & \multicolumn{1}{l|}{} & \multicolumn{3}{c|}{\textbf{Alt+Sim}}                                                   \\ \cline{3-5} \cline{7-9} \cline{11-13} \cline{15-17}
                                 &  & \multicolumn{1}{c|}{\textbf{Median}} & \multicolumn{1}{c|}{\textbf{Q25}} & \multicolumn{1}{c|}{\textbf{Q75}} & \multicolumn{1}{l|}{} & \multicolumn{1}{c|}{\textbf{Median}} & \multicolumn{1}{c|}{\textbf{Q25}} & \multicolumn{1}{c|}{\textbf{Q75}} & \multicolumn{1}{l|}{} & \multicolumn{1}{c|}{\textbf{Median}} & \multicolumn{1}{c|}{\textbf{Q25}} & \multicolumn{1}{c|}{\textbf{Q75}} & \multicolumn{1}{l|}{} & \multicolumn{1}{c|}{\textbf{Median}} & \multicolumn{1}{c|}{\textbf{Q25}} & \textbf{Q75} \\ \hline
6                                &  & \multicolumn{1}{c|}{12862.78}        & \multicolumn{1}{c|}{12856.2}      & \multicolumn{1}{c|}{12889.49}     & \multicolumn{1}{c|}{} & \multicolumn{1}{c|}{0.1272}          & \multicolumn{1}{c|}{0.1272}       & \multicolumn{1}{c|}{0.1653}       & \multicolumn{1}{c|}{} & \multicolumn{1}{c|}{\textbf{0.1187}} & \multicolumn{1}{c|}{0.1102}       & \multicolumn{1}{c|}{0.1780}       & \multicolumn{1}{c|}{} & \multicolumn{1}{c|}{\textbf{0.1187}} & \multicolumn{1}{c|}{0.1187}       & 0.1272       \\
8                                &  & \multicolumn{1}{c|}{7111.32}         & \multicolumn{1}{c|}{7110.25}      & \multicolumn{1}{c|}{7114.96}      & \multicolumn{1}{c|}{} & \multicolumn{1}{c|}{0.0454}          & \multicolumn{1}{c|}{0.0454}       & \multicolumn{1}{c|}{0.0936}       & \multicolumn{1}{c|}{} & \multicolumn{1}{c|}{\textbf{0.0340}} & \multicolumn{1}{c|}{0.0227}       & \multicolumn{1}{c|}{0.0993}       & \multicolumn{1}{c|}{} & \multicolumn{1}{c|}{\textbf{0.0340}} & \multicolumn{1}{c|}{0.0340}       & 0.0369       \\
10                               &  & \multicolumn{1}{c|}{359.87}          & \multicolumn{1}{c|}{358.91}       & \multicolumn{1}{c|}{362.73}       & \multicolumn{1}{c|}{} & \multicolumn{1}{c|}{0.1802}          & \multicolumn{1}{c|}{0.0515}       & \multicolumn{1}{c|}{0.5828}       & \multicolumn{1}{c|}{} & \multicolumn{1}{c|}{0.3015}          & \multicolumn{1}{c|}{0.0827}       & \multicolumn{1}{c|}{1.7409}       & \multicolumn{1}{c|}{} & \multicolumn{1}{c|}{\textbf{0.0441}} & \multicolumn{1}{c|}{0.0441}       & 0.0515       \\
12                               &  & \multicolumn{1}{c|}{315.23}          & \multicolumn{1}{c|}{314.98}       & \multicolumn{1}{c|}{319.3}        & \multicolumn{1}{c|}{} & \multicolumn{1}{c|}{3.1634}          & \multicolumn{1}{c|}{0.5950}       & \multicolumn{1}{c|}{4.0234}       & \multicolumn{1}{c|}{} & \multicolumn{1}{c|}{3.1575}          & \multicolumn{1}{c|}{0.6244}       & \multicolumn{1}{c|}{4.7009}       & \multicolumn{1}{c|}{} & \multicolumn{1}{c|}{\textbf{1.5905}} & \multicolumn{1}{c|}{0.0353}       & 3.1516       \\
14                               &  & \multicolumn{1}{c|}{940.51}          & \multicolumn{1}{c|}{933.67}       & \multicolumn{1}{c|}{956.65}       & \multicolumn{1}{c|}{} & \multicolumn{1}{c|}{0.0621}          & \multicolumn{1}{c|}{0.0532}       & \multicolumn{1}{c|}{0.1685}       & \multicolumn{1}{c|}{} & \multicolumn{1}{c|}{0.1286}          & \multicolumn{1}{c|}{0.0532}       & \multicolumn{1}{c|}{0.7404}       & \multicolumn{1}{c|}{} & \multicolumn{1}{c|}{\textbf{0.0532}} & \multicolumn{1}{c|}{0.0532}       & 0.0532       \\
16                               &  & \multicolumn{1}{c|}{426.07}          & \multicolumn{1}{c|}{423.47}       & \multicolumn{1}{c|}{433.41}       & \multicolumn{1}{c|}{} & \multicolumn{1}{c|}{0.1299}          & \multicolumn{1}{c|}{0.0649}       & \multicolumn{1}{c|}{0.6021}       & \multicolumn{1}{c|}{} & \multicolumn{1}{c|}{1.0507}          & \multicolumn{1}{c|}{0.1033}       & \multicolumn{1}{c|}{3.0192}       & \multicolumn{1}{c|}{} & \multicolumn{1}{c|}{\textbf{0.0472}} & \multicolumn{1}{c|}{0.0472}       & 0.0590       \\
18                               &  & \multicolumn{1}{c|}{118.38}          & \multicolumn{1}{c|}{117.59}       & \multicolumn{1}{c|}{119.3}        & \multicolumn{1}{c|}{} & \multicolumn{1}{c|}{0.1390}          & \multicolumn{1}{c|}{0.0821}       & \multicolumn{1}{c|}{0.3032}       & \multicolumn{1}{c|}{} & \multicolumn{1}{c|}{3.7105}          & \multicolumn{1}{c|}{0.7981}       & \multicolumn{1}{c|}{9.5521}       & \multicolumn{1}{c|}{} & \multicolumn{1}{c|}{\textbf{0.0674}} & \multicolumn{1}{c|}{0.0674}       & 0.0758       \\ \hline
\end{tabular}
\caption{Relative Error for 3 layer YZ linear ansatz in $U_q$ and XY Hamiltonian DLA gates in \Ug}
\end{subtable}

\begin{subtable}{\linewidth}

\begin{tabular}{|c|l|ccccccccccccccc|}
\hline
\multirow{3}{*}{\textbf{Qubits}} &  & \multicolumn{15}{c|}{\textbf{Relative Error (units of $10^{-5}$)}}                                                                                                                                                                                                                                                                                                                                                                                                                                                                  \\ \cline{3-17}
                                 &  & \multicolumn{3}{c|}{\textbf{\g}}                                                                             & \multicolumn{1}{l|}{} & \multicolumn{3}{c|}{\textbf{Full-PSR}}                                                                       & \multicolumn{1}{l|}{} & \multicolumn{3}{c|}{\textbf{Alternate}}                                                                      & \multicolumn{1}{l|}{} & \multicolumn{3}{c|}{\textbf{Alt+Sim}}                                                   \\ \cline{3-5} \cline{7-9} \cline{11-13} \cline{15-17}
                                 &  & \multicolumn{1}{c|}{\textbf{Median}} & \multicolumn{1}{c|}{\textbf{Q25}} & \multicolumn{1}{c|}{\textbf{Q75}} & \multicolumn{1}{l|}{} & \multicolumn{1}{c|}{\textbf{Median}} & \multicolumn{1}{c|}{\textbf{Q25}} & \multicolumn{1}{c|}{\textbf{Q75}} & \multicolumn{1}{l|}{} & \multicolumn{1}{c|}{\textbf{Median}} & \multicolumn{1}{c|}{\textbf{Q25}} & \multicolumn{1}{c|}{\textbf{Q75}} & \multicolumn{1}{l|}{} & \multicolumn{1}{c|}{\textbf{Median}} & \multicolumn{1}{c|}{\textbf{Q25}} & \textbf{Q75} \\ \hline
6                                &  & \multicolumn{1}{c|}{12862.78}        & \multicolumn{1}{c|}{12856.2}      & \multicolumn{1}{c|}{12889.49}     & \multicolumn{1}{c|}{} & \multicolumn{1}{c|}{0.5676}          & \multicolumn{1}{c|}{0.1611}       & \multicolumn{1}{c|}{1.4060}       & \multicolumn{1}{c|}{} & \multicolumn{1}{c|}{0.5760}          & \multicolumn{1}{c|}{0.1441}       & \multicolumn{1}{c|}{1.4949}       & \multicolumn{1}{c|}{} & \multicolumn{1}{c|}{\textbf{0.1272}} & \multicolumn{1}{c|}{0.1272}       & 0.1611       \\
8                                &  & \multicolumn{1}{c|}{7111.32}         & \multicolumn{1}{c|}{7110.25}      & \multicolumn{1}{c|}{7114.96}      & \multicolumn{1}{c|}{} & \multicolumn{1}{c|}{0.1589}          & \multicolumn{1}{c|}{0.0681}       & \multicolumn{1}{c|}{0.7207}       & \multicolumn{1}{c|}{} & \multicolumn{1}{c|}{0.5845}          & \multicolumn{1}{c|}{0.1475}       & \multicolumn{1}{c|}{1.7165}       & \multicolumn{1}{c|}{} & \multicolumn{1}{c|}{\textbf{0.0454}} & \multicolumn{1}{c|}{0.0340}       & 0.0567       \\
10                               &  & \multicolumn{1}{c|}{359.87}          & \multicolumn{1}{c|}{358.91}       & \multicolumn{1}{c|}{362.73}       & \multicolumn{1}{c|}{} & \multicolumn{1}{c|}{0.4706}          & \multicolumn{1}{c|}{0.1893}       & \multicolumn{1}{c|}{0.9890}       & \multicolumn{1}{c|}{} & \multicolumn{1}{c|}{1.4413}          & \multicolumn{1}{c|}{0.3750}       & \multicolumn{1}{c|}{3.7428}       & \multicolumn{1}{c|}{} & \multicolumn{1}{c|}{\textbf{0.0588}} & \multicolumn{1}{c|}{0.0515}       & 0.0735       \\
12                               &  & \multicolumn{1}{c|}{315.23}          & \multicolumn{1}{c|}{314.98}       & \multicolumn{1}{c|}{319.3}        & \multicolumn{1}{c|}{} & \multicolumn{1}{c|}{1.6377}          & \multicolumn{1}{c|}{0.1885}       & \multicolumn{1}{c|}{3.2753}       & \multicolumn{1}{c|}{} & \multicolumn{1}{c|}{3.6346}          & \multicolumn{1}{c|}{1.3578}       & \multicolumn{1}{c|}{5.5020}       & \multicolumn{1}{c|}{} & \multicolumn{1}{c|}{\textbf{1.4786}} & \multicolumn{1}{c|}{0.0353}       & 3.1575       \\
14                               &  & \multicolumn{1}{c|}{940.51}          & \multicolumn{1}{c|}{933.67}       & \multicolumn{1}{c|}{956.65}       & \multicolumn{1}{c|}{} & \multicolumn{1}{c|}{0.1330}          & \multicolumn{1}{c|}{0.0709}       & \multicolumn{1}{c|}{0.2416}       & \multicolumn{1}{c|}{} & \multicolumn{1}{c|}{2.9881}          & \multicolumn{1}{c|}{1.1970}       & \multicolumn{1}{c|}{3.9945}       & \multicolumn{1}{c|}{} & \multicolumn{1}{c|}{\textbf{0.0532}} & \multicolumn{1}{c|}{0.0532}       & 0.0621       \\
16                               &  & \multicolumn{1}{c|}{426.07}          & \multicolumn{1}{c|}{423.47}       & \multicolumn{1}{c|}{433.41}       & \multicolumn{1}{c|}{} & \multicolumn{1}{c|}{0.1417}          & \multicolumn{1}{c|}{0.1181}       & \multicolumn{1}{c|}{0.4014}       & \multicolumn{1}{c|}{} & \multicolumn{1}{c|}{5.3951}          & \multicolumn{1}{c|}{2.3257}       & \multicolumn{1}{c|}{8.6062}       & \multicolumn{1}{c|}{} & \multicolumn{1}{c|}{\textbf{0.0590}} & \multicolumn{1}{c|}{0.0472}       & 0.0708       \\
18                               &  & \multicolumn{1}{c|}{118.38}          & \multicolumn{1}{c|}{117.59}       & \multicolumn{1}{c|}{119.3}        & \multicolumn{1}{c|}{} & \multicolumn{1}{c|}{0.1137}          & \multicolumn{1}{c|}{0.0927}       & \multicolumn{1}{c|}{0.1853}       & \multicolumn{1}{c|}{} & \multicolumn{1}{c|}{4.9487}          & \multicolumn{1}{c|}{2.4343}       & \multicolumn{1}{c|}{15.1262}      & \multicolumn{1}{c|}{} & \multicolumn{1}{c|}{\textbf{0.0842}} & \multicolumn{1}{c|}{0.0674}       & 0.1032       \\ \hline
\end{tabular}
\caption{Relative Error for 6 layer YZ linear ansatz in $U_q$ and XY Hamiltonian DLA gates in \Ug}
\end{subtable}

\begin{subtable}{\linewidth}

\begin{tabular}{|c|l|ccclccccccccccc|}
\hline
\multirow{3}{*}{\textbf{Qubits}} &  & \multicolumn{15}{c|}{\textbf{Relative Error (units of $10^{-5}$)}}                                                                                                                                                                                                                                                                                                                                                                                                                                                                  \\ \cline{3-17}
                                 &  & \multicolumn{3}{c|}{\textbf{\g}}                                                              & \multicolumn{1}{l|}{} & \multicolumn{3}{c|}{\textbf{Full-PSR}}                                                                       & \multicolumn{1}{l|}{} & \multicolumn{3}{c|}{\textbf{Alternate}}                                                                      & \multicolumn{1}{l|}{} & \multicolumn{3}{c|}{\textbf{Alt+Sim}}                                                   \\ \cline{3-5} \cline{7-9} \cline{11-13} \cline{15-17}
                                 &  & \multicolumn{1}{c|}{\textbf{Median}} & \multicolumn{1}{c|}{\textbf{Q25}} & \multicolumn{1}{c|}{\textbf{Q75}} & \multicolumn{1}{l|}{} & \multicolumn{1}{c|}{\textbf{Median}} & \multicolumn{1}{c|}{\textbf{Q25}} & \multicolumn{1}{c|}{\textbf{Q75}} & \multicolumn{1}{l|}{} & \multicolumn{1}{c|}{\textbf{Median}} & \multicolumn{1}{c|}{\textbf{Q25}} & \multicolumn{1}{c|}{\textbf{Q75}} & \multicolumn{1}{l|}{} & \multicolumn{1}{c|}{\textbf{Median}} & \multicolumn{1}{c|}{\textbf{Q25}} & \textbf{Q75} \\ \hline
6                                &  & \multicolumn{1}{c|}{12862.78}        & \multicolumn{1}{c|}{12856.2}      & \multicolumn{1}{c|}{12889.49}     & \multicolumn{1}{l|}{} & \multicolumn{1}{c|}{0.5337}          & \multicolumn{1}{c|}{0.1272}       & \multicolumn{1}{c|}{4.7936}       & \multicolumn{1}{c|}{} & \multicolumn{1}{c|}{2.3969}          & \multicolumn{1}{c|}{0.4342}       & \multicolumn{1}{c|}{5.1028}       & \multicolumn{1}{c|}{} & \multicolumn{1}{c|}{\textbf{0.1272}} & \multicolumn{1}{c|}{0.1187}       & 0.3220       \\
8                                &  & \multicolumn{1}{c|}{7111.32}         & \multicolumn{1}{c|}{7110.25}      & \multicolumn{1}{c|}{7114.96}      & \multicolumn{1}{l|}{} & \multicolumn{1}{c|}{0.5107}          & \multicolumn{1}{c|}{0.1759}       & \multicolumn{1}{c|}{1.7534}       & \multicolumn{1}{c|}{} & \multicolumn{1}{c|}{2.7351}          & \multicolumn{1}{c|}{0.6242}       & \multicolumn{1}{c|}{6.0377}       & \multicolumn{1}{c|}{} & \multicolumn{1}{c|}{\textbf{0.0681}} & \multicolumn{1}{c|}{0.0454}       & 0.0965       \\
10                               &  & \multicolumn{1}{c|}{359.87}          & \multicolumn{1}{c|}{358.91}       & \multicolumn{1}{c|}{362.73}       & \multicolumn{1}{l|}{} & \multicolumn{1}{c|}{0.4559}          & \multicolumn{1}{c|}{0.2390}       & \multicolumn{1}{c|}{0.9798}       & \multicolumn{1}{c|}{} & \multicolumn{1}{c|}{2.8788}          & \multicolumn{1}{c|}{1.1104}       & \multicolumn{1}{c|}{5.4323}       & \multicolumn{1}{c|}{} & \multicolumn{1}{c|}{\textbf{0.0735}} & \multicolumn{1}{c|}{0.0552}       & 0.1177       \\
12                               &  & \multicolumn{1}{c|}{315.23}          & \multicolumn{1}{c|}{314.98}       & \multicolumn{1}{c|}{319.3}        & \multicolumn{1}{l|}{} & \multicolumn{1}{c|}{2.0206}          & \multicolumn{1}{c|}{0.2121}       & \multicolumn{1}{c|}{3.5522}       & \multicolumn{1}{c|}{} & \multicolumn{1}{c|}{4.7009}          & \multicolumn{1}{c|}{3.1811}       & \multicolumn{1}{c|}{7.9762}       & \multicolumn{1}{c|}{} & \multicolumn{1}{c|}{\textbf{1.5905}} & \multicolumn{1}{c|}{0.0589}       & 3.1693       \\
14                               &  & \multicolumn{1}{c|}{940.51}          & \multicolumn{1}{c|}{933.67}       & \multicolumn{1}{c|}{956.65}       & \multicolumn{1}{l|}{} & \multicolumn{1}{c|}{0.1064}          & \multicolumn{1}{c|}{0.0709}       & \multicolumn{1}{c|}{0.1862}       & \multicolumn{1}{c|}{} & \multicolumn{1}{c|}{4.2472}          & \multicolumn{1}{c|}{2.5581}       & \multicolumn{1}{c|}{5.7191}       & \multicolumn{1}{c|}{} & \multicolumn{1}{c|}{\textbf{0.0709}} & \multicolumn{1}{c|}{0.0621}       & 0.1131       \\
16                               &  & \multicolumn{1}{c|}{426.07}          & \multicolumn{1}{c|}{423.47}       & \multicolumn{1}{c|}{433.41}       & \multicolumn{1}{l|}{} & \multicolumn{1}{c|}{0.1062}          & \multicolumn{1}{c|}{0.0590}       & \multicolumn{1}{c|}{0.2774}       & \multicolumn{1}{c|}{} & \multicolumn{1}{c|}{5.2593}          & \multicolumn{1}{c|}{3.0399}       & \multicolumn{1}{c|}{12.2541}      & \multicolumn{1}{c|}{} & \multicolumn{1}{c|}{\textbf{0.0826}} & \multicolumn{1}{c|}{0.0708}       & 0.1062       \\
18                               &  & \multicolumn{1}{c|}{118.38}          & \multicolumn{1}{c|}{117.59}       & \multicolumn{1}{c|}{119.3}        & \multicolumn{1}{l|}{} & \multicolumn{1}{c|}{0.1390}          & \multicolumn{1}{c|}{0.0863}       & \multicolumn{1}{c|}{0.2843}       & \multicolumn{1}{c|}{} & \multicolumn{1}{c|}{6.6966}          & \multicolumn{1}{c|}{4.5486}       & \multicolumn{1}{c|}{13.9533}      & \multicolumn{1}{c|}{} & \multicolumn{1}{c|}{\textbf{0.1011}} & \multicolumn{1}{c|}{0.0842}       & 0.1137       \\ \hline
\end{tabular}

\caption{Relative Error for 9 layer YZ linear ansatz in $U_q$ and XY Hamiltonian DLA gates in \Ug}
\end{subtable}

\caption{\raggedright\textbf{XY Hamiltonian VQE (Relative Error) : }The relative error median, 25\% quantile (Q25) and 75\% quantile (Q75) for Full-PSR, Alternate and Alternate + Simultaneous (referred to as Alt+Sim) for the configurations with 1, 3, 6, and 9 layers of YZ linear ansatz in $U_q$ and XY Hamiltonian DLA gates in \Ug.  The relative error for \g with only XY Hamiltonian DLA gates for each qubit count is shown for comparison, which is several orders of magnitude higher than the other methods. Alt+Sim has lower reported median relative error than Full-PSR in the displayed configurations. Alternate alone is less consistent and has higher median error in several deeper configurations.}
\label{vqe_xy_error}
\end{table}
\clearpage
\section{Gradient behavior for VQE on the XY Hamiltonian}

\begin{figure}[!tbh]
    \begin{subfigure}{0.45\linewidth}
        \includegraphics[width=1\linewidth]{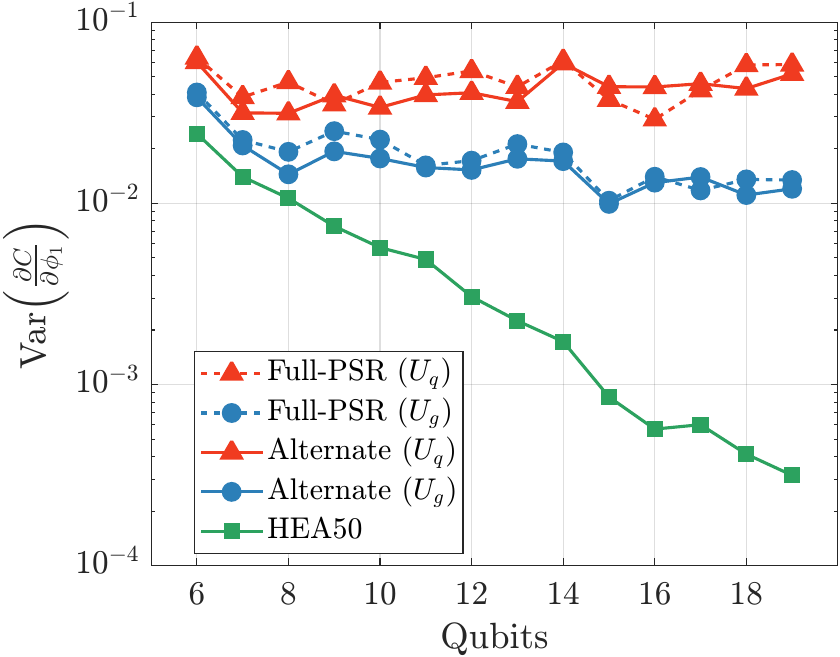}
        \caption{3 YZ linear layers in $U_q$}
    \end{subfigure}
        \begin{subfigure}{0.45\linewidth}
        \includegraphics[width=1\linewidth]{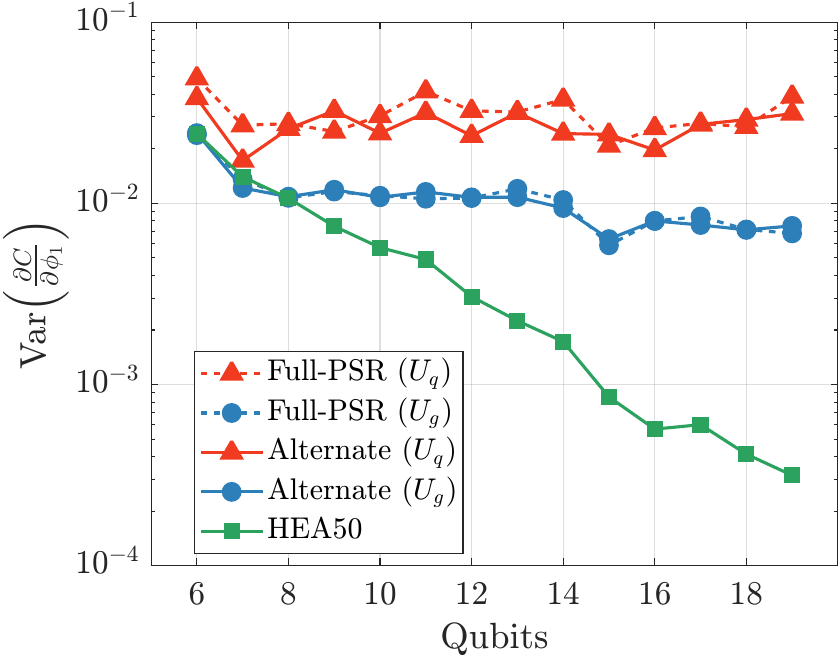}
        \caption{6 YZ linear layers in $U_q$}
    \end{subfigure}

    \begin{subfigure}{0.45\linewidth}
        \includegraphics[width=1\linewidth]{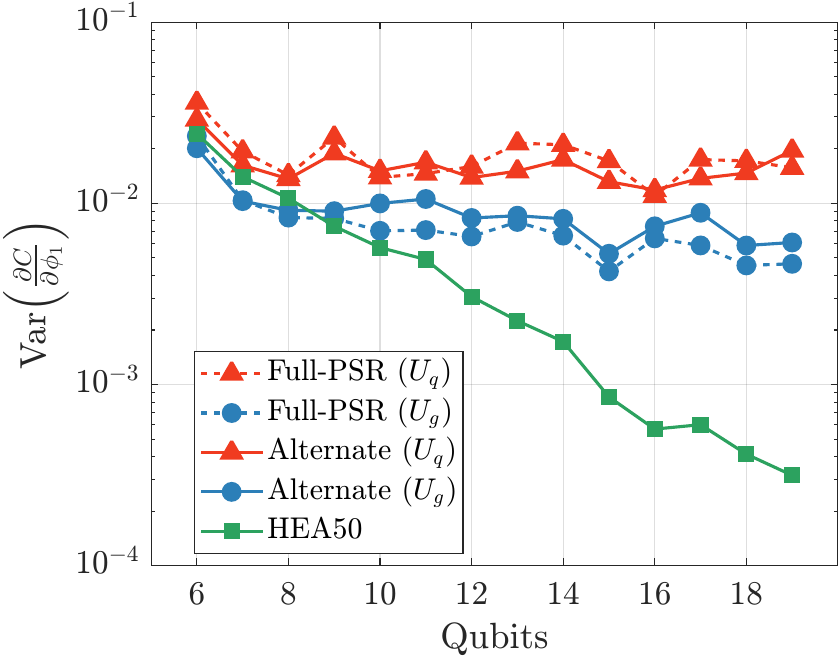}
        \caption{9 YZ linear layers in $U_q$}
    \end{subfigure}

    \caption{\raggedright \textbf{Gradients for XY Hamiltonian VQE:} Variance of the first parameter of $U_q$ (red) and $U_{\mathfrak{g}}$ (blue) are plotted with increasing qubits for both Full-PSR and our Alternate method for finding the ground state of XY Hamiltonian using VQE. The diagrams show the plots for 3(a), 6(b) and 9 (c) YZ linear layer in $U_q$. For comparison, the same task is performed using a deep HEA (referred to as HEA50) with standard PSR and variance of its first partial derivative is plotted in green. In both $U_q$ and \Ug, we notice a slower decay of gradients as compared to HEA50 over the displayed system sizes.}
    \label{vanishing_fig2}
\end{figure}

Here we plot the variance of the gradients for configurations with 3, 6, and 9 YZ linear layers in $U_q$ of HELIA, when used for VQE on XY Hamiltonian in Fig. \ref{vanishing_fig2}. In each case, we track the gradient of the first parameter from each of $U_q$ and $U_{\mathfrak{g}}$ and plot its variance during training (Fig.\ref{vanishing_fig2}). Each datapoint is obtained after evaluating the variance from the full training of 64 independent trials.

For deep-circuit reference, we plot a deep hardware efficient ansatz (labeled HEA50) in green in Fig.\ref{vanishing_fig2}. Across the displayed fixed-depth configurations, HELIA shows a clear improvement in terms of slow decay in gradients for all the $U_q$ configurations. The gradients for $U_q$ and \Ug demonstrate a much slower decay compared to the HEA50 reference, showing potential for training large-qubit models. Importantly, the comparison is not matched in circuit complexity and does not establish asymptotic gradient scaling or general barren-plateau mitigation.

\section{ \texorpdfstring{$\mathfrak{g}$-Purity}{g-Purity} of Hardware Efficient Ansatz at various circuit depths}\label{purity_sec}

 Using results from Ref.~\cite{ragone2023unified} and considering $\mathfrak{g}$ to be simple, we obtain Eq.\eqref{BP_eq}. The $\mathfrak{g}$-Purity is defined as,
 \begin{equation}\label{purity_eq}
     \mathcal{P}_\mathfrak{g}(\hat{O)} = \sum_{i=1 }^{dim(\mathfrak{g})}  \Big|\Tr[\hat{B}_i^\dagger O]\Big|^2
 \end{equation}
where $\{B_i\}_{i=1}^{dim(\mathfrak{g})}$ forms an orthonormal basis on $\mathfrak{g}$ with respect to the Hilbert-Schmidt inner product $\langle A,B\rangle = \Tr (A^\dagger B)$. Intuitively, this represents how much of the operator overlaps with $\mathfrak{g}$.

\begin{figure}[H]
    \centering
    \includegraphics[width=0.45\linewidth]{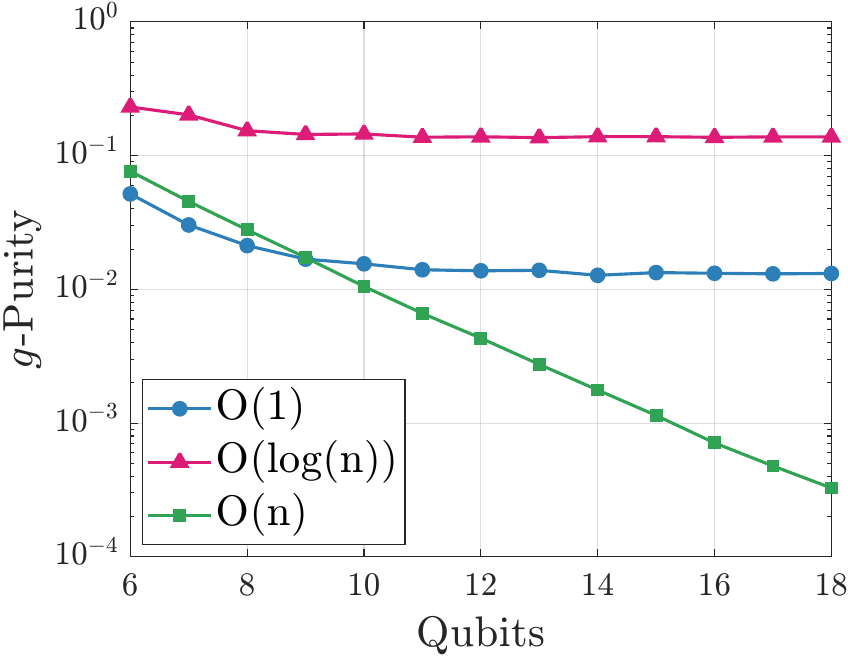}
    \caption{Average $\mathfrak{g}$-Purity of HEA for $\mathfrak{g}_{XY}$ DLA (defined in Eq.\,\eqref{purity_eq}) with increasing qubits and varying circuit depths. For constant ($O(1)$) and logarithmic depth in qubits ($O(log(n))$), the decay is less than exponential (shown in blue and purple respectively). For linear depth ($O(n)$ in green), $\mathfrak{g}$-Purity decreases exponentially. Each point in the plot corresponds to an average of 10000 independent parameter samples. }
    \label{purity}
\end{figure}

Based on Eq.\,\eqref{BP_eq}, we see that in order to have sub-exponential decay of gradients, $\mathcal{P}_\mathfrak{g}(\rho_q(\theta))$ or $g$-Purity plays an important role. In Fig.\,\ref{purity}, we numerically evaluate it for varying circuit depths considering the $\mathfrak{g}_{XY}$ DLA. Both constant and logarithmic depth circuits show sub-exponential decay while linear depth circuit has exponential decay. This indicates that HELIA with logarithmic depth HEA could also potentially show slow gradient decay. However, we keep that direction open for future exploration. Importantly, these observations provide a finite-size diagnostic of the input to the structured block; they do not directly establish the scaling of parameter derivatives or guarantee barren-plateau avoidance for all qubits.

\section{Example instances of \texorpdfstring{\g}{g-sim} reaching accurate solution without quantum hardware}\label{odd_gsim}

\begin{figure}[!htp]
    \begin{subfigure}{0.48\linewidth}
       \includegraphics[width = 1\linewidth]{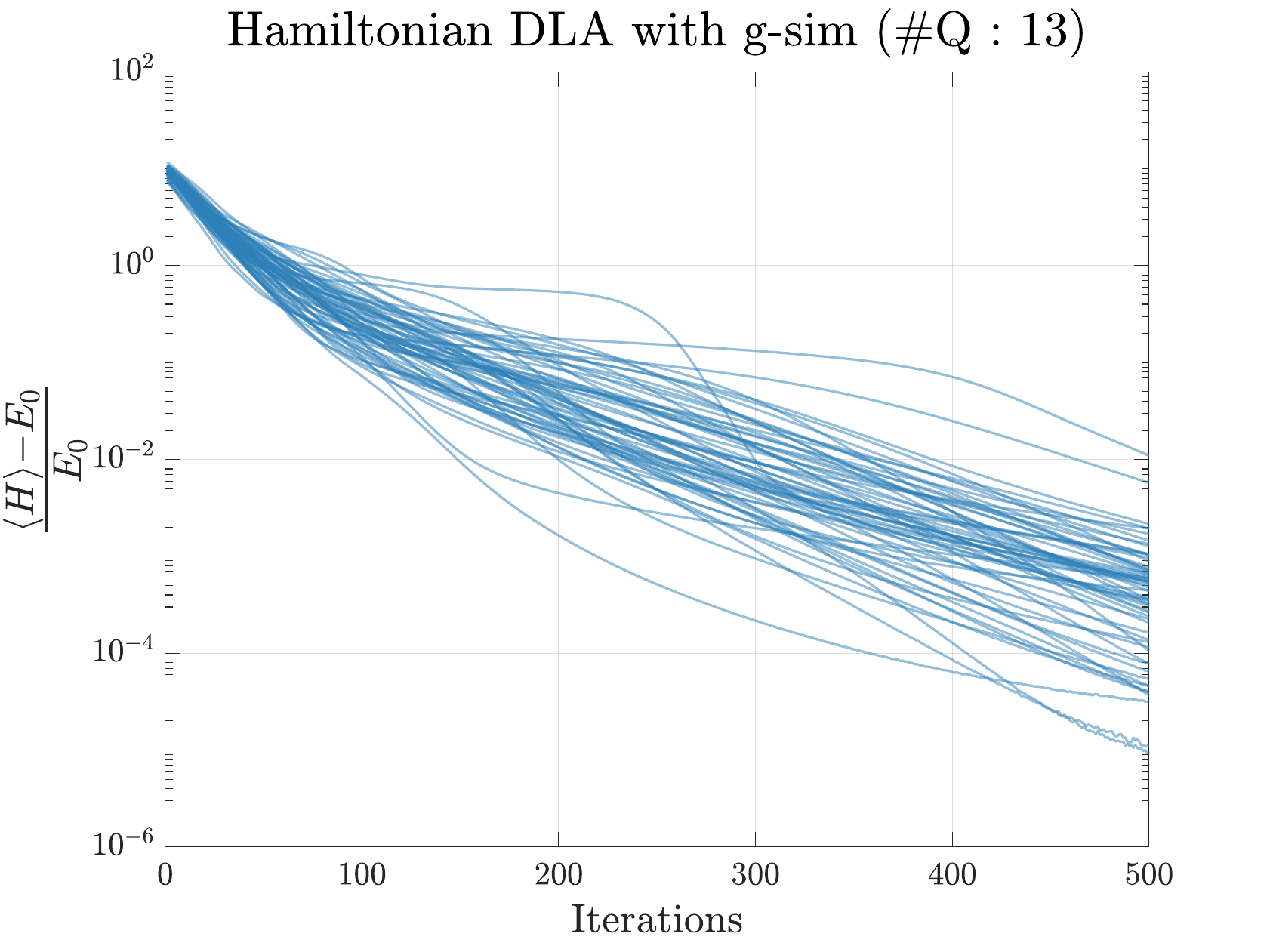}
        \caption{XY Hamiltonian with 13 qubits}
        \label{gsim_13}
    \end{subfigure}
    \begin{subfigure}{.48\linewidth}
        \includegraphics[width = 1\linewidth]{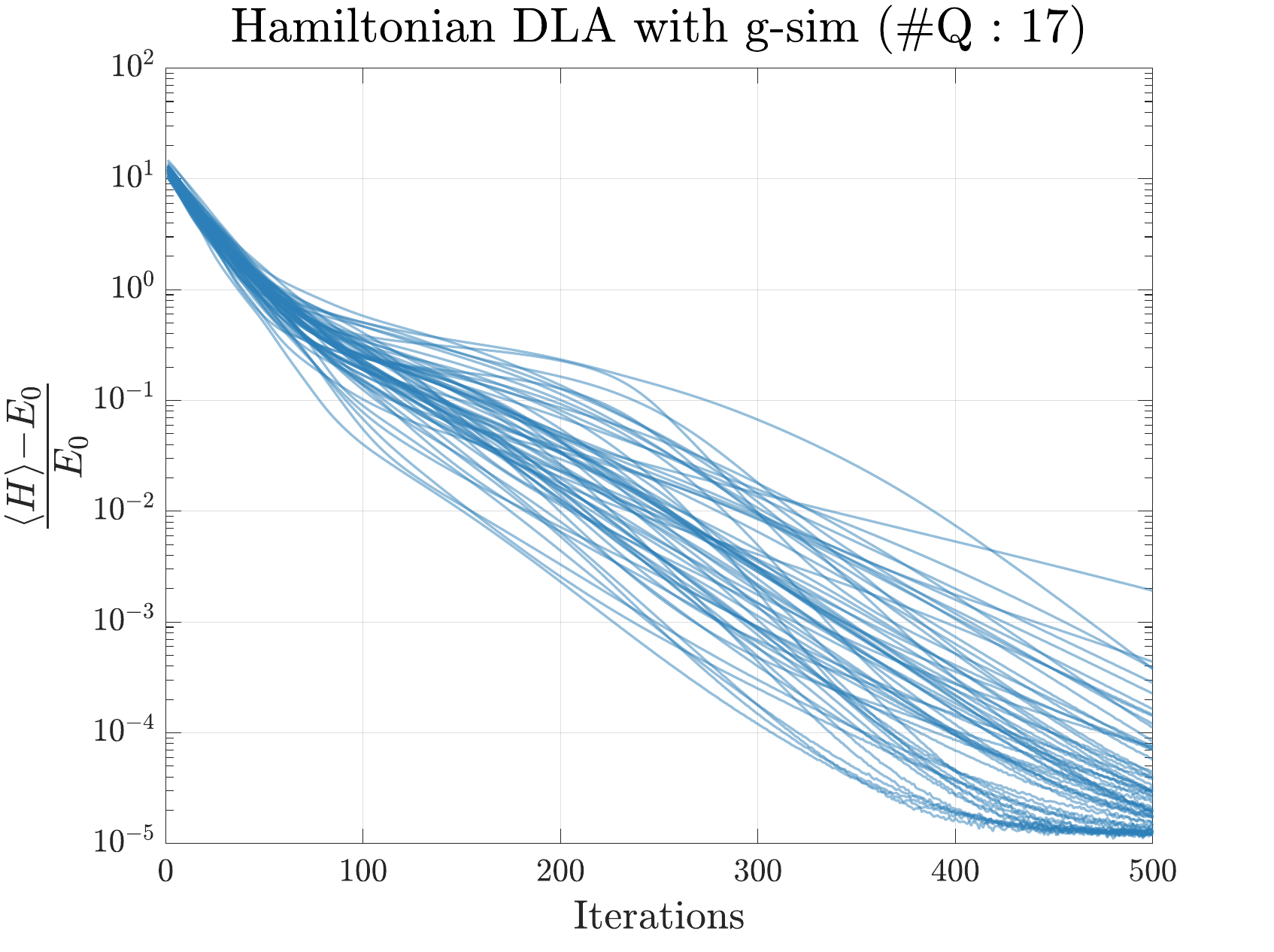}
        \caption{XY Hamiltonian with 17 qubits}
        \label{gsim_17}
    \end{subfigure}
    \caption{\raggedright Relative-error-versus-iteration curve for finding the ground state of an XY Hamiltonian using only \g. For 13 (\ref{gsim_13}) and 17 (\ref{gsim_17}) qubits, \g is enough to approach the ground state and does not require our protocol.}
    \label{odd_gsim_fig}
\end{figure}

Although our aim has been to elevate \g to a hybrid scheme with a more expressive circuit, it might not be necessary for a given task. In fact, the limited search space might be sufficient to reach a target solution, as shown in the examples below. We use \g to find the ground state of the XY Hamiltonian (Eq. \ref{xy_hamil}) for the 13- and 17-qubit cases, and the results are shown in Fig.\ref{odd_gsim_fig}. Clearly, \g is able to converge to good solutions (relative error $\sim 10^{-3}-10^{-5}$) without adding any additional gates.

\section{VQE for TFIM Hamiltonian}\label{tfim_vqe}
We also implement our training schemes for VQE using a different Hamiltonian example, in order to demonstrate that the improvements are not restricted to a specific choice. For this, we use the Transverse Field Ising Model (TFIM) Hamiltonian,
\begin{equation}
    H_{TFIM} = \sum_{i=0}^{N-2} \alpha_i X_i X_{i+1} + \sum_{j=0}^{N-1} \beta_j Z_j
\end{equation}
which has a \textit{poly}-DLA scaling as $2n^2 - n$ for $n$ qubits. All parameters associated with the Hamiltonian terms are initialized independently according to a uniform distribution over the interval $[-1,1]$. We measure the improvement in terms of Relative Error, Success and QPU calls reduction as defined in \ref{result}.

We tabulate the numerical results comparing \g, Full-PSR, Alternate and Alternate+Simultaneous in Table \ref{vqe_tfim_successqpu} and \ref{vqe_tfim_error} and provide plots for visualization in Fig.\,\ref{vqe_tfim_plot_all}. Except for the 12-qubit case, where \g is able to produce accurate solutions (similar to the scenario mentioned in App.\ref{odd_gsim}), our proposed Alternate+Simultaneous method generally achieves a lower relative error and higher success rate than the other protocols, while almost always reducing the QPU calls necessary compared to Full-PSR.

\begin{figure*}[tbh!]
    \begin{subfigure}{1\linewidth}
        \includegraphics[width = 1\linewidth]{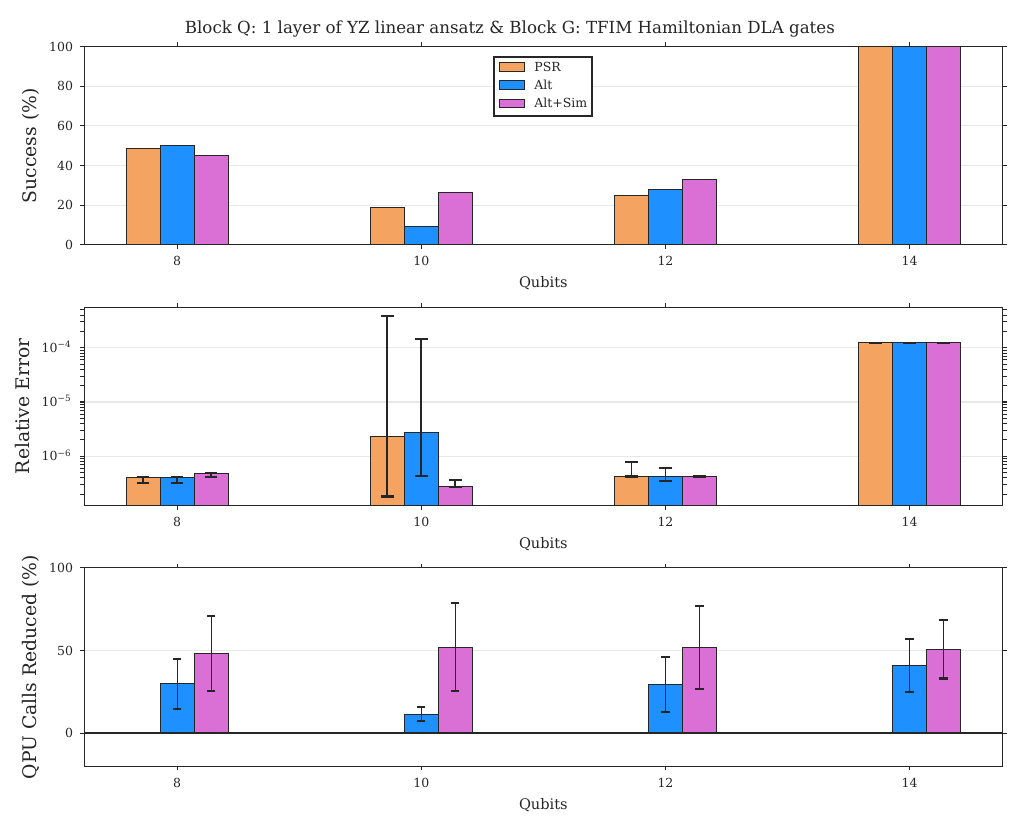}
        \caption{Success, Relative Error and QPU Call Reduction for 1 YZ linear layer in $U_q$}
    \end{subfigure}
    \end{figure*}
\begin{figure*}\ContinuedFloat
    \begin{subfigure}{1\linewidth}
        \includegraphics[width = 1\linewidth]{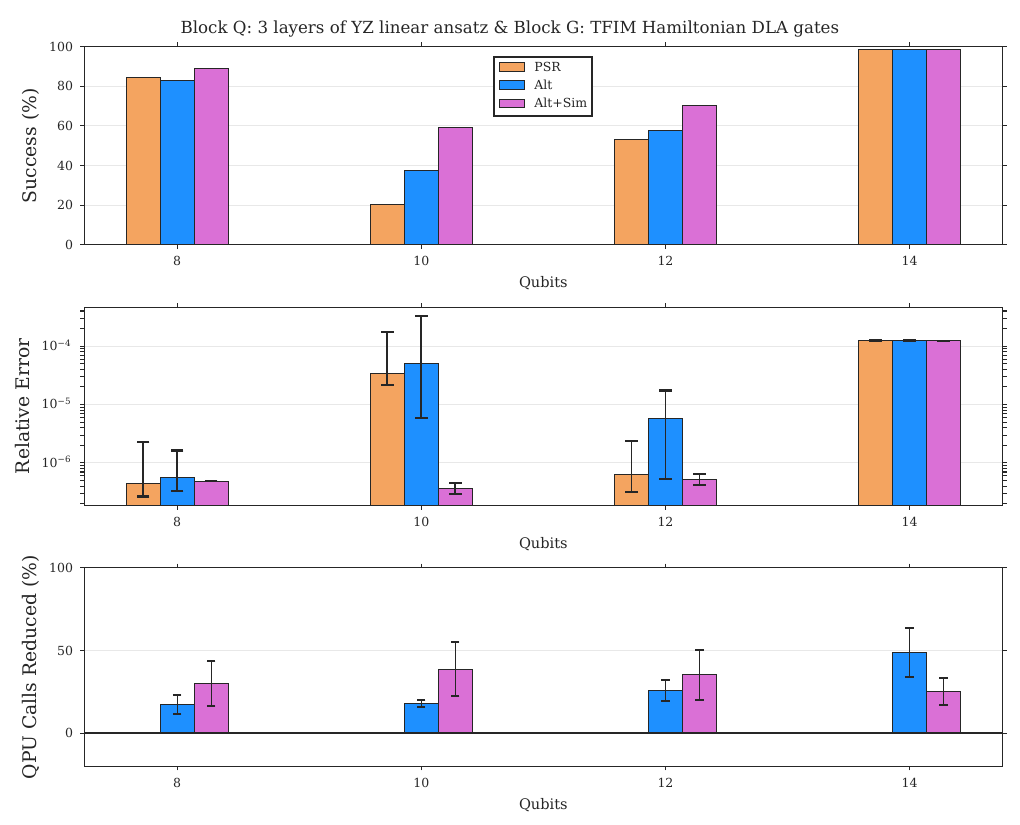}
        \caption{Success, Relative Error and QPU Call Reduction for 3 YZ linear layers in $U_q$}
    \end{subfigure}
\end{figure*}

\begin{figure*}\ContinuedFloat
    \begin{subfigure}{1\linewidth}
        \includegraphics[width = 1\linewidth]{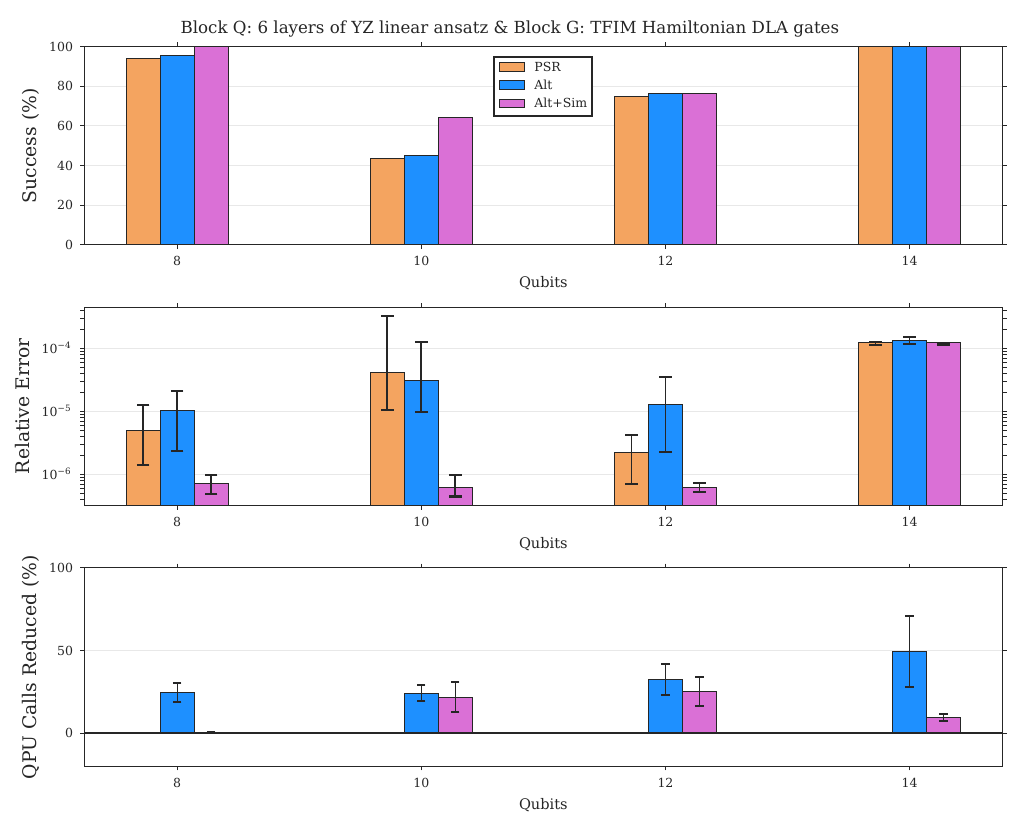}
        \caption{Success, Relative Error and QPU Call Reduction for 6 YZ linear layers in $U_q$}
    \end{subfigure}
    \end{figure*}
\begin{figure*}\ContinuedFloat
    \begin{subfigure}{1\linewidth}
        \includegraphics[width = 1\linewidth]{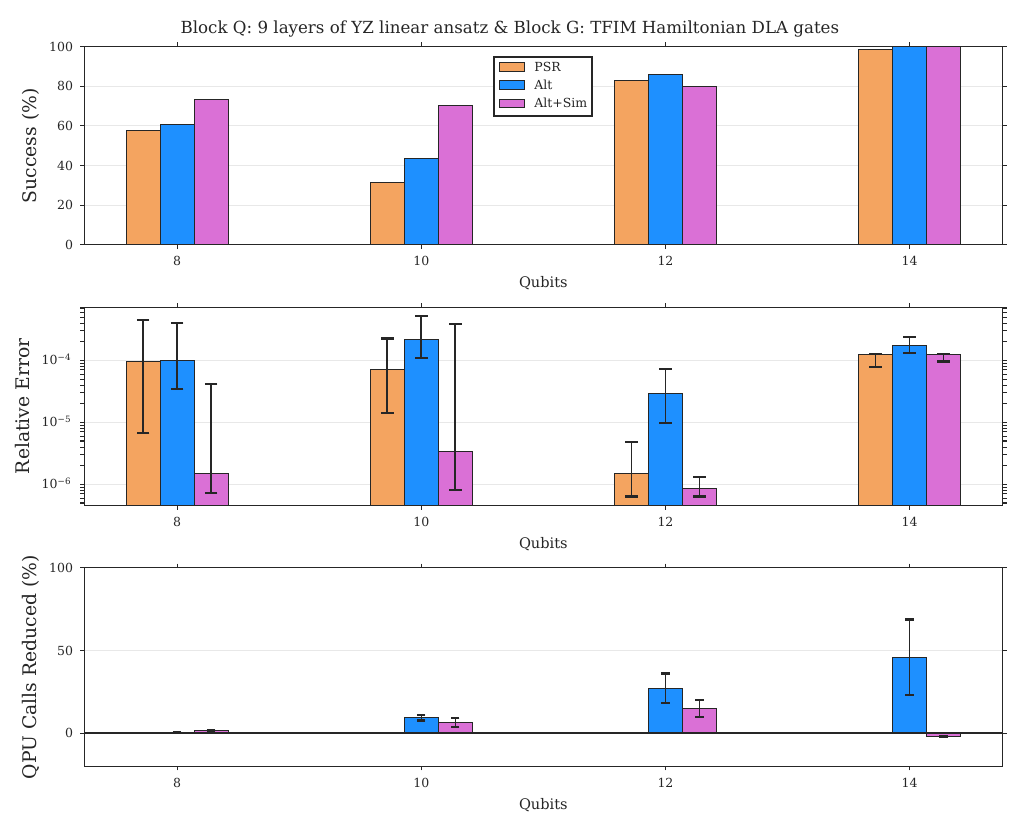}
        \caption{Success, Relative Error and QPU Call Reduction for 9 YZ linear layers in $U_q$}
    \end{subfigure}

    \caption{ \raggedright \textbf{TFIM VQE results for $n=8,10,12,14$ qubits} using (a) one, (b) three, (c) six, and (d) nine YZ-linear layers in $U_q$ (referred to as Block Q), with TFIM Lie-algebra-supported gates in $U_{\mathfrak{g}}$ (referred to as Block G). Full-PSR (orange), Alternate (blue), and Alternate followed by Simultaneous training (Alt+Sim, purple) are compared. \textbf{On the first row}, we report success rate at the relative-error threshold $10^{-3}$. \textbf{The middle row} shows successful-trial median relative errors, with black intervals indicating the 25th--75th percentiles The Alt+Sim method gives comparable or lower relative error than PSR for all qubits. \textbf{The bottom row} shows mean QPU Calls reduction relative to Full-PSR, with black error bars indicating one standard deviation. Positive reductions indicate fewer logical evaluations; negative reductions indicate more. Resource reductions are positive in all displayed settings except Alt+Sim at 14 qubits and nine layers.}
    \label{vqe_tfim_plot_all}
\end{figure*}

\begin{figure*}
    \begin{subfigure}{0.45\linewidth}
        \includegraphics[width = 1\linewidth]{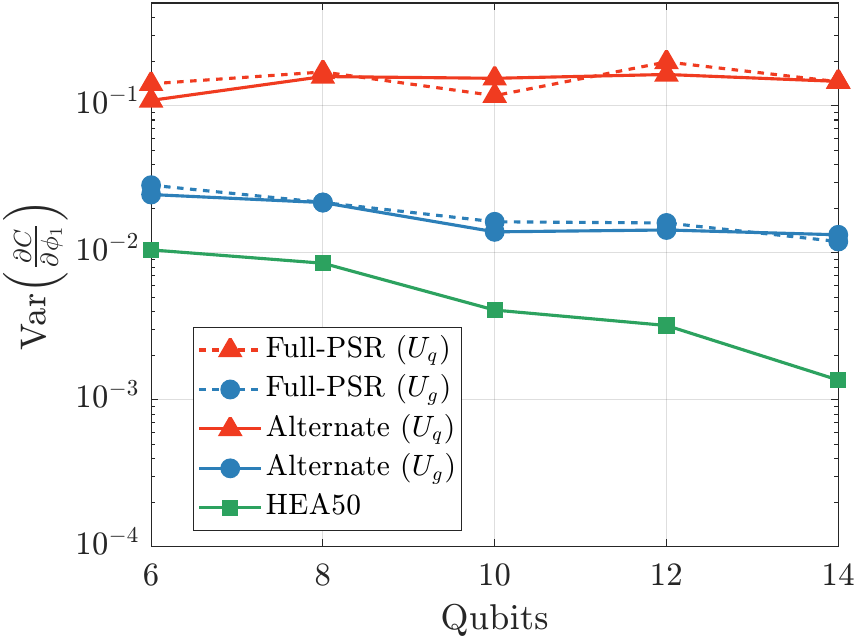}
        \caption{1 YZ linear layer in $U_q$}
    \end{subfigure}
        \begin{subfigure}{0.45\linewidth}
        \includegraphics[width = 1\linewidth]{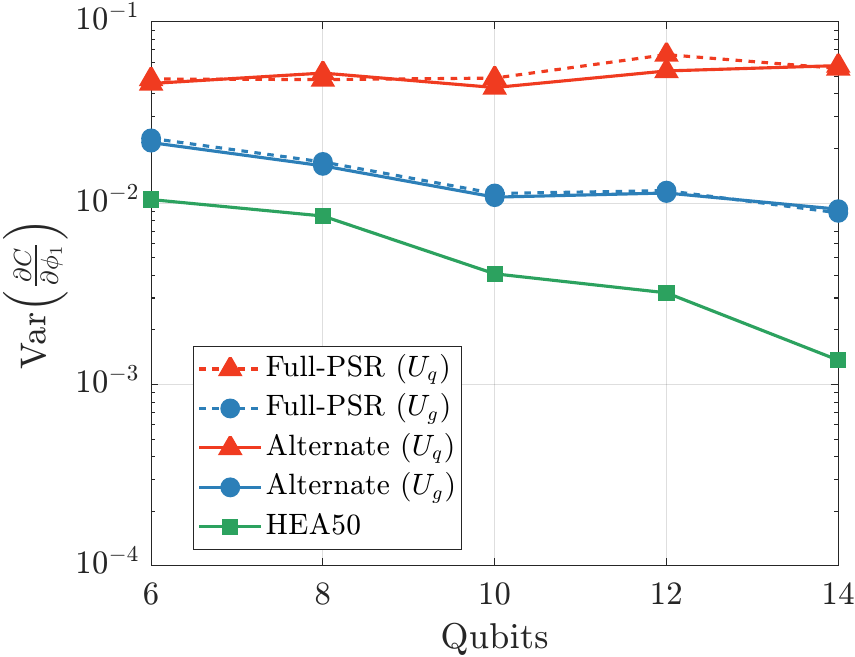}
        \caption{3 YZ linear layers in $U_q$}
    \end{subfigure}
        \begin{subfigure}{0.45\linewidth}
        \includegraphics[width = 1\linewidth]{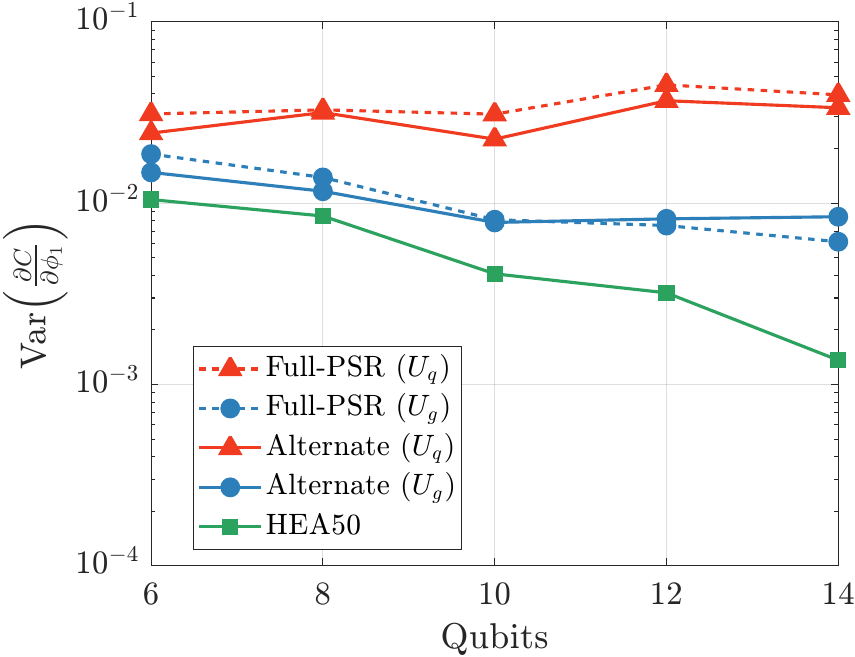}
        \caption{6 YZ linear layer in $U_q$}
    \end{subfigure}
        \begin{subfigure}{0.45\linewidth}
        \includegraphics[width = 1\linewidth]{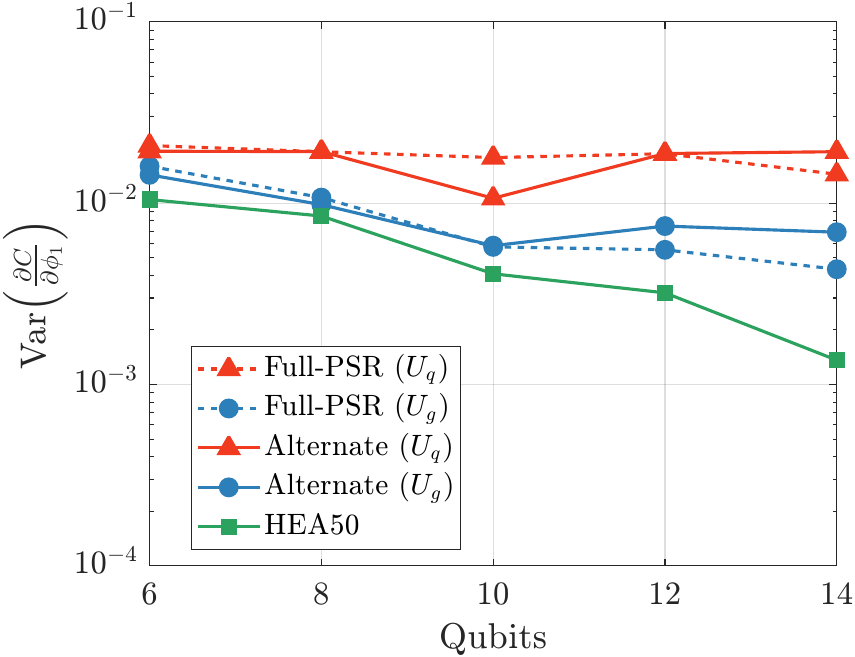}
        \caption{9 YZ linear layer in $U_q$}
    \end{subfigure}

    \caption{\raggedright \textbf{Gradients for TFIM Hamiltonian VQE:} Variance of the first parameter from $U_q$ (red) and $U_{\mathfrak{g}}$ (blue) are plotted with increasing qubits for both Full-PSR and our Alternate method for finding the ground state of TFIM Hamiltonian using VQE. The diagrams also show the plots for 1 (a), 3 (b), 6(c), 9 (d) YZ linear layers in $U_q$ . For comparison, the same task is performed using a deep HEA (referred to as HEA50) with standard PSR and variance of its first partial derivative is plotted in green. The green curve is the same in all of the plots and provided as a reference for our improved gradients. The plot shows that our chosen ansatz is able to preserve higher magnitude of gradients as compared to HEA50 over the displayed system sizes.}
    \label{grad_tfim}
\end{figure*}

\begin{resultbox}{Additional polynomial-DLA benchmark}
Across 8--14 qubits, Alt+Sim generally improves or preserves success and relative error while reducing logical QPU calls by up to $52.11\%$. The broadly positive trend across the tested configurations suggests that similar benefits may persist at larger qubit counts, although one setting increases QPU usage and further validation is required.
\end{resultbox}

\begin{table}
\begin{tabular}{cc}
    \begin{minipage}{.45\linewidth}

\begin{subtable}{\linewidth}

\begin{tabular}{|c|ccc|}
\hline
\multirow{2}{*}{\textbf{Qubits}} & \multicolumn{3}{c|}{\textbf{Success (\%)}}                                                          \\ \cline{2-4}
                                 & \multicolumn{1}{c|}{\textbf{Full-PSR}} & \multicolumn{1}{c|}{\textbf{Alternate}} & \textbf{Alt+Sim} \\ \hline
8                                & \multicolumn{1}{c|}{48.44}             & \multicolumn{1}{c|}{\textbf{50.00}}     & 45.31            \\
10                               & \multicolumn{1}{c|}{18.75}             & \multicolumn{1}{c|}{9.38}               & \textbf{26.56}   \\
12                               & \multicolumn{1}{c|}{25.00}             & \multicolumn{1}{c|}{28.12}              & \textbf{32.81}   \\
14                               & \multicolumn{1}{c|}{\textbf{100.00}}   & \multicolumn{1}{c|}{\textbf{100.00}}    & \textbf{100.00}  \\ \hline
\end{tabular}
\caption{Success rate for 1 layer YZ linear ansatz in $U_q$ and TFIM Hamiltonian DLA gates in \Ug}
\end{subtable}

\begin{subtable}{\linewidth}

\begin{tabular}{|c|ccc|}
\hline
\multirow{2}{*}{\textbf{Qubits}} & \multicolumn{3}{c|}{\textbf{Success (\%)}}                                                          \\ \cline{2-4}
                                 & \multicolumn{1}{c|}{\textbf{Full-PSR}} & \multicolumn{1}{c|}{\textbf{Alternate}} & \textbf{Alt+Sim} \\ \hline
8                                & \multicolumn{1}{c|}{84.38}             & \multicolumn{1}{c|}{82.81}              & \textbf{89.06}   \\
10                               & \multicolumn{1}{c|}{20.31}             & \multicolumn{1}{c|}{37.50}              & \textbf{59.38}   \\
12                               & \multicolumn{1}{c|}{53.12}             & \multicolumn{1}{c|}{57.81}              & \textbf{70.31}   \\
14                               & \multicolumn{1}{c|}{\textbf{98.44}}    & \multicolumn{1}{c|}{\textbf{98.44}}     & \textbf{98.44}   \\ \hline
\end{tabular}

\caption{Success rate for 3 layer YZ linear ansatz in $U_q$ and TFIM Hamiltonian DLA gates in \Ug}
\end{subtable}

\begin{subtable}{\linewidth}

\begin{tabular}{|c|ccc|}
\hline
\multirow{2}{*}{\textbf{Qubits}} & \multicolumn{3}{c|}{\textbf{Success (\%)}}                                                          \\ \cline{2-4}
                                 & \multicolumn{1}{c|}{\textbf{Full-PSR}} & \multicolumn{1}{c|}{\textbf{Alternate}} & \textbf{Alt+Sim} \\ \hline
8                                & \multicolumn{1}{c|}{93.75}             & \multicolumn{1}{c|}{95.31}              & \textbf{100}     \\
10                               & \multicolumn{1}{c|}{43.75}             & \multicolumn{1}{c|}{45.31}              & \textbf{64.06}   \\
12                               & \multicolumn{1}{c|}{75.00}             & \multicolumn{1}{c|}{\textbf{76.56}}     & \textbf{76.56}   \\
14                               & \multicolumn{1}{c|}{\textbf{100.00}}   & \multicolumn{1}{c|}{\textbf{100.00}}    & \textbf{100.00}  \\ \hline
\end{tabular}

\caption{Success rate for 6 layer YZ linear ansatz in $U_q$ and TFIM Hamiltonian DLA gates in \Ug}
\end{subtable}

\begin{subtable}{\linewidth}

\begin{tabular}{|c|ccc|}
\hline
\multirow{2}{*}{\textbf{Qubits}} & \multicolumn{3}{c|}{\textbf{Success (\%)}}                                                          \\ \cline{2-4}
                                 & \multicolumn{1}{c|}{\textbf{Full-PSR}} & \multicolumn{1}{c|}{\textbf{Alternate}} & \textbf{Alt+Sim} \\ \hline
8                                & \multicolumn{1}{c|}{57.81}             & \multicolumn{1}{c|}{60.94}              & \textbf{73.44}   \\
10                               & \multicolumn{1}{c|}{31.25}             & \multicolumn{1}{c|}{43.75}              & \textbf{70.31}   \\
12                               & \multicolumn{1}{c|}{82.81}             & \multicolumn{1}{c|}{\textbf{85.94}}     & 79.69            \\
14                               & \multicolumn{1}{c|}{98.44}             & \multicolumn{1}{c|}{\textbf{100.00}}    & \textbf{100.00}  \\ \hline
\end{tabular}

\caption{Success rate for 9 layer YZ linear ansatz in $U_q$ and TFIM Hamiltonian DLA gates in \Ug}

\end{subtable}
\end{minipage}

&
\begin{minipage}{0.45\linewidth}

\begin{subtable}{\linewidth}
\begin{tabular}{|c|cc|}
\hline
\multirow{2}{*}{\textbf{Qubits}} & \multicolumn{2}{c|}{\textbf{QPU Reduction (\%)}}           \\ \cline{2-3}
                                 & \multicolumn{1}{c|}{\textbf{Alternate}} & \textbf{Alt+Sim} \\ \hline
8                                & \multicolumn{1}{c|}{29.78 +/- 15.08}    & 48.09 +/- 22.71  \\
10                               & \multicolumn{1}{c|}{11.52 +/- 4.17}     & 52.11 +/- 26.46  \\
12                               & \multicolumn{1}{c|}{29.46 +/- 16.81}    & 51.80 +/- 25.37  \\
14                               & \multicolumn{1}{c|}{40.91 +/- 15.93}    & 50.66 +/- 17.60  \\ \hline
\end{tabular}

\caption{QPU Reduction for 1 layer YZ linear ansatz in $U_q$ and TFIM Hamiltonian DLA gates in \Ug}
\end{subtable}

\begin{subtable}{\linewidth}

\begin{tabular}{|c|cc|}
\hline
\multirow{2}{*}{\textbf{Qubits}} & \multicolumn{2}{c|}{\textbf{QPU Reduction (\%)}}           \\ \cline{2-3}
                                 & \multicolumn{1}{c|}{\textbf{Alternate}} & \textbf{Alt+Sim} \\ \hline
8                                & \multicolumn{1}{c|}{17.45 +/- 5.79}     & 30.08 +/- 13.40  \\
10                               & \multicolumn{1}{c|}{17.97 +/- 1.97}     & 38.72 +/- 16.46  \\
12                               & \multicolumn{1}{c|}{25.66 +/- 6.49}     & 35.29 +/- 15.03  \\
14                               & \multicolumn{1}{c|}{48.79 +/- 15.07}    & 25.09 +/- 8.19   \\ \hline
\end{tabular}

\caption{QPU Reduction for 3 layer YZ linear ansatz in $U_q$ and TFIM Hamiltonian DLA gates in \Ug}
\end{subtable}

\begin{subtable}{\linewidth}

\begin{tabular}{|c|cc|}
\hline
\multirow{2}{*}{\textbf{Qubits}} & \multicolumn{2}{c|}{\textbf{QPU Reduction (\%)}}           \\ \cline{2-3}
                                 & \multicolumn{1}{c|}{\textbf{Alternate}} & \textbf{Alt+Sim} \\ \hline
8                                & \multicolumn{1}{c|}{24.52 +/- 5.93}     & 0.64 +/- 0.15    \\
10                               & \multicolumn{1}{c|}{24.27 +/- 5.04}     & 21.82 +/- 9.20   \\
12                               & \multicolumn{1}{c|}{32.43 +/- 9.54}     & 25.35 +/- 8.84   \\
14                               & \multicolumn{1}{c|}{49.64 +/- 21.50}    & 9.29 +/- 2.21    \\ \hline
\end{tabular}

\caption{QPU Reduction for 6 layer YZ linear ansatz in $U_q$ and TFIM Hamiltonian DLA gates in \Ug}
\end{subtable}

\begin{subtable}{\linewidth}

\begin{tabular}{|c|cc|}
\hline
\multirow{2}{*}{\textbf{Qubits}} & \multicolumn{2}{c|}{\textbf{QPU Reduction (\%)}}           \\ \cline{2-3}
                                 & \multicolumn{1}{c|}{\textbf{Alternate}} & \textbf{Alt+Sim} \\ \hline
8                                & \multicolumn{1}{c|}{0.68 +/- 0.11}      & 1.39 +/- 0.54    \\
10                               & \multicolumn{1}{c|}{9.33 +/- 1.70}      & 6.55 +/- 2.57    \\
12                               & \multicolumn{1}{c|}{27.27 +/- 8.83}     & 14.90 +/- 5.16   \\
14                               & \multicolumn{1}{c|}{45.79 +/- 22.98}    & -1.98 +/- 0.37  \\ \hline
\end{tabular}
\caption{QPU Reduction for 9 layer YZ linear ansatz in $U_q$ and TFIM Hamiltonian DLA gates in \Ug}
\end{subtable}

\end{minipage}
\end{tabular}

\caption{\raggedright \textbf{TFIM Hamiltonian (Success and QPU Reduction)}: The success and QPU calls reduction are shown for the configurations with 1, 3, 6, and 9 layers of YZ linear ansatz in $U_q$ and TFIM Hamiltonian DLA gates in \Ug using Full-PSR, Alternate and Alternate + Simultaneous training (referred to as Alt+Sim). The QPU call reduction is measured relative to the Full-PSR method where all parameters are trained by standard PSR. The metrics show that our methods are generally better at reaching higher success rates and lower relative error (in Table \ref{vqe_tfim_error}), while QPU calls are reduced significantly.}
\label{vqe_tfim_successqpu}
\end{table}

\begin{table}[]
\begin{subtable}{\linewidth}

\begin{tabular}{|c|c|ccccccccccccccc|}
\hline
\multirow{3}{*}{\textbf{Qubits}} & \textbf{} & \multicolumn{15}{c|}{\textbf{Relative Error (units of $10^{-5}$)}}                                                                                                                                                                                                                                                                                                                                                                                                                                                                                               \\ \cline{3-17}
                                 & \textbf{} & \multicolumn{3}{c|}{\textbf{\g}}                                                                             & \multicolumn{1}{c|}{\textbf{}} & \multicolumn{3}{c|}{\textbf{Full-PSR}}                                                                        & \multicolumn{1}{c|}{\textbf{}} & \multicolumn{3}{c|}{\textbf{Alternate}}                                                                       & \multicolumn{1}{c|}{\textbf{}} & \multicolumn{3}{c|}{\textbf{Alt + Sim}}                                                 \\ \cline{3-5} \cline{7-9} \cline{11-13} \cline{15-17}
                                 & \textbf{} & \multicolumn{1}{c|}{\textbf{Median}} & \multicolumn{1}{c|}{\textbf{Q25}} & \multicolumn{1}{c|}{\textbf{Q75}} & \multicolumn{1}{c|}{\textbf{}} & \multicolumn{1}{c|}{\textbf{Median}}  & \multicolumn{1}{c|}{\textbf{Q25}} & \multicolumn{1}{c|}{\textbf{Q75}} & \multicolumn{1}{c|}{\textbf{}} & \multicolumn{1}{c|}{\textbf{Median}}  & \multicolumn{1}{c|}{\textbf{Q25}} & \multicolumn{1}{c|}{\textbf{Q75}} & \multicolumn{1}{c|}{\textbf{}} & \multicolumn{1}{c|}{\textbf{Median}} & \multicolumn{1}{c|}{\textbf{Q25}} & \textbf{Q75} \\ \hline
8                                &           & \multicolumn{1}{c|}{4034.16}         & \multicolumn{1}{c|}{4034.16}      & \multicolumn{1}{c|}{4034.17}      & \multicolumn{1}{c|}{}          & \multicolumn{1}{c|}{\textbf{0.0409}}  & \multicolumn{1}{c|}{0.0327}       & \multicolumn{1}{c|}{0.0409}       & \multicolumn{1}{c|}{}          & \multicolumn{1}{c|}{\textbf{0.0409}}  & \multicolumn{1}{c|}{0.0327}       & \multicolumn{1}{c|}{0.0409}       & \multicolumn{1}{c|}{}          & \multicolumn{1}{c|}{0.0491}          & \multicolumn{1}{c|}{0.0409}       & 0.0491       \\
10                               &           & \multicolumn{1}{c|}{287.65 }          & \multicolumn{1}{c|}{284.92}       & \multicolumn{1}{c|}{306.97}       & \multicolumn{1}{c|}{}          & \multicolumn{1}{c|}{0.2309}           & \multicolumn{1}{c|}{0.0181}       & \multicolumn{1}{c|}{38.2376}      & \multicolumn{1}{c|}{}          & \multicolumn{1}{c|}{0.2717}           & \multicolumn{1}{c|}{0.0430}       & \multicolumn{1}{c|}{14.3759}      & \multicolumn{1}{c|}{}          & \multicolumn{1}{c|}{\textbf{0.0272}} & \multicolumn{1}{c|}{0.0272}       & 0.0362       \\
12                               &           & \multicolumn{1}{c|}{\textbf{0.04}}   & \multicolumn{1}{c|}{0.04}         & \multicolumn{1}{c|}{0.04}         & \multicolumn{1}{c|}{}          & \multicolumn{1}{c|}{0.0424}           & \multicolumn{1}{c|}{0.0424}       & \multicolumn{1}{c|}{0.0795}       & \multicolumn{1}{c|}{}          & \multicolumn{1}{c|}{0.0424}           & \multicolumn{1}{c|}{0.0344}       & \multicolumn{1}{c|}{0.0609}       & \multicolumn{1}{c|}{}          & \multicolumn{1}{c|}{0.0424}          & \multicolumn{1}{c|}{0.0424}       & 0.0424       \\
14                               &           & \multicolumn{1}{c|}{24.8}            & \multicolumn{1}{c|}{24.8}         & \multicolumn{1}{c|}{24.8}         & \multicolumn{1}{c|}{}          & \multicolumn{1}{c|}{\textbf{12.3975}} & \multicolumn{1}{c|}{12.3975}      & \multicolumn{1}{c|}{12.4076}      & \multicolumn{1}{c|}{}          & \multicolumn{1}{c|}{\textbf{12.3975}} & \multicolumn{1}{c|}{12.3975}      & \multicolumn{1}{c|}{12.4304}      & \multicolumn{1}{c|}{}          & \multicolumn{1}{c|}{12.4178}         & \multicolumn{1}{c|}{12.4178}      & 12.4178      \\ \hline
\end{tabular}

\caption{Relative Error for 1 layer YZ linear ansatz in $U_q$ and TFIM Hamiltonian DLA gates in \Ug}
\end{subtable}

\begin{subtable}{\linewidth}
\begin{tabular}{|c|c|ccccccccccccccc|}
\hline
\multirow{3}{*}{\textbf{Qubits}} & \textbf{} & \multicolumn{15}{c|}{\textbf{Relative Error (units of $10^{-5}$)}}                                                                                                                                                                                                                                                                                                                                                                                                                                                                                              \\ \cline{3-17}
                                 & \textbf{} & \multicolumn{3}{c|}{\textbf{\g}}                                                                             & \multicolumn{1}{c|}{\textbf{}} & \multicolumn{3}{c|}{\textbf{Full-PSR}}                                                                        & \multicolumn{1}{c|}{\textbf{}} & \multicolumn{3}{c|}{\textbf{Alternate}}                                                                      & \multicolumn{1}{c|}{\textbf{}} & \multicolumn{3}{c|}{\textbf{Alt + Sim}}                                                 \\ \cline{3-5} \cline{7-9} \cline{11-13} \cline{15-17}
                                 & \textbf{} & \multicolumn{1}{c|}{\textbf{Median}} & \multicolumn{1}{c|}{\textbf{Q25}} & \multicolumn{1}{c|}{\textbf{Q75}} & \multicolumn{1}{c|}{\textbf{}} & \multicolumn{1}{c|}{\textbf{Median}}  & \multicolumn{1}{c|}{\textbf{Q25}} & \multicolumn{1}{c|}{\textbf{Q75}} & \multicolumn{1}{c|}{\textbf{}} & \multicolumn{1}{c|}{\textbf{Median}} & \multicolumn{1}{c|}{\textbf{Q25}} & \multicolumn{1}{c|}{\textbf{Q75}} & \multicolumn{1}{c|}{\textbf{}} & \multicolumn{1}{c|}{\textbf{Median}} & \multicolumn{1}{c|}{\textbf{Q25}} & \textbf{Q75} \\ \hline
8                                &           & \multicolumn{1}{c|}{4034.16}         & \multicolumn{1}{c|}{4034.16}      & \multicolumn{1}{c|}{4034.17}      & \multicolumn{1}{c|}{}          & \multicolumn{1}{c|}{0.0450}           & \multicolumn{1}{c|}{0.0266}       & \multicolumn{1}{c|}{0.2270}       & \multicolumn{1}{c|}{}          & \multicolumn{1}{c|}{0.0573}          & \multicolumn{1}{c|}{0.0327}       & \multicolumn{1}{c|}{0.1636}       & \multicolumn{1}{c|}{}          & \multicolumn{1}{c|}{\textbf{0.0491}} & \multicolumn{1}{c|}{0.0491}       & 0.0491       \\
10                               &           & \multicolumn{1}{c|}{287.65 }          & \multicolumn{1}{c|}{284.92}       & \multicolumn{1}{c|}{306.97}       & \multicolumn{1}{c|}{}          & \multicolumn{1}{c|}{3.3506}           & \multicolumn{1}{c|}{2.1643}       & \multicolumn{1}{c|}{17.4322}      & \multicolumn{1}{c|}{}          & \multicolumn{1}{c|}{4.9852}          & \multicolumn{1}{c|}{0.5818}       & \multicolumn{1}{c|}{32.6050}      & \multicolumn{1}{c|}{}          & \multicolumn{1}{c|}{\textbf{0.0362}} & \multicolumn{1}{c|}{0.0294}       & 0.0453       \\
12                               &           & \multicolumn{1}{c|}{\textbf{0.04}}   & \multicolumn{1}{c|}{0.04}         & \multicolumn{1}{c|}{0.04}         & \multicolumn{1}{c|}{}          & \multicolumn{1}{c|}{0.0636}           & \multicolumn{1}{c|}{0.0318}       & \multicolumn{1}{c|}{0.2358}       & \multicolumn{1}{c|}{}          & \multicolumn{1}{c|}{0.5723}          & \multicolumn{1}{c|}{0.0530}       & \multicolumn{1}{c|}{1.7380}       & \multicolumn{1}{c|}{}          & \multicolumn{1}{c|}{0.0530}          & \multicolumn{1}{c|}{0.0424}       & 0.0636       \\
14                               &           & \multicolumn{1}{c|}{24.8}            & \multicolumn{1}{c|}{24.8}         & \multicolumn{1}{c|}{24.8}         & \multicolumn{1}{c|}{}          & \multicolumn{1}{c|}{\textbf{12.4178}} & \multicolumn{1}{c|}{12.3874}      & \multicolumn{1}{c|}{12.5901}      & \multicolumn{1}{c|}{}          & \multicolumn{1}{c|}{12.4989}         & \multicolumn{1}{c|}{12.3975}      & \multicolumn{1}{c|}{12.8790}      & \multicolumn{1}{c|}{}          & \multicolumn{1}{c|}{12.4279}         & \multicolumn{1}{c|}{12.3975}      & 12.4381      \\ \hline
\end{tabular}
\caption{Relative Error for 3 layer YZ linear ansatz in $U_q$ and TFIM Hamiltonian DLA gates in \Ug}
\end{subtable}

\begin{subtable}{\linewidth}

\begin{tabular}{|c|c|ccccccccccccccc|}
\hline
\multirow{3}{*}{\textbf{Qubits}} & \textbf{} & \multicolumn{15}{c|}{\textbf{Relative Error (units of $10^{-5}$)}}                                                                                                                                                                                                                                                                                                                                                                                                                                                                                              \\ \cline{3-17}
                                 & \textbf{} & \multicolumn{3}{c|}{\textbf{\g}}                                                                             & \multicolumn{1}{c|}{\textbf{}} & \multicolumn{3}{c|}{\textbf{Full-PSR}}                                                                        & \multicolumn{1}{c|}{\textbf{}} & \multicolumn{3}{c|}{\textbf{Alternate}}                                                                      & \multicolumn{1}{c|}{\textbf{}} & \multicolumn{3}{c|}{\textbf{Alt + Sim}}                                                 \\ \cline{3-5} \cline{7-9} \cline{11-13} \cline{15-17}
                                 & \textbf{} & \multicolumn{1}{c|}{\textbf{Median}} & \multicolumn{1}{c|}{\textbf{Q25}} & \multicolumn{1}{c|}{\textbf{Q75}} & \multicolumn{1}{c|}{\textbf{}} & \multicolumn{1}{c|}{\textbf{Median}}  & \multicolumn{1}{c|}{\textbf{Q25}} & \multicolumn{1}{c|}{\textbf{Q75}} & \multicolumn{1}{c|}{\textbf{}} & \multicolumn{1}{c|}{\textbf{Median}} & \multicolumn{1}{c|}{\textbf{Q25}} & \multicolumn{1}{c|}{\textbf{Q75}} & \multicolumn{1}{c|}{\textbf{}} & \multicolumn{1}{c|}{\textbf{Median}} & \multicolumn{1}{c|}{\textbf{Q25}} & \textbf{Q75} \\ \hline
8                                &           & \multicolumn{1}{c|}{4034.16}         & \multicolumn{1}{c|}{4034.16}      & \multicolumn{1}{c|}{4034.17}      & \multicolumn{1}{c|}{}          & \multicolumn{1}{c|}{0.5031}           & \multicolumn{1}{c|}{0.1432}       & \multicolumn{1}{c|}{1.2640}       & \multicolumn{1}{c|}{}          & \multicolumn{1}{c|}{1.0553}          & \multicolumn{1}{c|}{0.2372}       & \multicolumn{1}{c|}{2.1598}       & \multicolumn{1}{c|}{}          & \multicolumn{1}{c|}{\textbf{0.0736}} & \multicolumn{1}{c|}{0.0491}       & 0.0982       \\
10                               &           & \multicolumn{1}{c|}{287.65}          & \multicolumn{1}{c|}{ 284.92}       & \multicolumn{1}{c|}{306.97}       & \multicolumn{1}{c|}{}          & \multicolumn{1}{c|}{4.1430}           & \multicolumn{1}{c|}{1.0731}       & \multicolumn{1}{c|}{32.6571}      & \multicolumn{1}{c|}{}          & \multicolumn{1}{c|}{3.1785}          & \multicolumn{1}{c|}{0.9961}       & \multicolumn{1}{c|}{12.7414}      & \multicolumn{1}{c|}{}          & \multicolumn{1}{c|}{\textbf{0.0634}} & \multicolumn{1}{c|}{0.0453}       & 0.0996       \\
12                               &           & \multicolumn{1}{c|}{\textbf{0.04}}   & \multicolumn{1}{c|}{0.04}         & \multicolumn{1}{c|}{0.04}         & \multicolumn{1}{c|}{}          & \multicolumn{1}{c|}{0.2278}           & \multicolumn{1}{c|}{0.0715}       & \multicolumn{1}{c|}{0.4292}       & \multicolumn{1}{c|}{}          & \multicolumn{1}{c|}{1.2929}          & \multicolumn{1}{c|}{0.2331}       & \multicolumn{1}{c|}{3.5501}       & \multicolumn{1}{c|}{}          & \multicolumn{1}{c|}{0.0636}          & \multicolumn{1}{c|}{0.0530}       & 0.0742       \\
14                               &           & \multicolumn{1}{c|}{24.8}            & \multicolumn{1}{c|}{24.8}         & \multicolumn{1}{c|}{24.8}         & \multicolumn{1}{c|}{}          & \multicolumn{1}{c|}{\textbf{12.4279}} & \multicolumn{1}{c|}{11.3585}      & \multicolumn{1}{c|}{12.5420}      & \multicolumn{1}{c|}{}          & \multicolumn{1}{c|}{13.3352}         & \multicolumn{1}{c|}{11.9059}      & \multicolumn{1}{c|}{15.0483}      & \multicolumn{1}{c|}{}          & \multicolumn{1}{c|}{12.4381}         & \multicolumn{1}{c|}{11.2799}      & 12.4786      \\ \hline
\end{tabular}

\caption{Relative Error for 6 layer YZ linear ansatz in $U_q$ and TFIM Hamiltonian DLA gates in \Ug}
\end{subtable}

\begin{subtable}{\linewidth}

\begin{tabular}{|c|c|ccccccccccccccc|}
\hline
\multirow{3}{*}{\textbf{Qubits}} & \textbf{} & \multicolumn{15}{c|}{\textbf{Relative Error (units of $10^{-5}$)}}                                                                                                                                                                                                                                                                                                                                                                                                                                                                                              \\ \cline{3-17}
                                 & \textbf{} & \multicolumn{3}{c|}{\textbf{\g}}                                                                             & \multicolumn{1}{c|}{\textbf{}} & \multicolumn{3}{c|}{\textbf{Full-PSR}}                                                                       & \multicolumn{1}{c|}{\textbf{}} & \multicolumn{3}{c|}{\textbf{Alternate}}                                                                      & \multicolumn{1}{c|}{\textbf{}} & \multicolumn{3}{c|}{\textbf{Alt + Sim}}                                                  \\ \cline{3-5} \cline{7-9} \cline{11-13} \cline{15-17}
                                 & \textbf{} & \multicolumn{1}{c|}{\textbf{Median}} & \multicolumn{1}{c|}{\textbf{Q25}} & \multicolumn{1}{c|}{\textbf{Q75}} & \multicolumn{1}{c|}{\textbf{}} & \multicolumn{1}{c|}{\textbf{Median}} & \multicolumn{1}{c|}{\textbf{Q25}} & \multicolumn{1}{c|}{\textbf{Q75}} & \multicolumn{1}{c|}{\textbf{}} & \multicolumn{1}{c|}{\textbf{Median}} & \multicolumn{1}{c|}{\textbf{Q25}} & \multicolumn{1}{c|}{\textbf{Q75}} & \multicolumn{1}{c|}{\textbf{}} & \multicolumn{1}{c|}{\textbf{Median}}  & \multicolumn{1}{c|}{\textbf{Q25}} & \textbf{Q75} \\ \hline
8                                &           & \multicolumn{1}{c|}{4034.16}         & \multicolumn{1}{c|}{4034.16}      & \multicolumn{1}{c|}{4034.17}      & \multicolumn{1}{c|}{}          & \multicolumn{1}{c|}{9.7762}          & \multicolumn{1}{c|}{0.6708}       & \multicolumn{1}{c|}{45.0443}      & \multicolumn{1}{c|}{}          & \multicolumn{1}{c|}{9.9071}          & \multicolumn{1}{c|}{3.4442}       & \multicolumn{1}{c|}{40.1194}      & \multicolumn{1}{c|}{}          & \multicolumn{1}{c|}{\textbf{0.1473}}  & \multicolumn{1}{c|}{0.0736}       & 4.1968       \\
10                               &           & \multicolumn{1}{c|}{287.65}          & \multicolumn{1}{c|}{284.92}       & \multicolumn{1}{c|}{306.97}       & \multicolumn{1}{c|}{}          & \multicolumn{1}{c|}{7.0227}          & \multicolumn{1}{c|}{1.3991}       & \multicolumn{1}{c|}{22.5464}      & \multicolumn{1}{c|}{}          & \multicolumn{1}{c|}{21.5390}         & \multicolumn{1}{c|}{10.7876}      & \multicolumn{1}{c|}{51.6152}      & \multicolumn{1}{c|}{}          & \multicolumn{1}{c|}{\textbf{0.3441}}  & \multicolumn{1}{c|}{0.0815}       & 38.5591      \\
12                               &           & \multicolumn{1}{c|}{\textbf{0.04}}   & \multicolumn{1}{c|}{0.04}         & \multicolumn{1}{c|}{0.04}         & \multicolumn{1}{c|}{}          & \multicolumn{1}{c|}{0.1484}          & \multicolumn{1}{c|}{0.0636}       & \multicolumn{1}{c|}{0.4875}       & \multicolumn{1}{c|}{}          & \multicolumn{1}{c|}{2.8719}          & \multicolumn{1}{c|}{0.9644}       & \multicolumn{1}{c|}{7.3757}       & \multicolumn{1}{c|}{}          & \multicolumn{1}{c|}{0.0848}           & \multicolumn{1}{c|}{0.0636}       & 0.1325       \\
14                               &           & \multicolumn{1}{c|}{24.8}            & \multicolumn{1}{c|}{24.8}         & \multicolumn{1}{c|}{24.8}         & \multicolumn{1}{c|}{}          & \multicolumn{1}{c|}{12.4685}         & \multicolumn{1}{c|}{7.7801}       & \multicolumn{1}{c|}{12.6256}      & \multicolumn{1}{c|}{}          & \multicolumn{1}{c|}{17.4254}         & \multicolumn{1}{c|}{13.2718}      & \multicolumn{1}{c|}{23.5684}      & \multicolumn{1}{c|}{}          & \multicolumn{1}{c|}{\textbf{12.4381}} & \multicolumn{1}{c|}{9.5845}       & 12.4887      \\ \hline
\end{tabular}

\caption{Relative Error for 9 layer YZ linear ansatz in $U_q$ and TFIM Hamiltonian DLA gates in \Ug}
\end{subtable}

\caption{\raggedright \textbf{TFIM Hamiltonian (Relative Error)}: The relative error median, 25\% quantile (Q25) and 75\% quantile (Q75) for Full-PSR, Alternate and Alternate + Simultaneous (referred to as Alt+Sim) for the configurations with 1, 3, 6, and 9 layers of YZ linear ansatz in $U_q$ and TFIM Hamiltonian DLA gates in \Ug.  The \g results are included as a separate restricted-ansatz comparison. Their accuracy depends strongly on the instance, including an accurate 12-qubit example. The relative performance of Alternate, Alt+Sim, and Full-PSR varies across the displayed qubit-depth configurations.}
\label{vqe_tfim_error}
\end{table}

We also provide plots of the gradient variance for the first parameter from each of $U_q$ and \Ug in Fig. \ref{grad_tfim}. Consistent with our previous example, the gradients decay more slowly with our choice of ansatz than with a deep, 50-layer hardware-efficient ansatz.

\SetKwComment{Comment}{/* }{ */}
\begin{algorithm}[h]
\DontPrintSemicolon
\SetKwFunction{poly}{\textbf{\textit{poly}-DLA Phase}}
\SetKwFunction{exp}{\textbf{\textit{Exponential}-DLA Phase}}
\caption{VQE for General Hamiltonian}\label{gen_hamil_alg}
\nonl\textbf{Input}: $T_{alt}, T_{sim}, T_{psr}, T_{full}, U_q(\mathbf{\theta_0}), H = \sum_i\beta_i P_i $\;
\nonl\textbf{Output}: $\theta^*,\phi^*$\;
\nonl\textbf{Initialize: } $\vec{\theta}_0 \sim \mathcal{N}(0,1); i=0$\;
\nonl\Begin{
    \nonl\poly\;
    Obtain a subset of operators $\{P_j\}_{j\in\mathcal{A}}$ from $H$ such that it forms a \textit{poly}-DLA $\mathfrak{g}$.\;
    Define $H_{poly} = \sum_{P_i \in \mathcal{A}} \beta_i P_i$, and unitary $U_{\mathfrak{g}}(\vec{\phi_0}) = e^{-iP_1\phi_1^0}\hdots e^{-iP_n\phi_n^0}$ \;
    Randomly initialize $ \vec{\phi}_0 \sim \mathcal{N}(0,1)$\;
    Define the cost function: $C_{poly}(\vec{\theta},\vec{\phi}) = \langle 0^{\otimes n}|U_q^\dagger(\vec{\theta})U_{\mathfrak{g}}^\dagger(\vec{\phi}) H_{poly} U_\mathfrak{g}(\vec{\phi}) U_q(\vec{\theta})|0^{\otimes n}\rangle$\;
    Run Alternate optimization given in Alg.\ref{alternate} for $T_{alt}$ iterations on cost function $C_{poly}(\vec{\theta},\vec{\phi})$\;
    Run Simultaneous optimization given in Alg.\ref{simultaneous} for $T_{sim}$ iterations on cost function $C_{poly}(\vec{\theta},\vec{\phi})$\;
    Extract the optimized parameters $(\vec{\theta}^*,\vec{\phi}^*)$\;
    \nonl\;
    \nonl\exp\;
    Define the full cost function, $C_{full}(\vec{\theta},\vec{\phi}) = \langle 0^{\otimes n}|U_q^\dagger(\vec{\theta})U_{\mathfrak{g}}^\dagger(\vec{\phi}) H U_\mathfrak{g}(\vec{\phi}) U_q(\vec{\theta})|0^{\otimes n}\rangle$ initialized at $(\vec{\theta}^*,\vec{\phi}^*)$\;
    Optimize only parameters $\vec{\theta}$ of $U_q(\vec{\theta})$ in $C_{full}(\vec{\theta},\vec{\phi})$ for $T_{psr}$ iterations using PSR\;
    Optimize full cost function $C_{full}(\vec{\theta},\vec{\phi})$ for $T_{full}$ using PSR\;
    Extract optimized parameters at the end of training $(\vec{\theta}^*,\vec{\phi}^*)$

}
\end{algorithm}

\section{VQE for LTFIM Hamiltonian (exponential-DLA) }\label{ltfim}

\begin{figure}
    \begin{subfigure}{0.45\linewidth}
              \includegraphics[width = 1\linewidth, height=179pt]{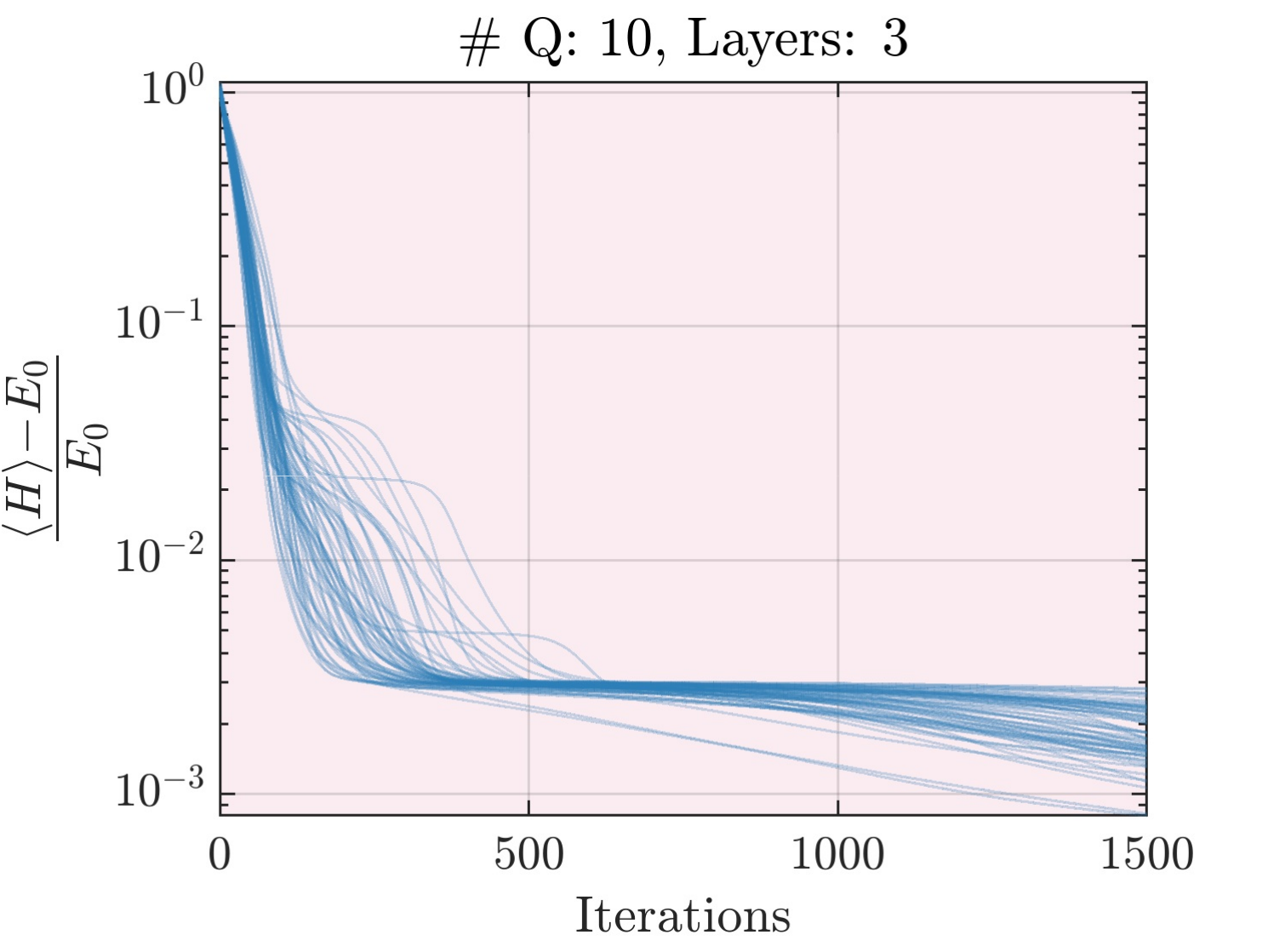}
        \caption{Trained using standard PSR}
    \end{subfigure}
        \begin{subfigure}{0.45\linewidth}
              \includegraphics[width = 1\linewidth, height=179pt]{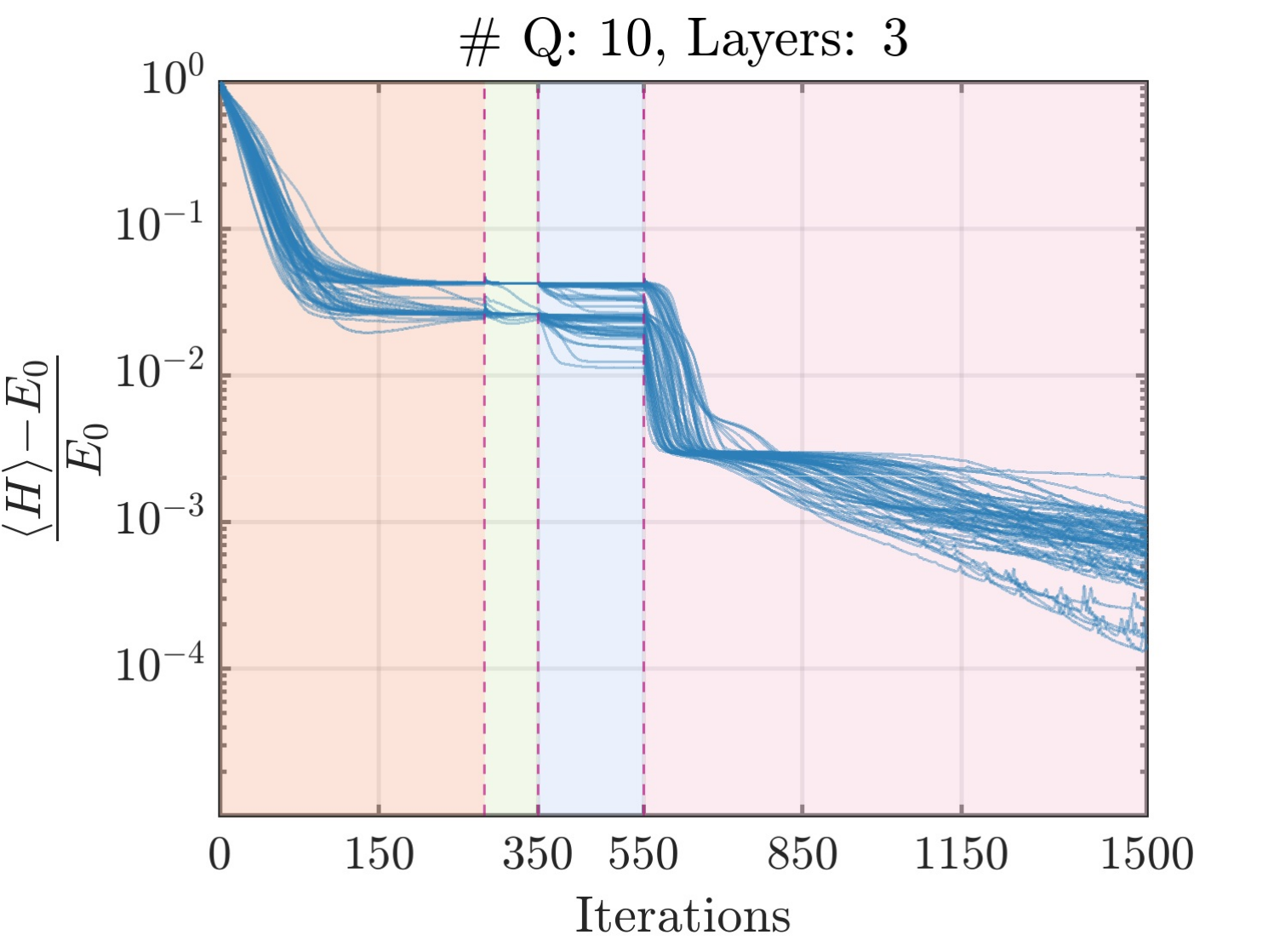}
        \caption{Pre-training using our method followed by PSR}
    \end{subfigure}

        \begin{subfigure}{0.45\linewidth}
              \includegraphics[width = 1\linewidth, height=179pt]{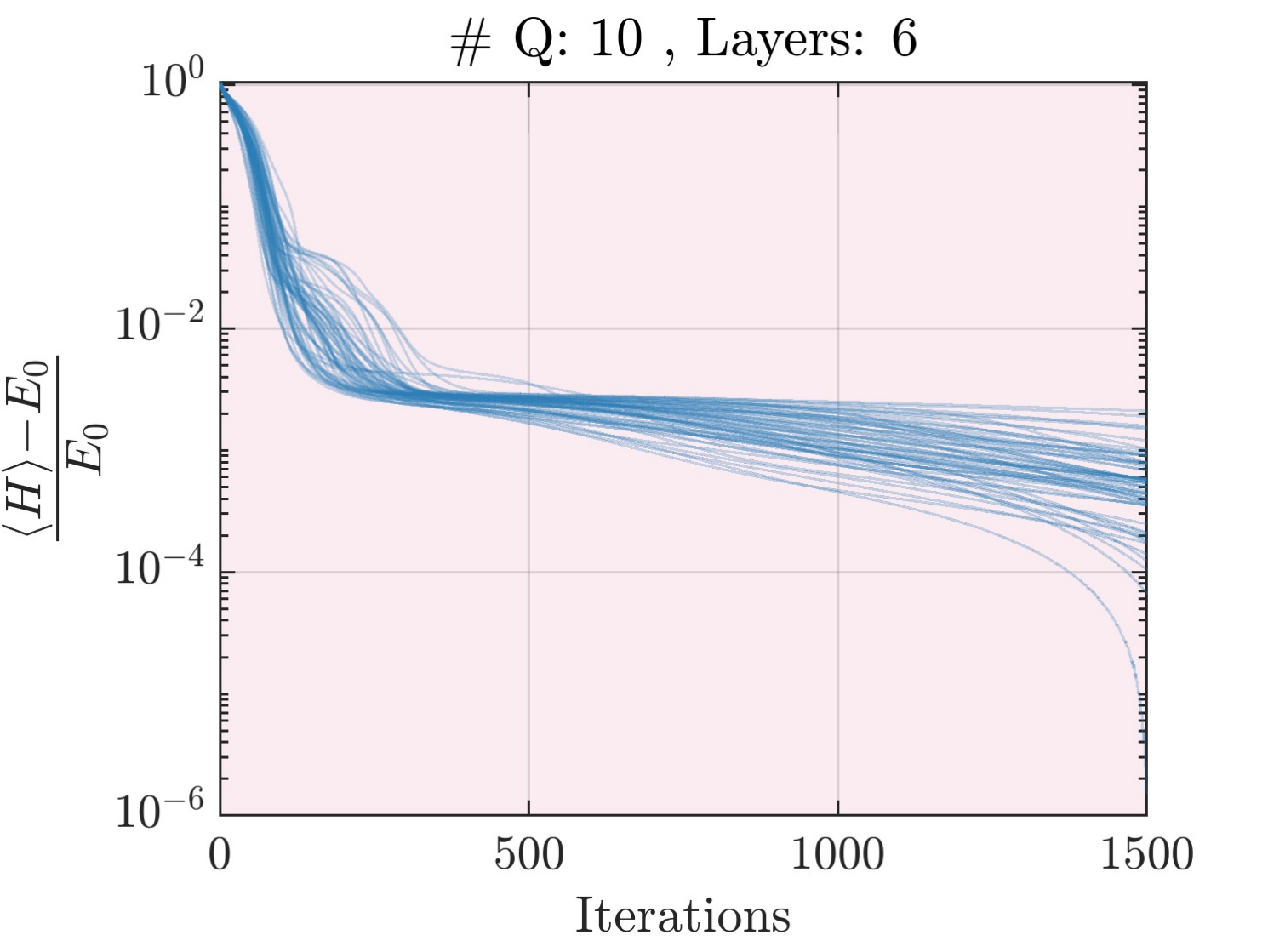}
        \caption{Trained using standard PSR}
    \end{subfigure}
        \begin{subfigure}{0.45\linewidth}
              \includegraphics[width = 1\linewidth, height=179pt]{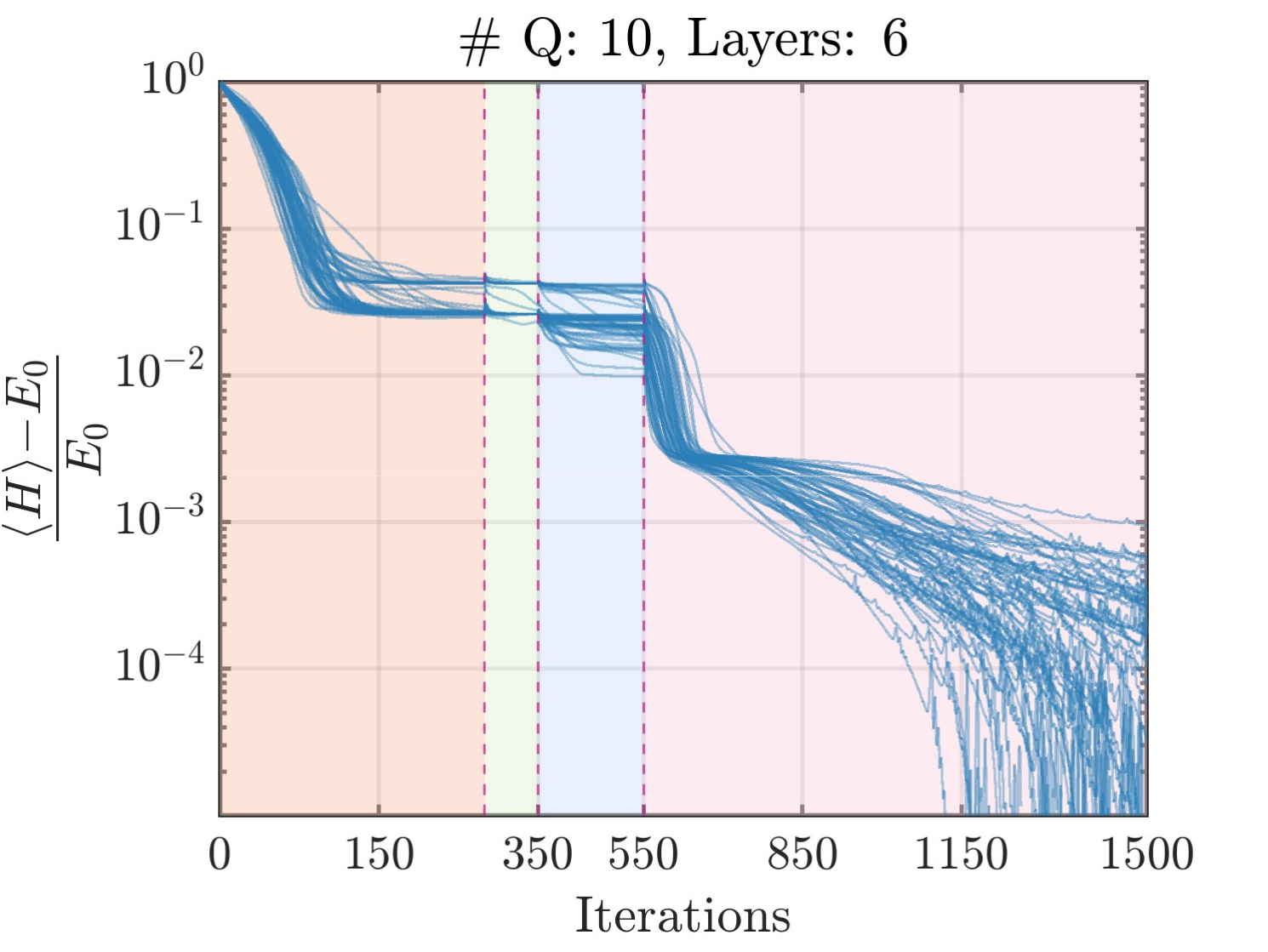}
        \caption{Pre-training using our method followed by PSR}
    \end{subfigure}

        \begin{subfigure}{0.45\linewidth}
              \includegraphics[width = 1\linewidth, height=179pt]{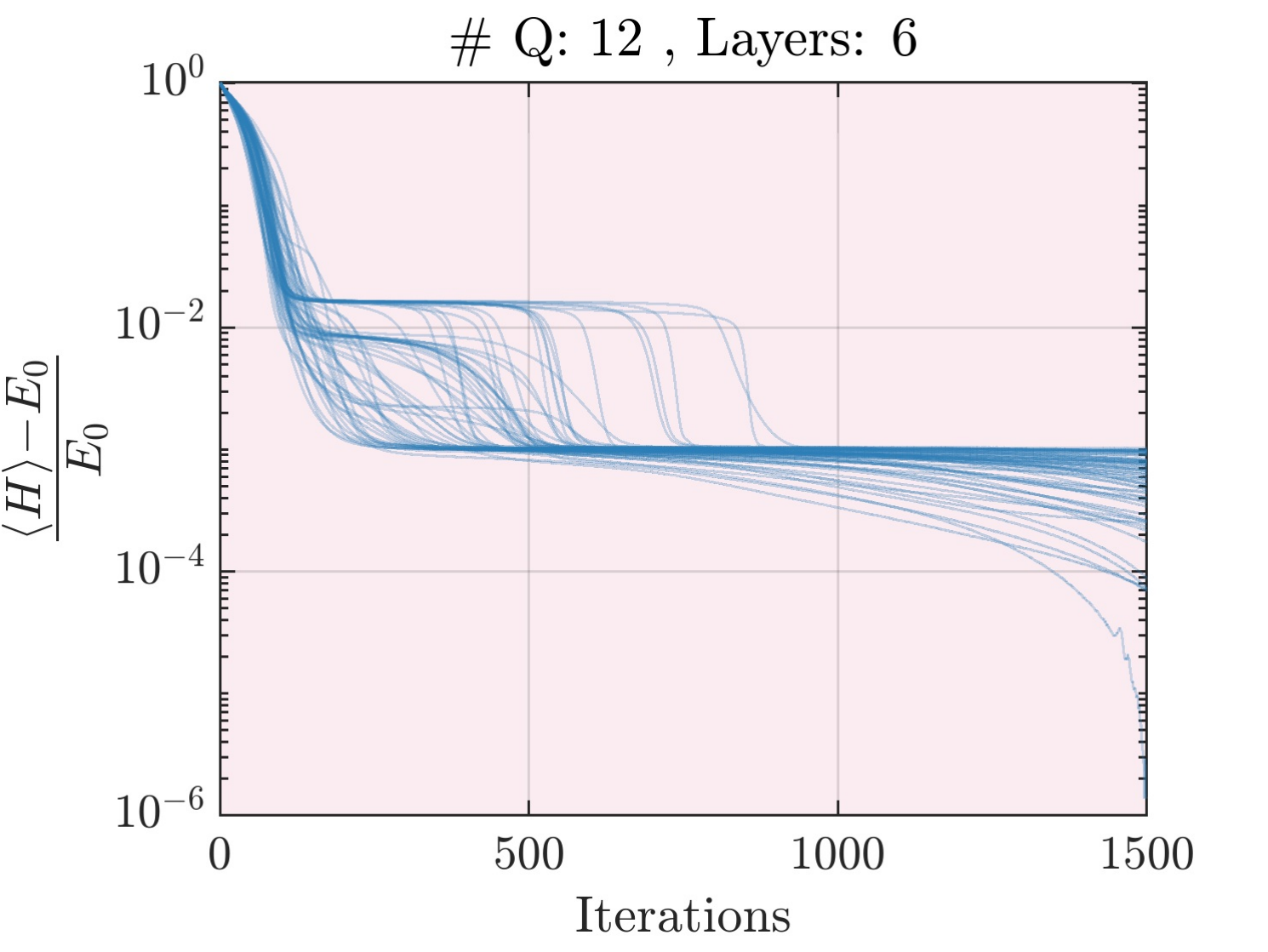}
        \caption{Trained using standard PSR}
    \end{subfigure}
        \begin{subfigure}{0.45\linewidth}
              \includegraphics[width = 1\linewidth, height=179pt]{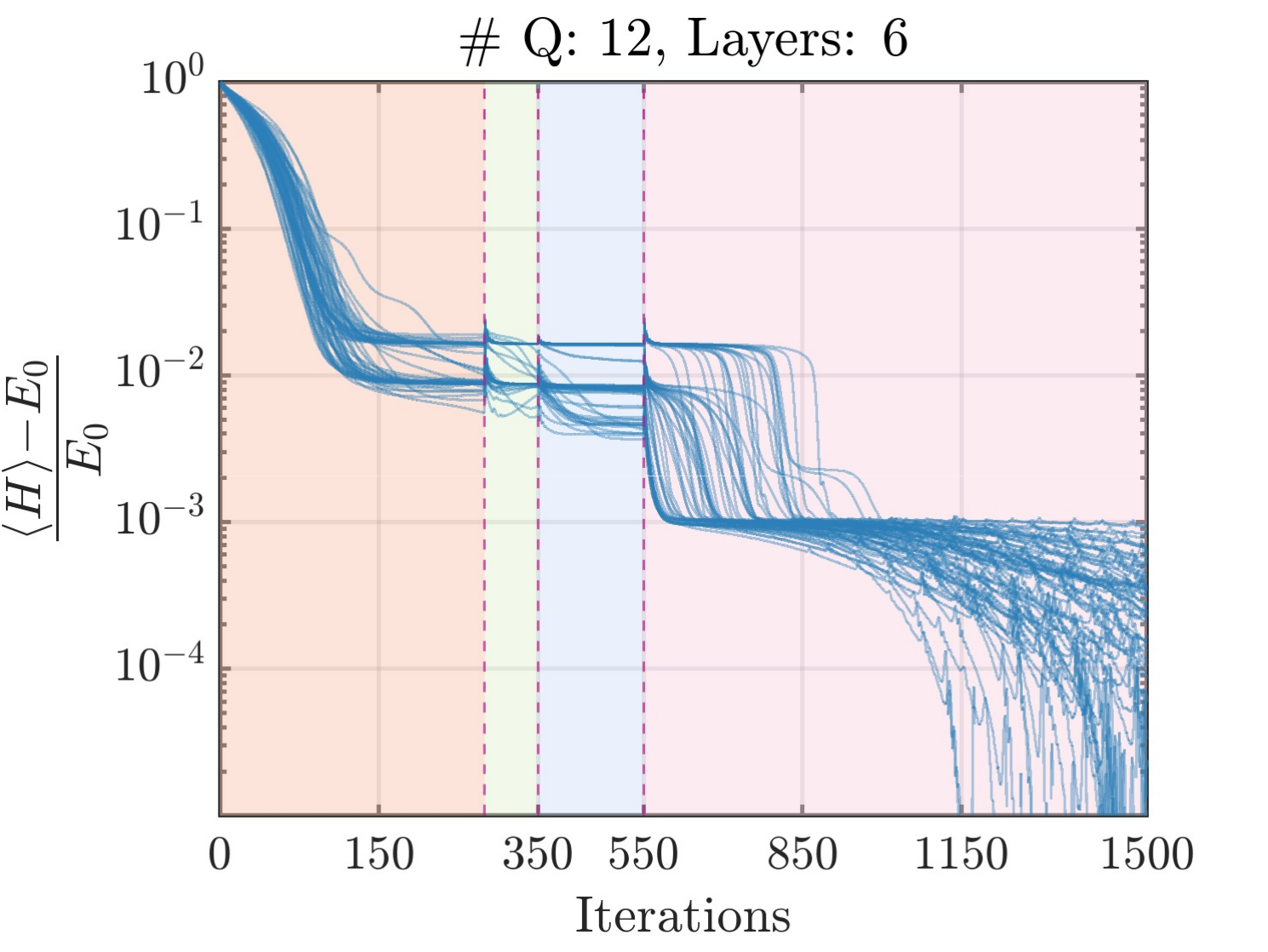}
        \caption{Pre-training using our method followed by PSR}
    \end{subfigure}
    \caption{ \raggedright Demonstration of PSR in (a), (c), and (e) and our method (b),(d),(f) to run VQE for 10- and 12-qubit LTFIM Hamiltonians (with an exponential DLA) using 3 and 6 YZ-linear layers in $U_q$. The orange and green regions show Alternate and Simultaneous training, respectively, for the restricted Hamiltonian, followed by training $U_q$ and the full circuit in the blue and pink regions, respectively, for the full LTFIM Hamiltonian. Our training methods in the initial phase allow VQE to reach much closer to the target ground state during the final training than using PSR for the full training, for similar total iteration counts. Across the displayed cases, QPU calls are reduced by approximately $10.71\%$--$15.00\%$.}
    \label{gen_hamil_img2}
\end{figure}

In this section, we provide the algorithm (Algo.\,\ref{gen_hamil_alg}) and plots for VQE on 10 and 12 qubit LTFIM Hamiltonian using PSR and our proposed method in Fig.\,\ref{gen_hamil_img2}. The relative error metric in Eq.\eqref{error} is shown for 3 and 6 YZ linear layers in $U_q$. Although the initial training phase is based on the reduced Hamiltonian, the plotted energy values correspond to expectation values of the full Hamiltonian.

\textbf{Experiment Details}: All parameters associated with the Hamiltonian terms are initialized independently according to a uniform distribution over the interval $[-1,1]$. $U_q$ is composed of 3 and 6 layers of YZ linear ansatz, while $U_{\mathfrak{g}}$ is generated from gates from $\mathfrak{g}_{TFIM}$ and the initial state is $\ket{0}^{\otimes n}$. In the shown examples, we run Alternate optimization for 250 iterations and Simultaneous optimization for 100 iterations with the reduced Hamiltonian, followed by 200 iterations of $U_q$ and 950 iterations of full circuit training. Similarly for Full-PSR training, we run 1500 iterations of PSR to match the number of total iterations in both methods. However, these choices are made ad-hoc and can be modified based on other stopping conditions, such as early stopping to further reduce QPU calls.

As shown in Fig.\ref{gen_hamil_img2}, the algorithm is indeed able to converge to good solutions by using our protocols in the first phase of the training. Compared to PSR, our protocol shows a lower average relative error. The QPU calls required for the 10-qubit Hamiltonian VQE using PSR is $4.5e+5$ and $6.3e+5$ for Layer 3 and 6 respectively, while our proposed method requires $3.825e+5$ ($\sim 15\%$ reduction) and $5.625e+5$ ($\sim 10.71\%$ reduction) QPU calls respectively. Similarly, for 12 qubits, PSR uses $8.28e+5$ for Layer 6 while our method requires $7.29e+5$ ($\sim 11.96\%$ reduction) respectively.

\begingroup
\section{LTFIM warm-start study with varying longitudinal coupling}\label{ltfim_finite_size}

We extend the trajectory-level LTFIM example above to varying coupling strengths. Method 1, denoted Full-PSR, uses random initialization followed by 1,550 Full-PSR iterations on the complete Hamiltonian $H_1+H_2$. Method 2, denoted pre-training or warm-start, uses 250 Alternate iterations and 100 Simultaneous iterations on the reduced Hamiltonian $H_1$, followed by 200 $U_q$-only PSR iterations and 1,000 full-circuit PSR iterations on $H_1+H_2$. One Alternate iteration comprises one $U_q$ update followed by one $U_{\mathfrak{g}}$ update, as defined in Algorithm 1. Both protocols therefore use 1,550 optimizer iterations under this iteration-counting convention. The reduced Hamiltonian $H_1$ contains the nearest-neighbor $XX$ and local $Z$ terms, while the longitudinal terms are restored in the final stage. We vary the longitudinal coupling by multiplying the longitudinal terms by $\lambda$. The resulting Hamiltonian is
\begin{equation}
H(\lambda) = \sum_{j=0}^{n-2} \alpha_j X_jX_{j+1} + \sum_{j=0}^{n-1} \beta_j Z_j + \lambda\sum_{j=0}^{n-1} \gamma_j X_j.
\end{equation}
Its impact on the results at various coupling strengths is studied.

We demonstrate results for qubits $n=6$-$12$, constant depth $L=3$, logarithmic depth $L=\lceil\log_2 n\rceil$, and linear depth $L=n$. We use $\lambda\in\{0.5,1,2\}$ and 64 random trials per setting. Success means that the minimum nonnegative relative error is below $10^{-3}$. Successful-trial errors are summarized by their median and Q25-Q75 interval. QPU-call reduction compares cumulative QPU call counts at each method's best iteration, includes the pretraining calls, and is conditioned on paired successful trials.

The results for $\lambda =0.5, 1,2$ are shown in Fig. \ref{ltfim_finite_size_performance_lambda_05}, \ref{ltfim_finite_size_performance_lambda_1}, \ref{ltfim_finite_size_performance_lambda_2} respectively. Across the tested qubits, depth families and coupling strengths, pre-training has a comparable success rate to Full-PSR. The successful relative error via pre-training is comparable or lower than Full-PSR in all the settings. In the paired successful-trial comparisons, pre-training reduces the required QPU calls compared to Full-PSR, with the most prominent reduction being for the constant-depth family, reducing calls by up to $46\%$ (at 9 qubits and $\lambda=0.5$).

\begin{figure*}[tbh!]
    \centering
    \includegraphics[width=0.98\linewidth]{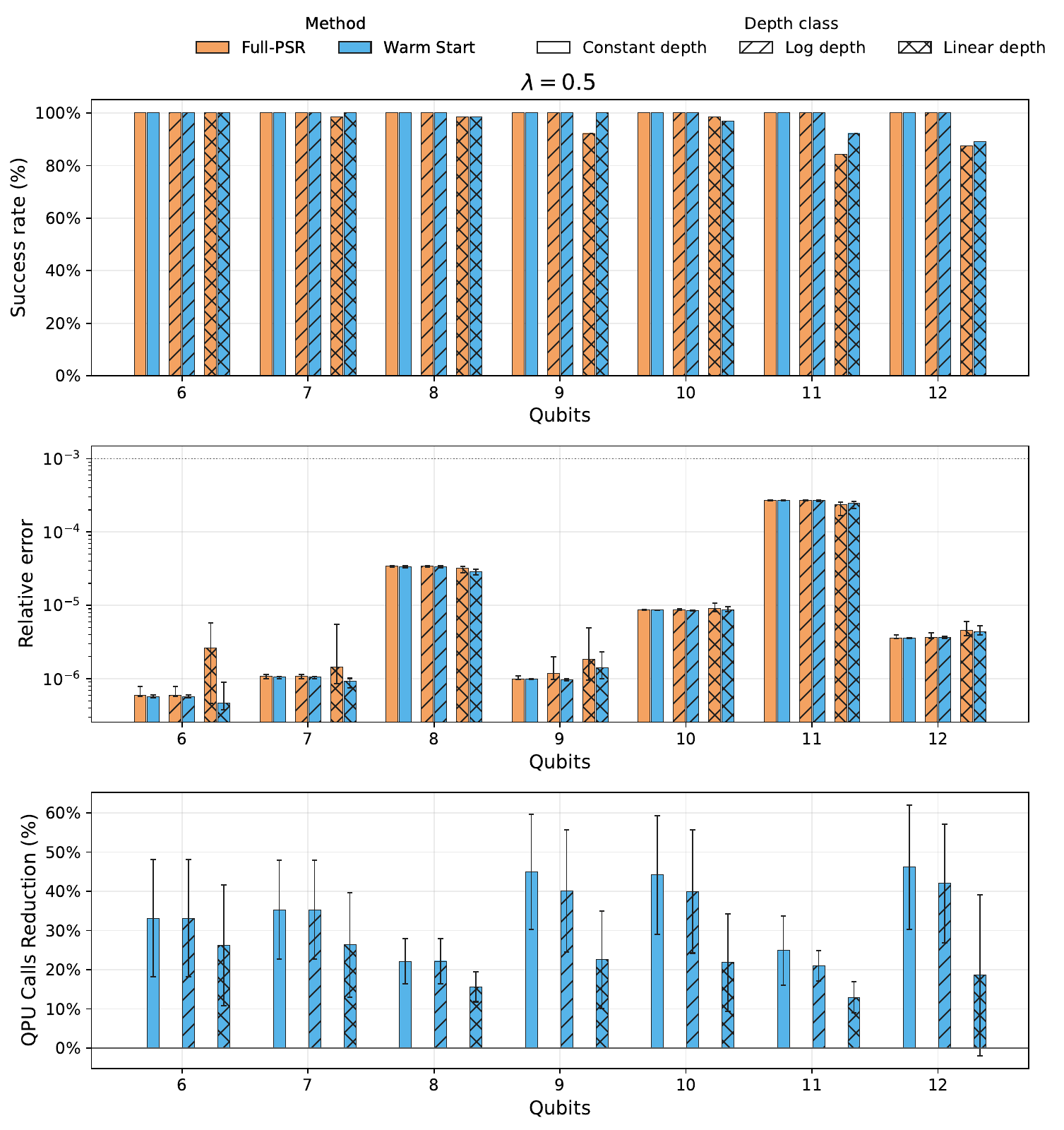}
    \caption{LTFIM warm-start performance at $\lambda=0.5$ using the protocol and metrics defined in Fig.~\ref{ltfim_finite_size_performance_lambda_1}.}
    \label{ltfim_finite_size_performance_lambda_05}
\end{figure*}

\begin{figure*}[tbh!]
    \centering
    \includegraphics[width=0.98\linewidth]{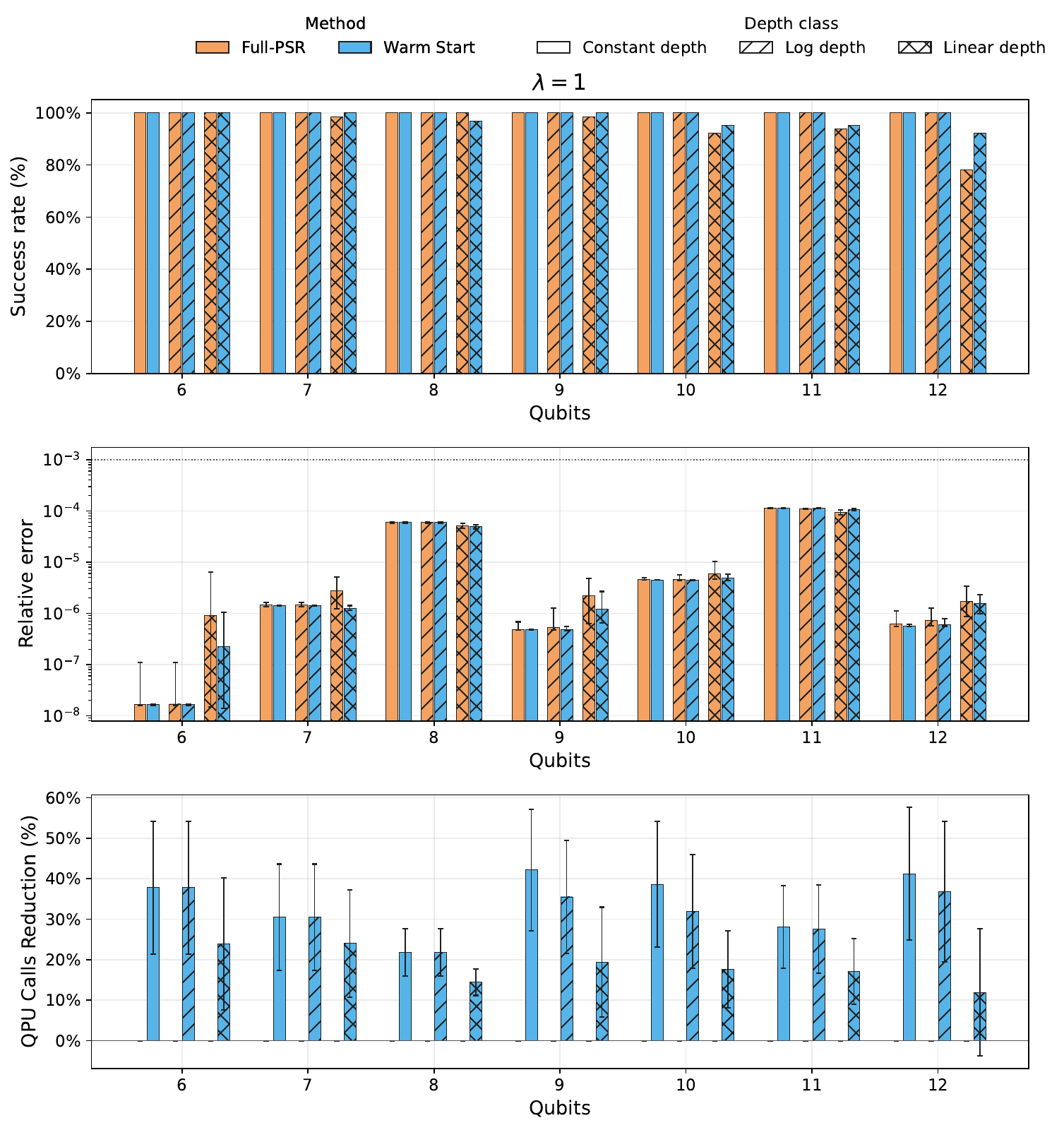}
    \caption{LTFIM warm-start performance at $\lambda=1$ using 64 paired seeds for $n=6$-$12$. The rows report the success fraction at relative error $10^{-3}$, the median best relative error among successful trials with Q25-Q75 intervals, and the mean paired QPU Calls reduction at the best iterate with one-standard-deviation error bars. The QPU-call estimate includes HELIA pretraining, and both methods use the same final circuit.}
    \label{ltfim_finite_size_performance_lambda_1}
\end{figure*}

\begin{figure*}[tbh!]
    \centering
    \includegraphics[width=0.98\linewidth]{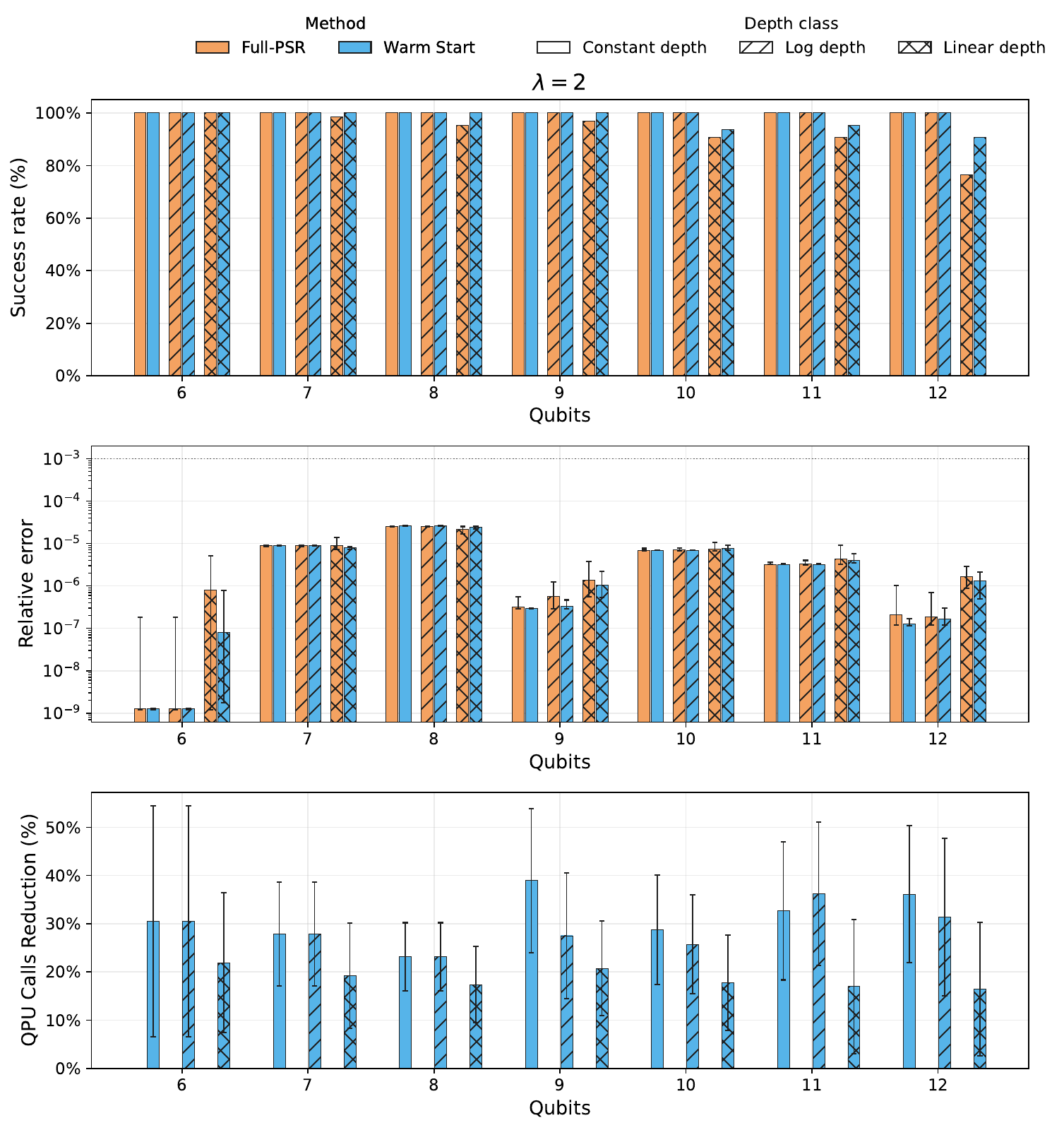}
    \caption{LTFIM warm-start performance at $\lambda=2$ using the protocol and metrics defined in Fig.~\ref{ltfim_finite_size_performance_lambda_1}.}
    \label{ltfim_finite_size_performance_lambda_2}
\end{figure*}

The gradient response is block-specific. Across the qubit counts displayed in Fig.~\ref{ltfim_finite_size_gradients}, depth families, and coupling strengths, the most pronounced changes are observed in the $U_\mathfrak{g}$ block associated with the pre-training protocol, while the response of the $U_q$ block is more heterogeneous. We cannot establish improved gradient performance with the pre-training phase: $U_\mathfrak{g}$ gradient variance is lower but does not show exponential decay, while $U_q$ follows the scaling of random initialization. Since our results on slower gradient decay with HELIA hold only when the measurement operator resides in \textit{poly}-DLA, we cannot make similar claims here when considering the full Hamiltonian, as is also reflected by the numerical experiments. However, taken together with the comparable success rate and reduced QPU-call requirements reported above, the results indicate that the warm-start initialization guides the subsequent optimization toward more favorable regions of the parameter landscape.

\begin{figure*}[tbh!]
    \centering
    \includegraphics[width=0.98\linewidth]{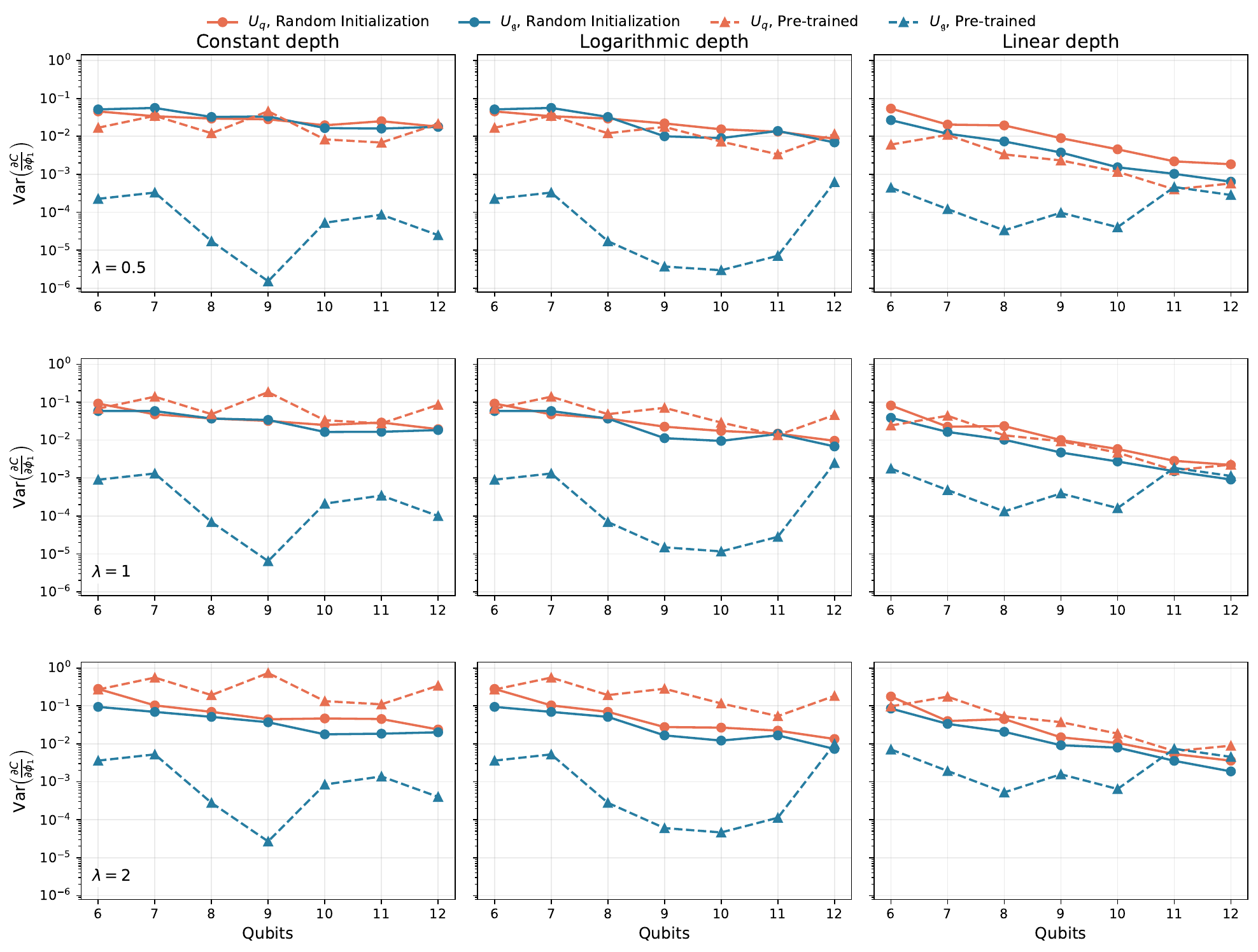}
    \caption{Population variance of the first gradient component for the LTFIM study. Rows correspond to $\lambda=0.5$, $1$, and $2$, and columns show constant, logarithmic, and linear depth families. Random-initialization and warm-start variances are shown separately for $U_q$ and $U_{\mathfrak{g}}$ over the 64 trials in each setting.}
    \label{ltfim_finite_size_gradients}
\end{figure*}
\endgroup

\begingroup

\section{Collective-spin orbit-algebra benchmark}\label{collective_spin}

We test the HELIA training procedure in a permutation-invariant operator space using the Pauli-orbit representation developed in Ref.~\cite{barligea2026enabling}. This representation provides a symmetry-adapted description of collective operators without explicitly storing their full Pauli-string expansions.

\paragraph{Operator representation.} Let $U_\pi$ denote the unitary that permutes the $n$ qubits according to $\pi\in S_n$. We define the Hermitian orbit average
\begin{equation} A_{p,q,r} = \frac{1}{n!} \sum_{\pi\in S_n} U_\pi \left( X^{\otimes p}\otimes Y^{\otimes q}\otimes Z^{\otimes r}\otimes I^{\otimes(n-p-q-r)} \right) U_\pi^\dagger, \label{eq:collective_orbit_average} \end{equation}
where $p,q,r\geq 0$ and $p+q+r\leq n$. The corresponding skew-Hermitian orbit element in the convention of Ref.\cite{barligea2026enabling} is $iA_{p,q,r}$. Equivalently, if $\mathcal{O}_{p,q,r}$ is the set of distinct Pauli strings with these counts, then
\begin{equation} A_{p,q,r} = \frac{1}{M_{p,q,r}} \sum_{P\in\mathcal{O}_{p,q,r}}P, \qquad M_{p,q,r} = \frac{n!}{p!\,q!\,r!\,(n-p-q-r)!}.
\end{equation}
Thus, the convention is an average over distinct strings. Distinct orbit averages are Hilbert--Schmidt orthogonal, with
\begin{equation} \operatorname{Tr} \left(A_{p,q,r}^\dagger A_{p',q',r'}\right) = \delta_{(p,q,r),(p',q',r')} \frac{2^n}{M_{p,q,r}}.
\end{equation}
They are therefore not generally unit-normalized. Excluding the identity orbit $A_{0,0,0}$, these operators span the traceless permutation-invariant Hermitian operator space
\begin{equation} \mathcal{V}_n = \operatorname{span}_{\mathbb{R}} \left\{ A_{p,q,r}: 1\leq p+q+r\leq n \right\}, \qquad d=\dim(\mathcal{V}_n)=\binom{n+3}{3}-1.
\end{equation}
The working dimensions are consequently $d=83$ for $n=6$ and $d=164$ for $n=8$. Here $d$ denotes the ambient operator-space dimension, not necessarily the dimension of the DLA generated by the selected circuit generators. We obtain the expectation values of the complete nonidentity orbit basis on the intermediate state prepared by $U_q$. The resulting $d$-component expectation vector is used for classical propagation and gradient evaluation. Thus, $d$ determines the logical intermediate-state measurement count in this implementation. The ambient space is invariant under conjugation by permutation-invariant unitaries, allowing the relevant observables to be propagated within this space.

\paragraph{Benchmark Hamiltonian.} The six- and eight-qubit instances use the fully connected collective-spin Hamiltonian \begin{equation} \begin{aligned} H ={}& cI +\frac{1}{2}\sum_{i=1}^{n} \left(h_xX_i+h_yY_i+h_zZ_i\right)\\ &+\frac{J_{xy}}{2} \sum_{1\leq i<j\leq n} \left(X_iY_j+Y_iX_j\right)\\ &+\frac{J_{zz}}{2} \sum_{1\leq i<j\leq n}Z_iZ_j. \end{aligned} \label{eq:collective_benchmark_hamiltonian} \end{equation} For each instance, the five coefficients $h_x,h_y,h_z,J_{xy},J_{zz}$ are sampled independently and uniformly from $[-1,1]$. Each sampled coefficient is shared across all sites or pairs in its corresponding sum, preserving permutation invariance. The scalar $c$ is an identity-energy offset.

In the orbit-average convention defined above, the same Hamiltonian is \begin{equation} \begin{aligned} H ={}& cI +\frac{n}{2} \left( h_xA_{1,0,0} +h_yA_{0,1,0} +h_zA_{0,0,1} \right)\\ &+\frac{J_{xy}n(n-1)}{2}A_{1,1,0} +\frac{J_{zz}n(n-1)}{4}A_{0,0,2}. \end{aligned} \label{eq:collective_benchmark_orbit_expansion} \end{equation}

\paragraph{Structured circuit block.} The $U_{\mathfrak{g}}$ block uses five Hermitian orbit generators,
\begin{equation} \mathcal{G} = \left\{ A_{0,0,1}, A_{0,0,2}, A_{0,1,0}, A_{1,0,0}, A_{1,1,0} \right\}, \end{equation}
through parameterized rotations generated by these operators. We do not use every element of the ambient orbit basis as an independently parameterized gate. The block therefore has $m=5$ trainable parameters, whereas the classical propagation uses a working operator space of dimension $d$. The Lie algebra generated by the selected gates is \begin{equation} \mathfrak{g}_{\mathrm{sel}} = \operatorname{Lie} \left\{ iA:A\in\mathcal{G} \right\} \subseteq i\mathcal{V}_n. \end{equation}

\paragraph{Training protocol.} We compare Full-PSR, Alternate, and Alternate followed by Simultaneous training (Alt+Sim) for $n=6$ and $n=8$, using one, three, six, and nine YZ-linear layers in $U_q$. Each configuration uses 64 trials and a success threshold of $10^{-2}$. Full-PSR uses 1000 joint parameter updates. Alternate uses 1000 Alternate iterations, each comprising a $U_q$ update followed by a $U_{\mathfrak{g}}$ update. Alt+Sim uses 500 Alternate iterations followed by 500 Simultaneous iterations. For the classical branch, the required orbit-observable expectation values are evaluated on the intermediate state prepared by $U_q$. The subsequent action of $U_{\mathfrak{g}}$ and the derivatives with respect to its parameters are evaluated in the orbit representation. The parameters are initialized randomly and the ground-state energy is obtained by exact diagonalization. The optimizer settings are kept the same as the earlier numerics.

\paragraph{Logical resource accounting.}
For $L$ YZ-linear layers, the hardware-efficient block $U_q$ contains $p=2nL$ trainable rotation parameters. Its gradient therefore requires $2p=4nL$ two-sided parameter-shift cost evaluations per optimization step. The unshifted forward cost evaluation is excluded from this accounting. For the five collective generators in $U_{\mathfrak{g}}$, let $s_a$ denote the number of shifted cost evaluations used for the derivative of parameter $a$. The generalized shift counts depend on the spectra of the individual orbit generators and the number of unique positive eigenvalue gaps~\cite{wierichs2022general}. The implemented generalized parameter-shift rules require \begin{equation} S_g(n):=\sum_{a=1}^{5}s_a = \begin{cases} 146, & n=6,\\ 284, & n=8. \end{cases} \end{equation} Consequently, the Full-PSR count per optimization step is \begin{equation} N_{\mathrm{Full\text{-}PSR}}(n,L) = 4nL+S_g(n). \end{equation} The hybrid procedures instead obtain the $d$ orbit expectation values required for classical propagation of the structured block, giving \begin{equation} N_{\mathrm{hybrid}}(n,L) = 4nL+d, \qquad d=\binom{n+3}{3}-1. \end{equation} The resulting per-step counts are summarized below. \begin{center} \begin{tabular}{ccrrr} \hline $n$ & $L$ & $U_q$ shifts & Full-PSR & Hybrid\\ \hline 6 & 1 & 24 & 170 & 107\\ 6 & 3 & 72 & 218 & 155\\ 6 & 6 & 144 & 290 & 227\\ 6 & 9 & 216 & 362 & 299\\ \hline 8 & 1 & 32 & 316 & 196\\ 8 & 3 & 96 & 380 & 260\\ 8 & 6 & 192 & 476 & 356\\ 8 & 9 & 288 & 572 & 452\\ \hline \end{tabular} \end{center}

 Counts are accumulated up to each trial's best recorded iterate. Each shifted cost evaluation and each intermediate-state orbit expectation is counted as one logical evaluation.

Alt+Sim matches or exceeds the Full-PSR success rate in all eight qubit-depth settings and strictly improves it in six. The largest increase occurs for three layers and eight qubits, where success rises from $68.75\%$ for Full-PSR to $90.62\%$ for Alt+Sim. At nine layers, the success-filtered median relative errors for Alt+Sim are $8.83\times10^{-4}$ and $5.55\times10^{-4}$ for six and eight qubits, compared with $2.29\times10^{-3}$ and $1.88\times10^{-3}$ for Full-PSR. The QPU Calls reduction for Alt+Sim ranges from $17.5\%$ to $38.0\%$ for successful trials. It should be noted that translating these logical-observable QPU-call estimates into hardware-level measurement cost requires a specified measurement basis and grouping strategy, which we ignore here. In particular, estimating an orbit average need not correspond to a single physical measurement setting.

\begin{figure*}[tbh!]
    \centering
    \begin{subfigure}[t]{0.49\linewidth}
        \centering
        \includegraphics[width=1\linewidth]{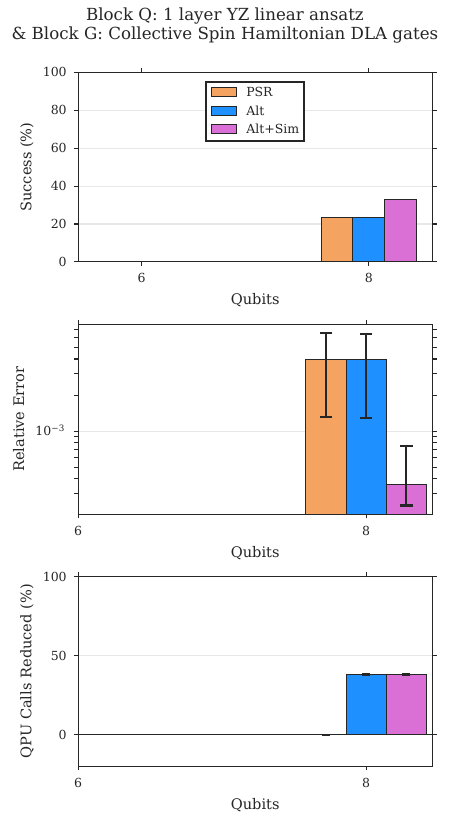}
        \caption{One layer in $U_q$.}
        \label{collective_spin_figure_one}
    \end{subfigure}\hfill
    \begin{subfigure}[t]{0.49\linewidth}
        \centering
        \includegraphics[width=1\linewidth]{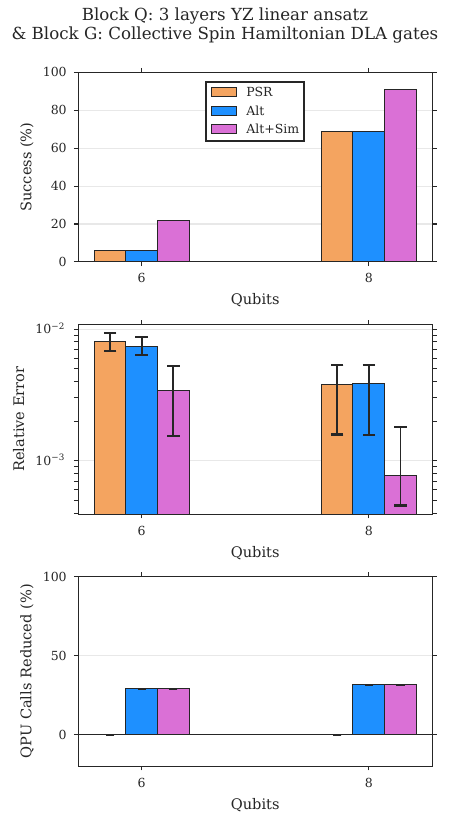}
        \caption{Three layers in $U_q$.}
        \label{collective_spin_figure_three}
    \end{subfigure}
    \caption{Collective-spin orbit-algebra benchmark at one and three layers for six and eight qubits, with 64 trials per setting and success threshold $10^{-2}$. The rows show the success fraction, success-filtered median relative error with Q25-Q75 intervals, and QPU Calls reduction relative to Full-PSR for successful trials. Reductions are accumulated to each successful trial's best iterate; error bars show the corresponding standard deviation. Missing bars indicate that none of the runs reached below the success threshold.}
    \label{collective_spin_figures_shallow}
\end{figure*}

\begin{figure*}[tbh!]
    \centering
    \begin{subfigure}[t]{0.49\linewidth}
        \centering
        \includegraphics[width=1\linewidth]{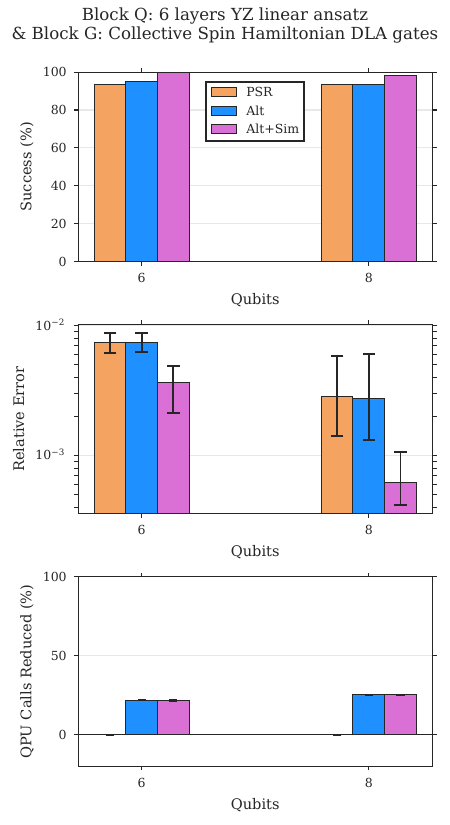}
        \caption{Six layers in $U_q$.}
        \label{collective_spin_figure_six}
    \end{subfigure}\hfill
    \begin{subfigure}[t]{0.49\linewidth}
        \centering
        \includegraphics[width=1\linewidth]{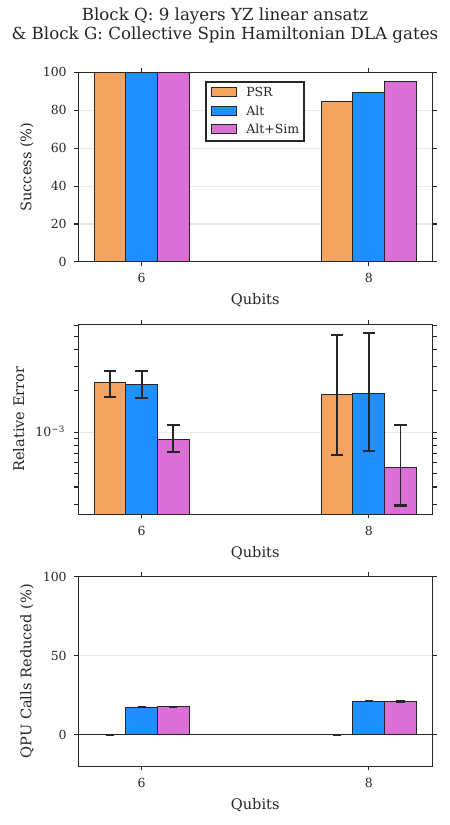}
        \caption{Nine layers in $U_q$.}
        \label{collective_spin_figure_nine}
    \end{subfigure}
    \caption{Collective-spin orbit-algebra benchmark at six and nine layers using the protocol and metrics defined in Fig.~\ref{collective_spin_figures_shallow}.}
    \label{collective_spin_figures_deep}
\end{figure*}

This benchmark establishes portability of the hybrid update procedure beyond free-fermionic \g constructions. It does not establish classical hardness of ground-state estimation: the fully permutation-invariant Hamiltonian can be decomposed into total-spin sectors of polynomial dimension. Instead, the benchmark shows that efficient symmetry-adapted observable propagation can support the proposed hybrid training procedure and reduce logical QPU evaluations in a non-free-fermionic setting. It does not establish quantum advantage, hardware-level measurement savings, or favorable end-to-end scaling.
\begin{resultbox}{Symmetry-adapted extension}
Alt+Sim matches or exceeds Full-PSR success in all eight tested configurations and reduces logical QPU calls by $17.5\%$--$38.0\%$. This demonstrates finite-size portability of HELIA beyond free-fermionic \g constructions. It does not imply that the permutation-invariant ground-state problem is classically hard, nor does it establish physical measurement savings, favorable asymptotic scaling, hardware performance, or quantum advantage.
\end{resultbox}
\endgroup

\section{Experimental details of VQE on LiH and Classification task}\label{liH_class_app}
\textbf{LiH Experimental details}: $U_q$ is composed of 6 layers of YZ linear ansatz, while $U_{\mathfrak{g}}$ is generated from gates from the above DLA, and the initial state $\ket{0}^{\otimes n}$. In the shown examples, we run Alternate optimization for 250 iterations and Simultaneous optimization for 100 iterations with the reduced Hamiltonian, followed by 200 iterations of $U_q$ and 900 iterations of full circuit training. Similarly for Full-PSR training, we run 1450 iterations of PSR to match the number of total iterations in both methods. These ad hoc choices can potentially be further optimized to reduce resources and improve accuracy.

 \textbf{Classification Experiment Details:} We construct HELIA with 9 YZ linear layers in $U_q$. $U_{\mathfrak{g}}$ is constructed from one of the 3 DLA choices mentioned above, and we repeat the experiment for each of the choices. The measurement operator is chosen randomly from the corresponding DLA and fixed at the beginning of the experiment. For each of these configurations of DLA, we run 20 independent trials and compare test accuracy.

\end{document}